\documentclass{article}
\usepackage{amssymb}
\usepackage{graphicx}
\usepackage{xcolor}
\usepackage{amsfonts}
\def\ligne#1{\hbox to\hsize{#1}}
\def\leurre{\noindent\leftskip0pt\small\baselineskip 10pt}
\newtheorem{thm}{\textbf{Theorem}}
\newtheorem{fig}{\textbf{Figure}}
\newtheorem{tab}{\textbf{Table}}

\author{Maurice {\sc Margenstern}}
\title{A strongly universal cellular automaton on the pentagrid with six states and
Moore neighbourhood.}
\begin{document}
\maketitle

\begin{abstract}
In this paper, we prove that there is a strongly universal cellular automaton on the 
pentagrid with six states. The rules of the cellular automaton are rotation invariant and
the cellular automaton works with a Moore neighbourhood for each cell. The cellular automaton
requires $1072$ rules.
\end{abstract}

\section{Introduction}~\label{intro}

    In many papers, the author studied the possibility to construct universal cellular
automata in tilings of the hyperbolic plane, a few ones in the hyperbolic $3D$ space. 
Most often, the constructed cellular automaton was weakly universal. By {\it weakly 
universal}, we mean that the automaton is able to simulate a universal device starting
from an infinite initial configuration. However, the initial configuration should not be
arbitrary. It was the case that it was periodic outside a large enough circle, in fact
it was periodic outside such a circle in two different directions as far as the simulated
device was a two-registered machine. In almost all papers, the considered tiling of the
hyperbolic plane was either the pentagrid or the heptagrid, {\it i.e.} the tessellation
$\{5,4\}$, $\{7,3\}$ respectively. Both tessellations live in the hyperbolic plane only.
In the pentagrid, the basic tile is a regular convex pentagon with right angles. In the
heptagrid, it is a regular convex heptagon with the angle 
\hbox{$\displaystyle{{2\pi}\over3}$} between consecutive sides. Below, Figure~\ref{penta}
provides us with a representation of the pentagrid in the Poincar\'e's disc model of the
hyperbolic plane.

   In the left-hand side picture, we can see five tiles which are counter-clockwise
numbered from~1 up to~5, those tiles being the neighbours of a tile which we call
the {\bf central tile} for convenience. Indeed, there is no central tile in the heptagrid
as there is no central point in the hyperbolic plane. We can see the disc model as a
window over the hyperbolic plane, as if we were flying over that plane in an abstract
spacecraft. The centre of the circle is the point on which our attention is focused while
the circle itself is our horizon. Accordingly, the central tile is the tile which is 
central with respect to the area under our consideration. The same picture also shows us 
five arcs of circle issued from a vertex of the central tile, continuing a side of that tile. 
Those rays delimit a right angle~$S$, and we call {\bf sector headed by~$1$} the set of 
tiles whose centre are inside~$S$, and tile~1 is called the {\bf head} of that sector. Similarly, 
we can define sectors $i$ with \hbox{$i\in\{2..5\}$} each of them having the tile~$i$ as its head.
\vskip 10pt
\vtop{
\ligne{\hfill
\includegraphics[scale=0.45]{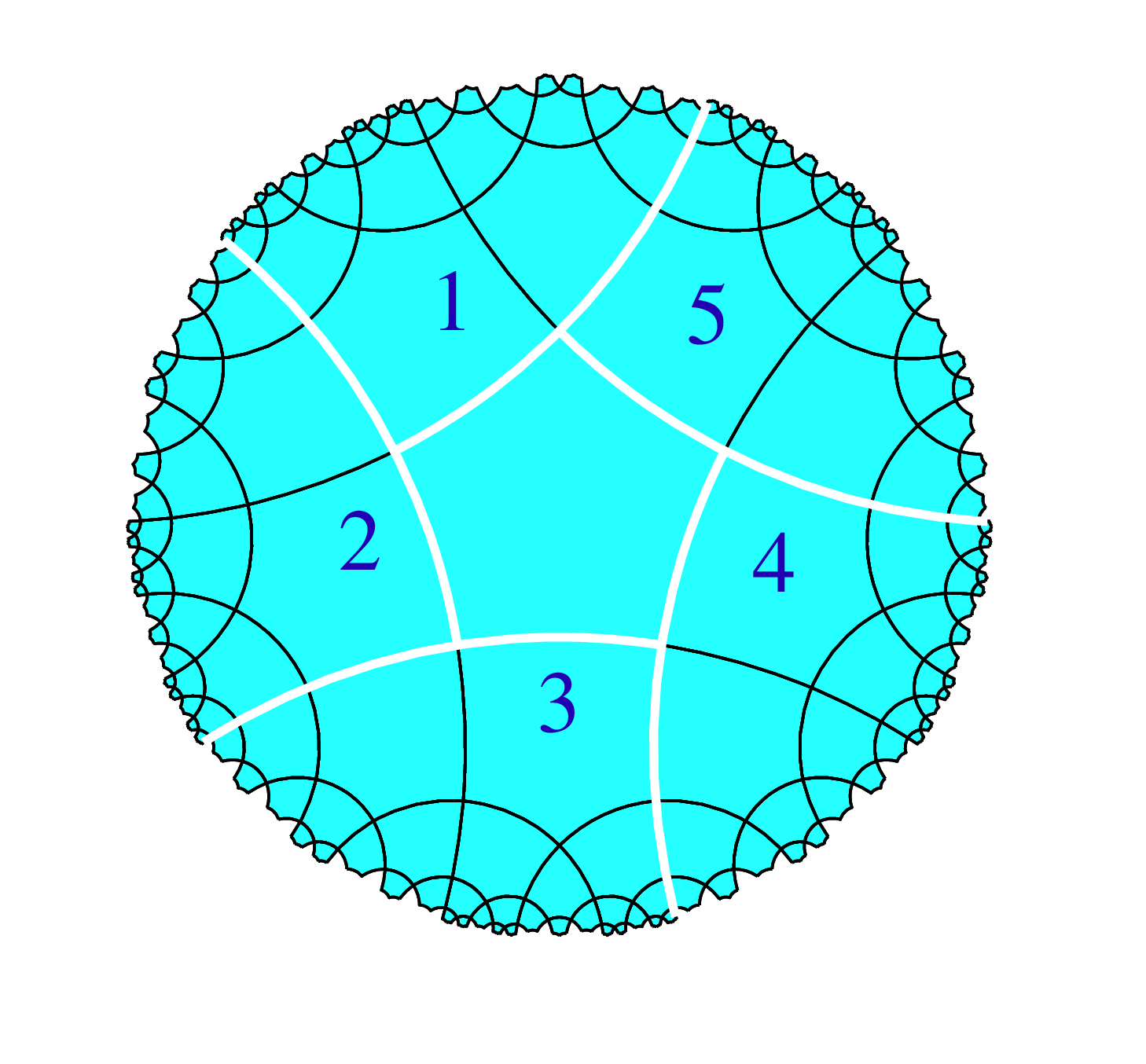}
\hskip-20pt
\includegraphics[scale=0.45]{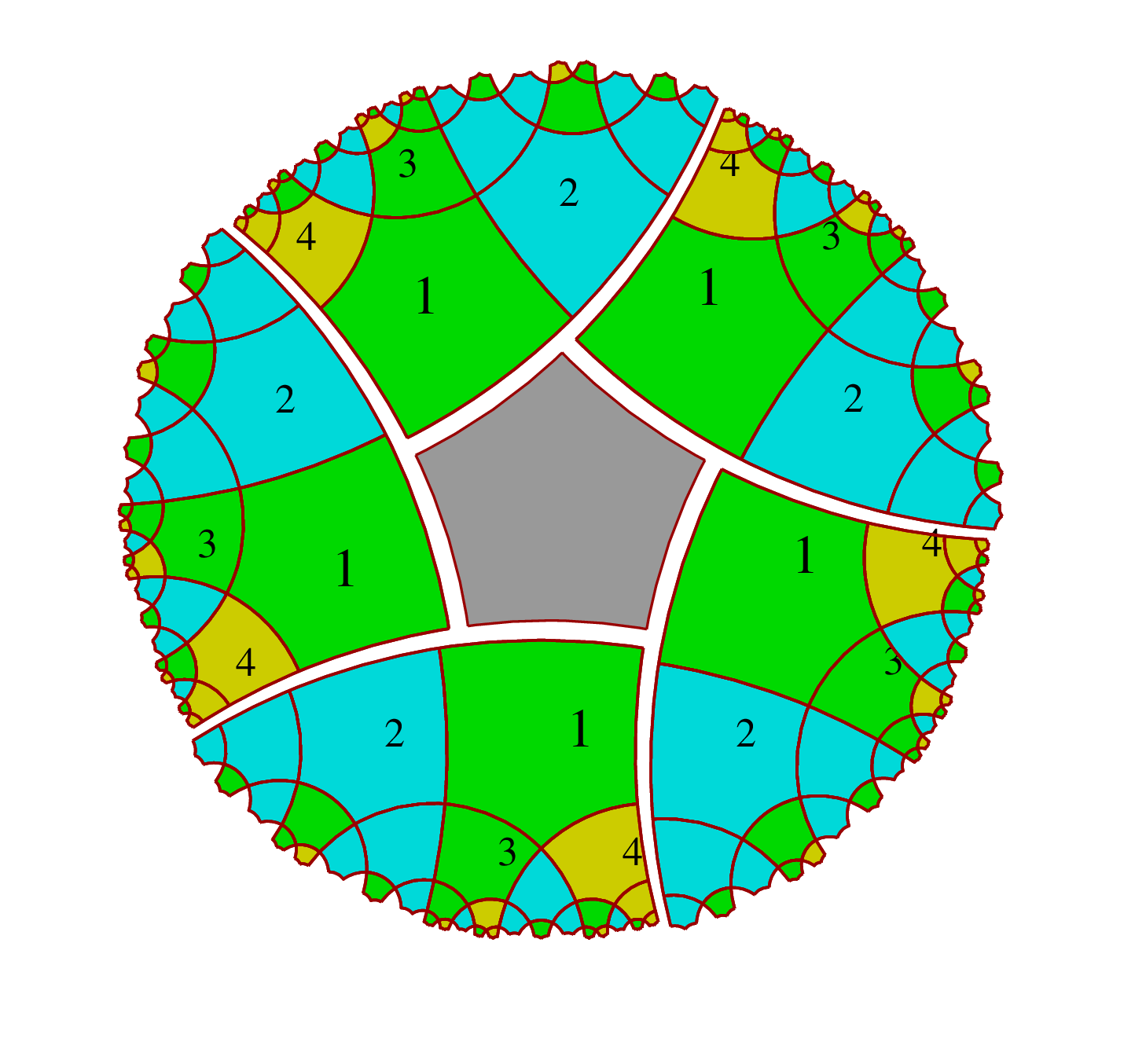} 
\hfill}
\vspace{-15pt}
\begin{fig}\label{penta}
\leurre
To left: the pentagrid; to right: showing the tree structure of the sectors around a central tile
indicated by the left-hand side picture.
\end{fig}
}

   In the right-hand side picture, we can see the central tile and the five sectors
which surround it: they are defined by the rays shown on the left-hand side
picture of Figure~\ref{penta}. 
In each sector we can see four tiles numbered from 1 up to 4: tile~1 is the head of the sector 
while tiles~2, 3 and~4 are the sons of tile~1, constituting the first level in the sector.
In each sector we have three kind of tiles: green, yellow and blue ones. More precisely,
the tiles of a sector are coloured with three colours: green, blue and yellow. Blue tiles have 
two neighbours which are blue and green in that order and green and yellow tiles have three
neighbours which are blue, green and yellow in that order, turning counter clockwise around the 
tile. These indications define rules
from which we can build a tree which spans the tiles of the sector. Also the numbering of the tiles
can be continued level after level and from left to right on each level. More indication
on that topic and on the way to locate the tiles can be found in~\cite{mmbook1,mmJUCS}.

   Now that the global setting is given, we shall proceed as follows: 
Section~\ref{scenario} indicates the main lines of the implementation which is precisely
described in Subsection~\ref{newrailway}. At last, Section~\ref{srules} gives us the rules
followed by the automaton. 
The figures illustrating the various configurations considered in Subsection~\ref{newrailway} 
were established from pieces of figures drawn by a computer program which applied the rules of 
the automaton displayed by Table~\ref{t_rules} to an appropriate window in each of the above 
mentioned configurations. The computer program also checked that the set of rules is coherent.

   That allows us to prove the following property:

\begin{thm}\label{letheo}
There is a strongly universal cellular automaton on the pentagrid with six states whose rules are
rotation invariant. The cellular automaton is truly planar and for each cell the neighbourhood is
considered in Moore sense. The automaton makes use of $1072$ rules.
\end{thm}

   The number of rules of the automaton is much higher than those for cellular automata
on the heptagrid. That number is also a consequence of the Moore neighbourhood used by the cells.
We remind the reader that the Moore neighbourhood of a cell~$c$ consists of the cells whose support
shares at least a vertex with the support of~$c$. Accordingly a Moore neighbourhood consists of
10 neighbours instead of 5 of them in an ordinary neighbourhood. If needed we distinguish the
{\bf full} neighbours of~$c$, those which share a side with~$c$, from the {\bf partial} neighbours
which share a vertex only with~$c$.

\section{Main lines of the computation}\label{scenario}

   The first paper about a universal cellular automaton in the pentagrid, the 
tessellation $\{5,4\}$ of the hyperbolic plane, was \cite{fhmmTCS}. This cellular 
automaton was also rotation invariant, at each step of the computation, the set of non 
quiescent states had infinitely many cycles: we shall say that it is a truly planar 
cellular automaton. That automaton had 22~states. That result was improved by a cellular 
automaton with 9~states in~\cite{mmysPPL}. Recently, it was improved with 5~states, 
see~\cite{mmpenta5st}. A bit later, I proved that in the heptagrid, the tessellation 
$\{7,3\}$ of the hyperbolic plane, there is a weakly universal cellular automaton with 
three states which is rotation invariant and which is truly planar, \cite{mmhepta3st}. 
Later, I improved the result down to two states but the rules are no more rotation 
invariant, see~\cite{mmpaper2st}. Paper \cite{JAC2010} constructs three cellular
automata which are strongly universal and rotation invariant: one in the pentagrid, one 
in the heptagrid, one in the tessellation \hbox{$\{5,3,4\}$} of the hyperbolic 
$3D$-space. By strongly universal we mean that the initial configuration is finite, 
{\it i.e.} it lies within a large enough circle.

    In the present paper, we borrow the ideas of~\cite{mmarXiv23v3}. 

    Our automaton simulates a two-registered machine which is enough to entail the strong 
universality as far as a two-registered machine can simulate any Turing machine as proved 
in~\cite{minsky}. In our simulation, each register consists of a finite sequence $\sigma$ of tiles
where each tile of~$\sigma$ is a full neighbour of two other tiles of~$\sigma$, two tiles 
of~$\sigma$ being excepted which have a single full neighbour in~$\sigma$. One of those
exceptional tiles is the beginning of the register, the other is its end. Moreover, the tiles 
of~$\sigma$ have a side which lie on a line~$\ell$ which is specific to~$\sigma$.
At each time of the computation, the register is a finite sequence of tiles. That sequence is 
appended one element to its end each time it is incremented by the application of an appropriate 
instruction of the register machine program. That appended element becomes the new end of the
register. Symmetrically, the sequence is removed the ending tile each time it is
decremented, with the exception of the case when the register is empty. The new end of $\sigma$
is then the neighbour in $\sigma$ of the removed tile.

    The simulation is based on the railway model devised in~\cite{stewart} revisited
by the implementations given in the author's paper~\cite{mmarXiv23v2}.
Sub-section~\ref{railway} describes the main structures of the model. In 
Sub-section~\ref{newrailway} we indicate the new features used in the present simulation.
Those new features, introduced in~\cite{mmarXiv21}, allowed us to reduce the number of auxiliary 
structures used to simulated the apparatus needed by the railway model as described 
in~\cite{stewart}.

\subsection{The railway model}\label{railway}

   The railway model of~\cite{stewart} lives in the Euclidean plane. It consists of
{\bf tracks}, {\bf crossings} and {\bf switches} and the configuration of all switches at time~$t$
defines the configuration of the computation at that time. There are three kinds of
switches, illustrated by Figure~\ref{switches}. The changes of the switch configurations
are performed by a locomotive which runs over the circuit defined by the tracks and their
connections organised by the switches.

A switch gathers three tracks $a$, $b$ and~$c$ at a point. In an active crossing,
the locomotive goes from~$a$ to either~$b$ or~$c$. In a passive crossing, it goes
either from~$b$ or~$c$ to~$a$. 

\vskip 10pt
\vtop{
\ligne{\hfill
\includegraphics[scale=0.8]{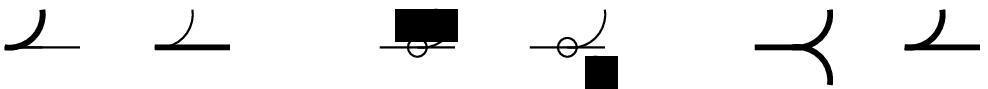}
\hfill}
\begin{fig}\label{switches}
\leurre
The switches used in the railway circuit of the model. To left, the fixed switch, in the
middle, the flip-flop switch, to right the memory switch. In the flip-flop switch, the 
bullet indicates which track has to be taken.
\end{fig}
}

In the fixed switch, the locomotive goes from~$a$ to always the same track: either always to~$b$ 
or always to~$c$. 
In any switch, the track followed by the locomotive in an active crossing is called the
{\bf selected} track. The passive crossing of the fixed switch is
possible. The flip-flop switch is always crossed actively only. The crossing of a locomotive entails
the change of its selected track. The memory switch can be crossed actively or passively. Now, the 
selected track is the track taken by the locomotive in the last passive crossing. 

   As an example, we give here the circuit which stores a one-bit unit of information,
see Figure~\ref{basicelem}. The locomotive may enter the circuit either through the 
gate~$R$ or through the gate~$W$.

  If it enters through the gate~$R$ where a memory switch sits, it goes either through
the track marked with~1 or through the track marked with~0. When it crossed the switch
through track~1, 0, it leaves the unit through the gate~$E_1$, $E_0$ respectively.
Note that on both ways, there are fixed switch sending the locomotive to the appropriate
gate~$E_i$. If the locomotive enters the unit through the gate~$W$, it is sent to the 
gate~$R$, either through track~0 or track~1 from~$W$. Accordingly, the locomotive
arrives to~$R$ where it crosses the switch passively, leaving the unit through the 
gate~$E$ thanks to a fixed switch leading to that latter gate. When the locomotive 
took track~0, 1 from~$W$, the switch after that selects track~1, 0 respectively and the 
locomotive arrives at~$R$ through track~1, 0 of~$R$. The track are numbered according to 
the value stored in the unit. By definition, the unit is~0, 1 when both tracks from~$W$ 
and from~$R$ are~0, 1 respectively. So that, as seen from that study, the entry 
through~$R$ performs a reading of the unit while the entry through~$W$, changes the unit
from~0 to~1 or from~1 to~0: the entry through~$W$ should be used when it is needed to
change the content of the unit and only in that case. The structure works like a memory
which can be read or rewritten. It is the reason why call it the {\bf one-bit memory}.

\vtop{
\ligne{\hfill
\includegraphics[scale=0.6]{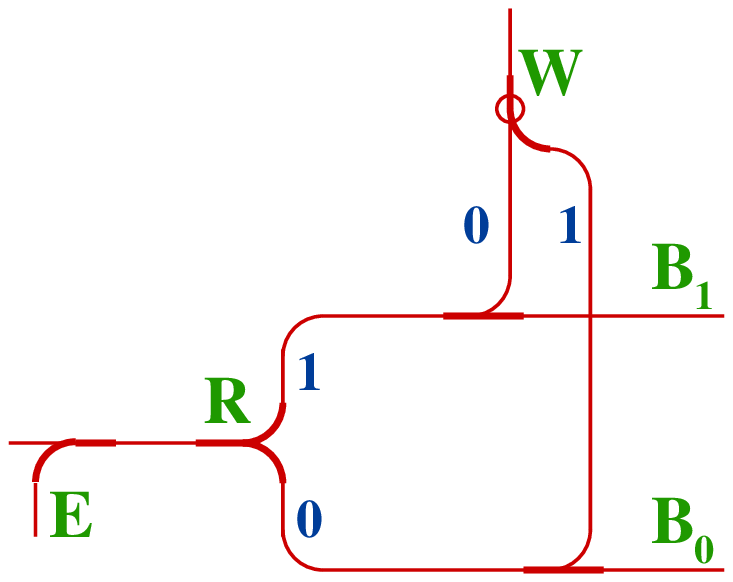}
\hfill}
\begin{fig}\label{basicelem}
\leurre
The basic element containing one bit of information.
\end{fig}
}

   We shall see how to combine one-bit memories in the next sub-section as far as we 
introduce several changes to the original setting for the reasons we indicate there.

\subsection{Tuning the railway model}\label{newrailway}

   We first look at the implementation of the tracks in Sub-subsection~\ref{sbbtracks}
and how it is possible to define the crossing of two tracks.
In Sub-subsection~\ref{sbbswitch} we see how the switches are implemented.
Then, in 
Sub-subsection~\ref{sbbunit}, we see how the one-bit memory is implemented in the new 
context and then, in Sub-section~\ref{sbbregdisc}, how we use it in various places. 

\subsubsection{The tracks}\label{sbbtracks}

    The tracks play a key role in the computation, as important as instructions and 
registers: indeed, they convey information without which any computation is impossible.
   
    As in~\cite{mmarXiv23v1}, the tracks are one-way. But as in~\cite{fhmmTCS} we define 
the tracks as cells with a particular colour. However, as far as the locomotive goes from the
program to the registers and from there back to the program, it should be possible for
the locomotive to move in opposite directions in some sense. But we keep to the one way constraint
as far as we impose a passive crossing for the fixed switch. As in~\cite{mmarXiv21}, the
memory switch is split into two different structures: one of them for an active passage, the other
for the passive passage.

   Although the rules are rotation invariant a word should be said about the way to
read the states of the neighbourhood. The problem is the numbering of the neighbours of a cell.
In the Euclidean case it is assumed that the north, south, east and west directions are defined
without further consideration. There are no such directions in the hyperbolic plane. However, the
'natural' structure of that plane is a tree, namely, the Fibonacci tree mentioned in 
Subsection~\ref{intro} and illustrated by the right-hand side part of Figure~\ref{penta}.
Consequently, we assume that each cell knows its father in the tree spanning the sector where it
lies. By convention, we say that the father is neighbour~1 and the others are increasingly 
numbered while counterclockwise turning around the cell from the father. There is some freedom
for cell~0: it has no father as far as its neighbours are the cell~1{} in each of the sectors
defined by those neighbours. We assume that for each cell, the neighbourhood is read in that way 
around the cell starting from neighbour~1. In Section~\ref{srules} we give more details about the
format of the rules.

As far as the way of reading the neighbours is fixed we can now indicate how the tracks are
defined. The increasing and decreasing of the registers must be along particularly convenient
paths. The only 'natural' way of following a track is, in our setting to follow a father or,
alternatively, once the father is known, the same son. We decide to have two kinds of tracks:
one kind we call {\bf verticals} are defined as follows: for each cell, its previous neighbour
of the track is its neighbour~1 and its next neighbour in the track is its neighbour~4. The
second kind of tracks are what we call {\bf horizontals}. In a sector, a horizontal is the set
of cells which are at the same distance from the head of the tree. There is a simple way to 
define a horizontal in a local way. Besides its numbering, it is 'natural' to assume that a
cell in a Fibonacci tree, knows the status of its supporting tile: either the tile is a white node 
of the tree, or its a black one. In terms of Subsection~\ref{intro}, green and yellow tiles are said
white while blue ones are said black. In a white tile $\tau$, the horizontal passing through 
$\tau$ connects its neighbours~2 and~5. In a black tile $\sigma$, the horizontal passing through 
$\sigma$ connects its neighbours~3 and~5. 

Consider four tiles pairwise adjacent, 
the third one being not in contact with the first one. The motion from the first tile to 
the third one can be symbolized as follows:
\def\ftt {\footnotesize\tt}
\vskip 5pt
\ligne{\hfill\ftt FWWW\hskip 30pt RFWW\hskip 30pt WRFW\hskip 30pt WWRF\hskip 30pt WWWR \hfill}

\noindent
and the reverse motion is given by:
\vskip 5pt
\ligne{\hfill\ftt WWWF\hskip 30pt WWFR\hskip 30pt WFRW\hskip 30pt FRWW\hskip 30pt RWWW\hfill}
\vskip 5pt
\def\FF{{\tt F}}
\def\RR{{\tt R}}
From that observation, we can deduce that a uniform colour for the tracks can accept a 
single direction, although it could accept both of them as far as the locomotive consists of two
contiguous cells, the front \FF{} and the rear \RR{}.


Consider two tiles~$U$ and~$V$. A {\bf path} from~$U$ to~$V$ is a finite
sequence of tiles \hbox{$\{T_i\}_{i\in[0..n]}$}, where \hbox{$T_0=U$}, \hbox{$T_n=V$}
and, for any $i$ in \hbox{$[1..n$$-$$1]$}, $T_i$ and $T_{i+1}$ share a common side.
In that case, $n$$-$1 is the {\bf length} of the path. The {\bf distance} from~$U$ to~$V$
is the smallest length of the paths from~$U$ to~$V$. A {\bf circle} around~$T$ of
radius~$n$ is the set of tiles~$U$ whose distance to~$T$ is~$n$. An arc of circle is
a path from a tile~$U$ of a circle~$\mathcal C$ to another tile~$V$ of~$\mathcal C$,
whose all tiles also belong to~$\mathcal C$. 
\def\BB{{\tt B}}
\def\GG{{\tt G}}
\def\YY{{\tt Y}}
\def\WW{{\tt W}}
\def\MM{{\tt M}}
\vskip 10pt
\vtop{
\ligne{\hfill
\includegraphics[scale=0.3]{secteurs_5_4.ps}
\hfill
\raise 20pt\hbox{\includegraphics[scale=0.4]{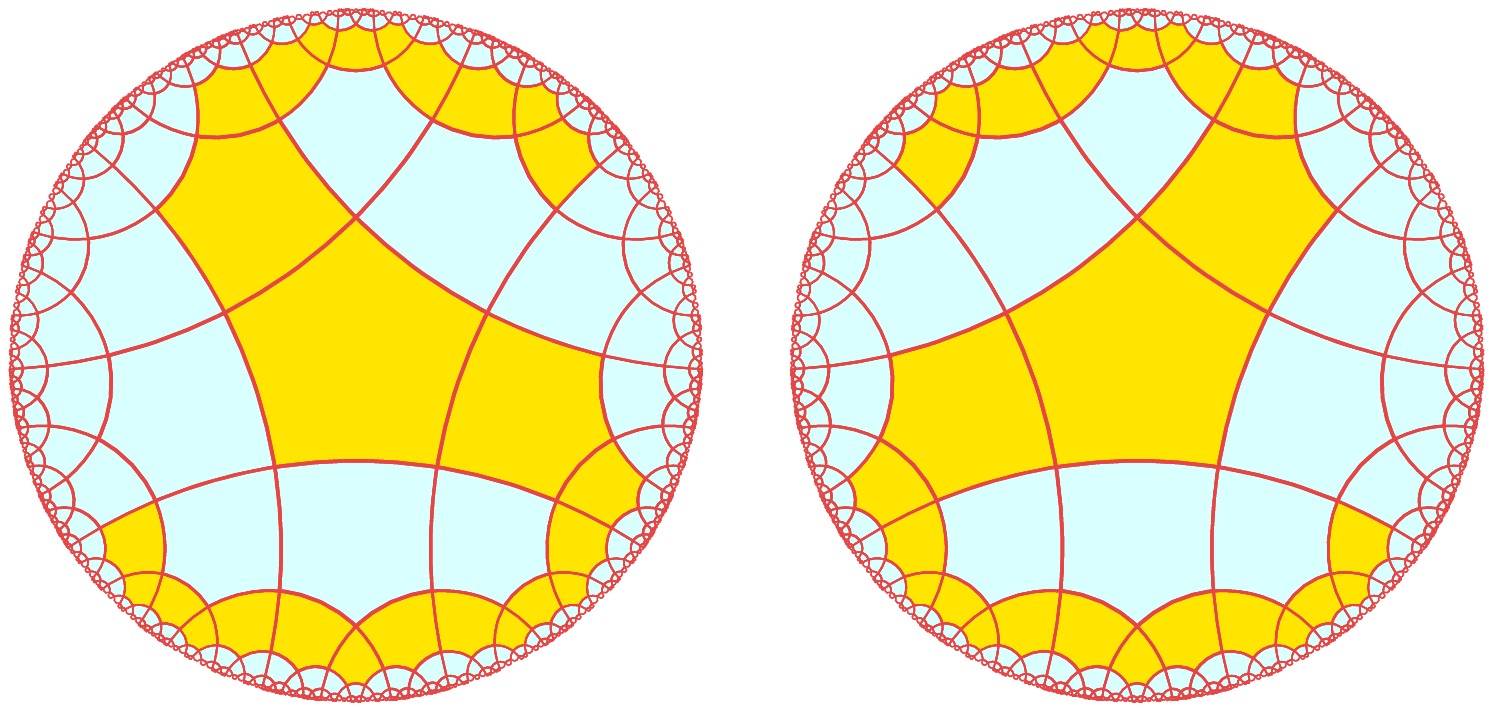}}
\hfill}
\vspace{-20pt}
\begin{fig}\label{f_voies}
\leurre
To right, the tracks used for testing the implementation. To left, the sectors in order to locate
the tiles of the tracks.
\end{fig}
}
\vskip 10pt
The single colour allows us to realize the required
circuit as far as horizontals and verticals constitute a kind of grid. It is especially the case if
the whole configuration lies in a single sector, its closest cell to the origin of the coordinates
is far enough for implementing the configurations we describe in Subsection~\ref{newrailway}.

\vskip 10pt
\vtop{
\ligne{\hfill
\includegraphics[scale=0.35]{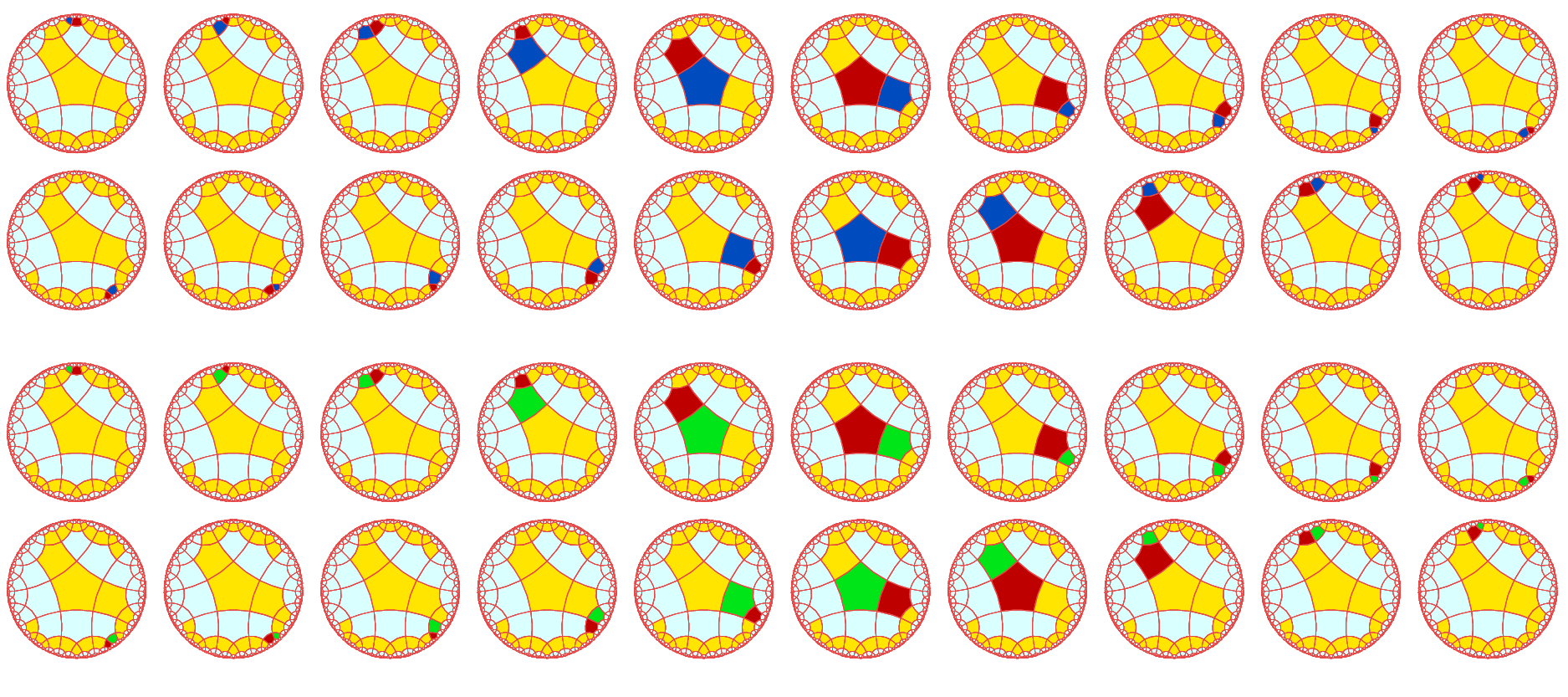}
\hfill}
\begin{fig}\label{f_voie_ess}
\leurre
Motion of the locomotives on a track which tests verticals and horizontals.
Top two rows: the \BB-locomotive, first row from top to bottom; second row, from bottom to top.
Bottom two rows, the \GG-locomotive, first from left to bottom and then from bottom to top.
\end{fig}
}

The colour of the tracks is fixed as \YY. However we shall indicate a track with respect to the 
colour of the locomotive which runs on that track. A \BB-, \GG-track is run by a locomotive whose
front is \BB-, \GG- respectively. We also say \BB-, \GG-locomotives respectively. Paths from a part
of the circuit to another part consists of pieces of verticals and horizontals connected together. 
On most parts of the circuit, the locomotive is \BB-. Later, we shall say that \BB- and \GG- are 
{\bf opposite colours}.

\vskip 10pt
\vtop{
\ligne{\hfill
\includegraphics[scale=0.35]{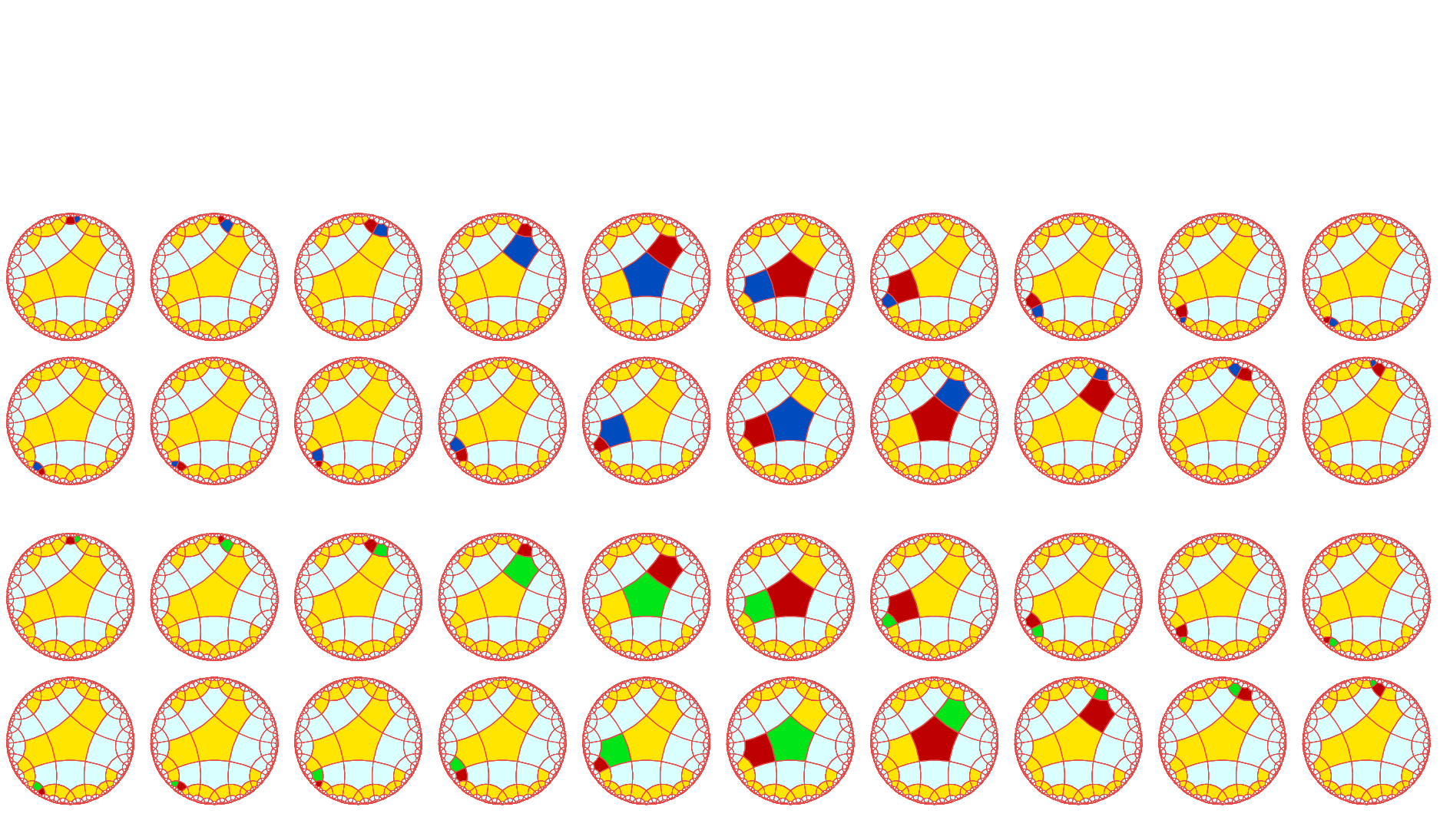}
\hfill}
\begin{fig}\label{f_voie_zed}
\leurre
Motion of the locomotives on a track which tests verticals and horizontals.
Top two rows: the \BB-locomotive, first row from top to bottom; second row, from bottom to top.
Bottom two rows, the \GG-locomotive, first from top to bottom and then from bottom to top.
\end{fig}
}

Figure~\ref{f_voies} represents two combinations of a vertical line with two horizontal lines.
In the left-hand side picture, the combination constitutes an 'S' which also allows us to test two
kinds of junctions of a vertical with a horizontal: the 'S' allows us to test a top left corner
together with a bottom right corner. The right-hand side picture of Figure~\ref{f_voies} represents
another combination which constitutes a 'Z'. That picture allow us to test the remaining
corners: top right one and bottom left one.
Use Figure~\ref{penta} in order to locate the tiles of the tracks. For the 'S' configuration, we
have from top to bottom, the following tiles; horizontal part in the top: (5,$i$) for $i$ ranging 
in \hbox{$\{$3,10,4$\}$} and (1,$i$) for $i$ in \hbox{$\{$5,15,6,18,7$\}$}; then the vertical part:
(1,3), (1,1), 0, (4,1) and (4,3). And last, the horizontal part in bottom: (4,$i$) for $i$ in
\hbox{$\{$7,18,6,15,5$\}$} and (3,$i$) with $i$ in \hbox{$\{$4,10,3,7,18,6,15,3$\}$}.
For the 'z' configuration, the tiles are, respectively: (1,$i$) for $i$ in 
\hbox {$\{$4,10,3,7,18,6,6,15,5$\}$}, then (5,4), (5,1), 0, (2,1) and (2,4) and, at last
(3,$i$) with $i$ in \hbox{$\{$5,15,6,18,73,10,4$\}$} and (4,$i$) with $i$ in 
\hbox{$\{$5,15,6,18,7$\}$}.

The locomotives, both \BB- and \GG- can run on those tracks
from (5,3) down to (3,5) in the 'S' and also in the reverse order, from (3,5) up to (5,3). For the
'Z', the run occurs from (1,4) down to (4,7) and also in the reverse order. Those motion have been 
tested by a computer program and it is illustrated by Figures~\ref{f_voie_ess}, \ref{f_voie_zed}
for the 'S'-, 'Z'-configuration respectively. As already noticed, those motions also test the 
various junctions of a vertical track with a horizontal one.

   We now turn to Sub-subsection~\ref{sbbauxil} where we describe the auxiliary structures
and the passive fixed switch used to implement the crossings and then the remaining switches whose 
implementation is discussed in Sub-subsection~\ref{sbbswitch}.

\subsubsection{Auxiliary structures and crossings}\label{sbbauxil}

The auxiliary structures we need are the fork, the changer and the filter. The changer and the 
filter are implemented in two versions : a \BB- and a \GG-one. The \BB, \GG-changer transforms a
\GG-, \BB-locomotive respectively into a \BB-, \GG-one respectively. A filter has a colour, either
\BB{} or \GG. It let the locomotive of the same colour as its one and it destroys a locomotive
having an opposite colour to its own one.

The figure illustrates the {\bf idle configuration} of those structure. We define an idle 
configuration of a structure to be a structure contained in a disc of radius~2 around the central
tile in which disc there is no locomotive.
\vskip 10pt
\vtop{
\ligne{\hfill
\includegraphics[scale=0.4]{secteurs_5_4.ps}
\raise 90pt
\hbox{
\includegraphics[scale=0.45]{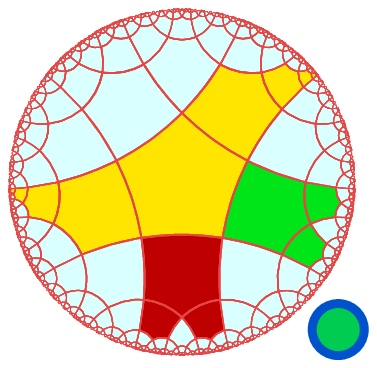}
\includegraphics[scale=0.45]{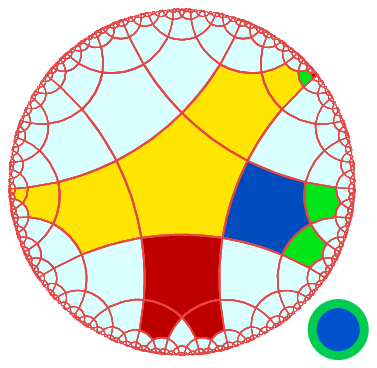}
}
\hskip -115pt\raise 30pt
\hbox{
\includegraphics[scale=0.45]{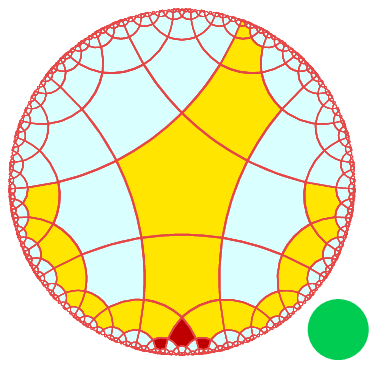}
\includegraphics[scale=0.45]{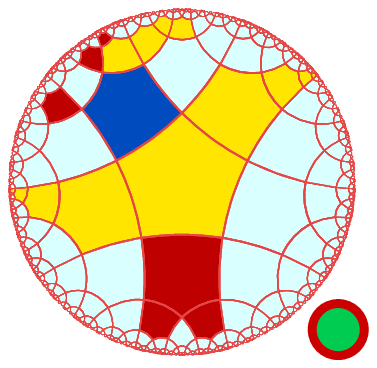}
\hskip-5pt
\includegraphics[scale=0.45]{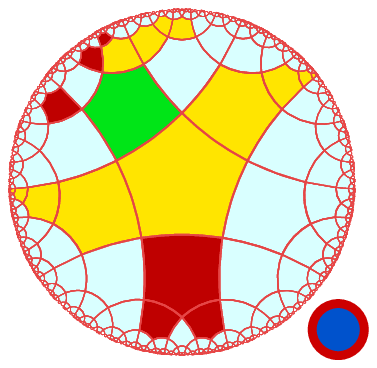}
}
\hfill}
\begin{fig}\label{f_auxil}
\leurre
To left, the sectors. To right, top line, the changer from \BB- to \GG-, the changer from \GG- 
to \BB. To right, bottom line, from left to right: the fork, the filter for \BB-, the filter for 
\GG-. With all pictures, the symbol which will represent them in further diagrams.
\end{fig}
}
\vskip 10pt

In the changers, a track crosses the figure through tiles (5,21), (5,8), (5,3), (5,1), 0, (3,1),
(3,3), (3,8) and (3,21). In (4,1) we have a \BB- or a \GG-tile and in (4,3) and (4,4) we have a
\GG-tile which fixes the tile in (4,1). We also have an \RR-tile at (3,1) with \RR-tiles too
at (3,3) and (3,4) to fix the tile in (3,1). In the filter, we have the same track and another one
with tiles (1,$i$) for $i$ in \hbox{$\{$3,7,18,6$\}$}. Moreover, there is a \BB- or a \GG-tile at
(1,1) together with \RR-tiles at (2,5), (1,10) and (1,9) to fix the tile in (1,1) and to allow the
locomotive a motion in tile (1,3). There are also \RR-tiles at (3,1), (3,3) and (3,4), the last two
ones fixing the tile (3,1). 

At last, we have the fork. It gathers three tracks around a \YY-tile. The arriving track passes 
through (5,33), (5,12), (5,4), (5,1), 0 and (3,1); a track leaves to right through (4,$i$) with $i$
in \hbox{$\{$5,15,6,17,7,3,10,4$\}$}; a track leaves to left through (3,$i$) with $i$ in 
\hbox{$\{$7,18,6,15,5$\}$} and (2,$i$) with $i$ in \hbox{$\{$4,10,3$\}$}.
As far as the rules of the cellular automaton are rotation invariant, the above fork is
enough to perform such a function in various configurations where the arriving and the leaving 
tracks may be different.

A computer program checked the motion of the locomotive through the structure illustrated by
the configuration of Figure~\ref{f_auxil}. It can be checked on Figure~\ref{f_fork} which
illustrates those various motions for a \BB-locomotive. Figure~\ref{f_fork} 
illustrates the motion for a \BB- and for a \GG-locomotive, to left and to right respectively on 
the figures.
\vskip 10pt
\vtop{
\ligne{\hfill
\includegraphics[scale=0.45]{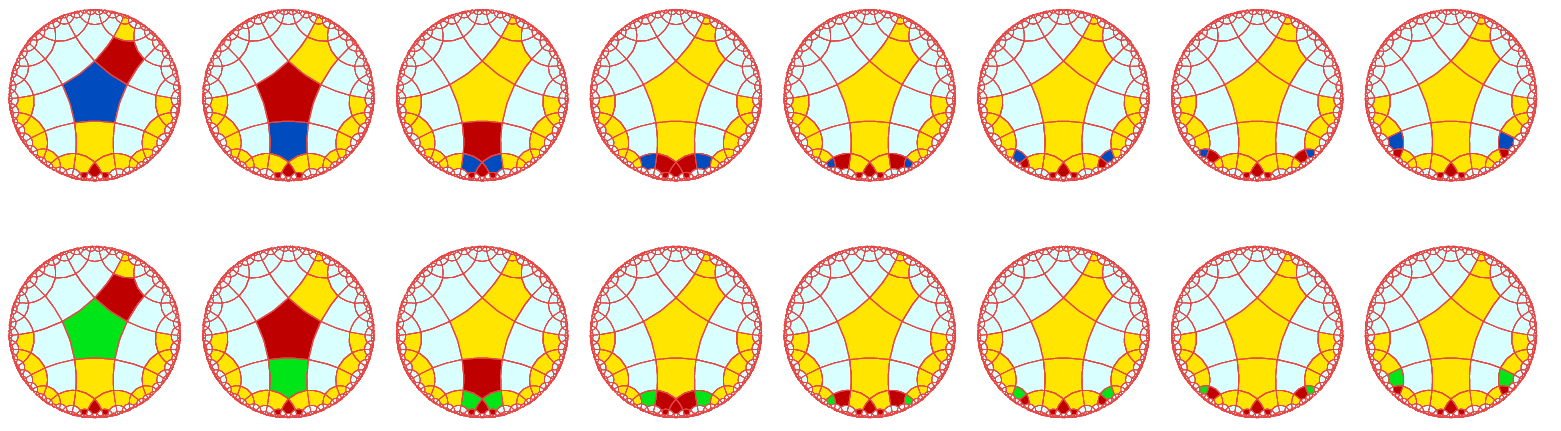}
\hfill}
\begin{fig}\label{f_fork}
\leurre
The locomotives through the fork. Top, the \BB-locomotive; bottom, the \GG-one.
\end{fig}
}

\vskip 10pt
The fixed switch is crossed passively only in the present implementation, 
following~\cite{mmarXiv21}. Figure~\ref{f_i_fix} illustrates the idle configuration of a fixed 
switch. However as far as the rules are rotation invariant, the configuration 
of Figure~\ref{f_i_fix} may be rotated by any angle ${2\pi k}\over 5$ with $k$ in 
\hbox{$\{$1..4$\}$}. 

\vskip 10pt
\vtop{
\ligne{\hfill
\includegraphics[scale=0.45]{secteurs_5_4.ps}
\hfill
\raise 30pt
\hbox{
\includegraphics[scale=1.2]{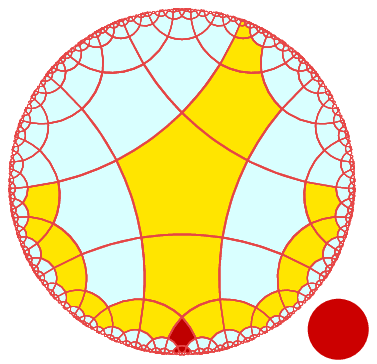}
}
\hfill}
\vspace{-20pt}
\begin{fig}\label{f_i_fix}
\leurre
To right, the configuration of the passive fixed switch. Rightmost red disc: representation of
a fixed switch in later figures.
\end{fig}
}
\vskip 10pt
Note the \RR-tile at (3,4) in Figure~\ref{f_i_fix} . The symmetry of the figure allows the 
locomotive to pass along one of its sides while it prevents the front of the locomotive 
to enter the branch starting from its side where the other branch of the switch arrives.

\vskip 5pt
The fork, the changers, the filters and the fixed switch can be assembled in a structure which 
performs the working of a crossing. 
\vskip 10pt
\vtop{
\ligne{\hfill
\includegraphics[scale=0.5]{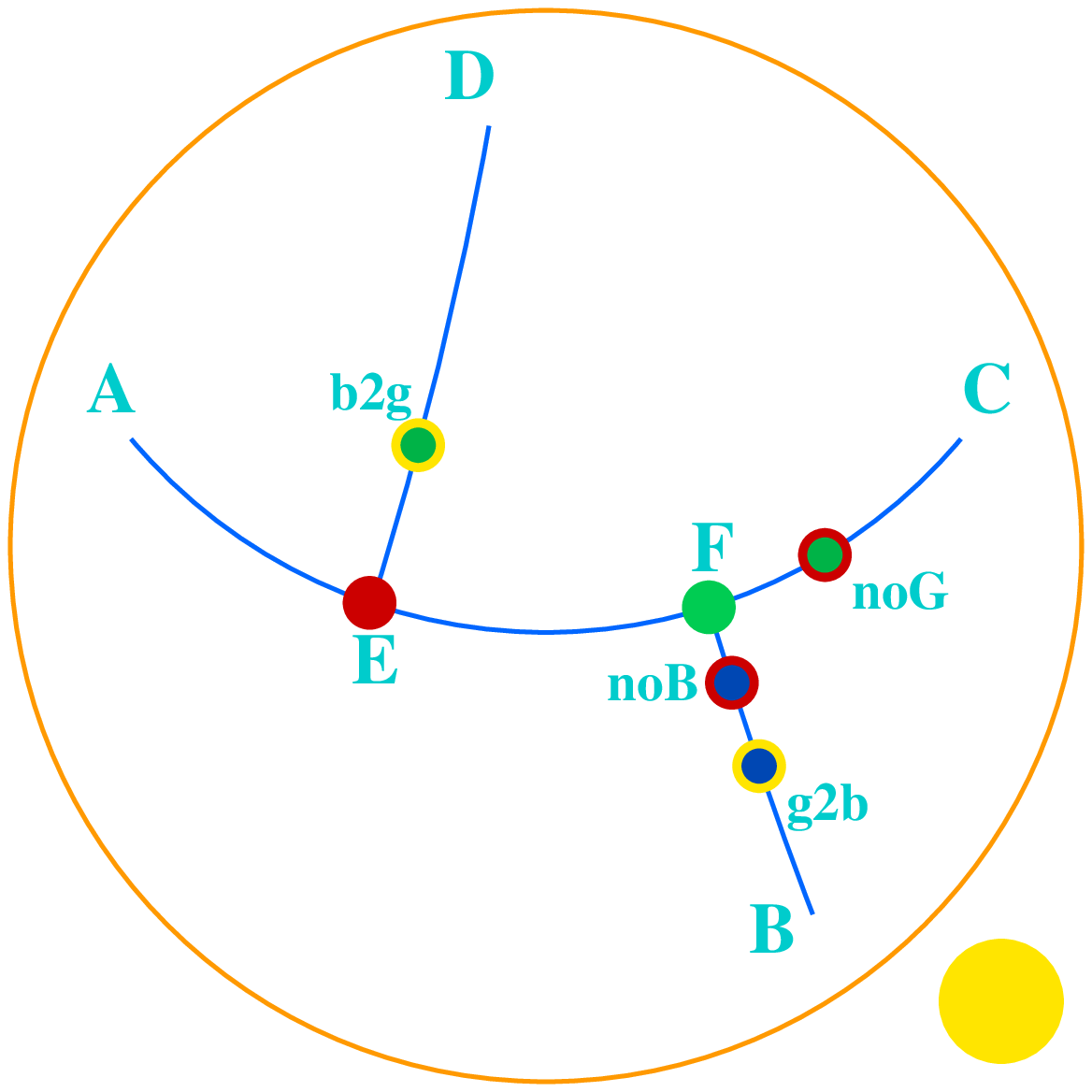}
\hfill}
\vspace{-5pt}
\begin{fig}\label{f_croise}
\leurre
Organisation of a crossing.
\end{fig}
}
\vskip 10pt
\vtop{
\ligne{\hfill
\includegraphics[scale=0.4]{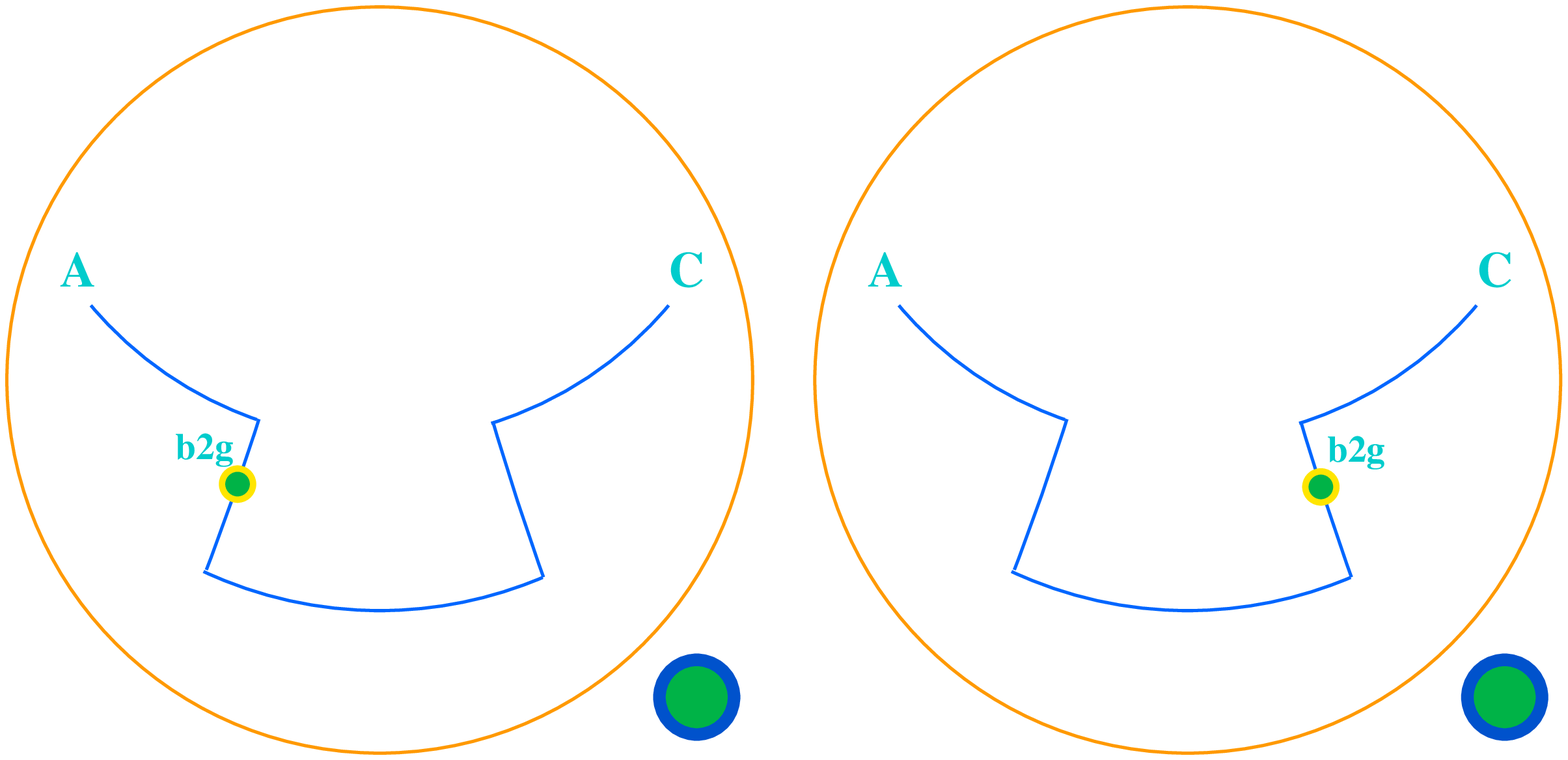}
\hfill}
\begin{fig}\label{f_flt_H}
\leurre
The filters for a horizontal line. To left, the motion goes from $A$ to~$C$. To right, it goes
from~$C$ to~$A$. At the bottom of the large disc, to right, the symbol of a \GG-converter.
\end{fig}
}

\vskip 5pt
That organisation is illustrated by Figure~\ref{f_croise}. A locomotive arrives at~$E$. If it comes 
from~$A$, $D$ it goes to~$C$, $B$ respectively. We assume that the locomotive are \BB-ones.
A locomotive coming from~$A$ crosses the fixed switch at~$E$ and is duplicated by the fork at~$F$
into two \BB-locomotives, one of them going to~$C$ and the other to~$B$. On the track from~$F$ 
to~$B$, the locomotive meets a \GG-filter which stops it. On the track from~$F$ to~$C$, the
\BB-locomotive meets a \BB-filter which let it go towards~$C$. The \BB-locomotive coming from~$D$
is converted to a \GG-one before reaching~$E$. From there, the locomotive goes to~$F$ where the
fork sitting there duplicates it into two copies. One of them goes to~$C$ where it is stopped by
a \BB-filter which sits on the track from~$F$ to~$C$. The other \GG-locomotive goes from~$F$ to~$B$.
It first meets the \GG-filter which let it go towards~$B$. A bit further towards~$B$, the 
\GG-locomotive meets a \BB-converter which changes it to a \BB-locomotive going towards~$B$.
Note that if a \GG-locomotive runs along $D$ to~$E$, there is no need of converters before~$E$ nor
after $F$ towards~$B$. It is assumed that a \BB-locomotive runs along the horizontal line. If it is
not the case, appropriate converters should be placed before~$E$ and after the \BB-filter placed
after~$F$. We remind the reader that tracks are one way. Accordingly, other possible crossings
can be obtained from the configuration of Figure~\ref{f_i_fix} by a suitable rotation or a symmetry
in a line, or a combination of both transformations.

From Figure~\ref{f_auxil}, we can see that the converters and the filters are assumed
to sit on a vertical line. Yet, we just mentioned the fact that it could be appropriate that filters
could also sit on a horizontal track. However, it is possible to set filters on a horizontal track
while the portion where a locomotive crosses it lies on a vertical track. Figure~\ref{f_flt_H}
illustrates how it can be organised. As indicated by the caption, there are two configurations
each one depending on the direction of the motion.

Before turning to the configuration of the fixed switch, we draw the attention of the reader on
Figures~\ref{f_cvcl} and~\ref{f_m_flt}. On Figure~\ref{f_cvcl}, we can see the change of colour
of a locomotive: on the top row of the figure, the \GG-converter changes a \BB-locomotive to a 
\GG-one. On the bottom row of the figure, the \BB-converter changes a \GG-locomotive to a \BB-one.
It is important to indicate that a converter should never be applied to a locomotive of the same
colour.

\vskip 10pt
\vtop{
\ligne{\hfill
\includegraphics[scale=0.5]{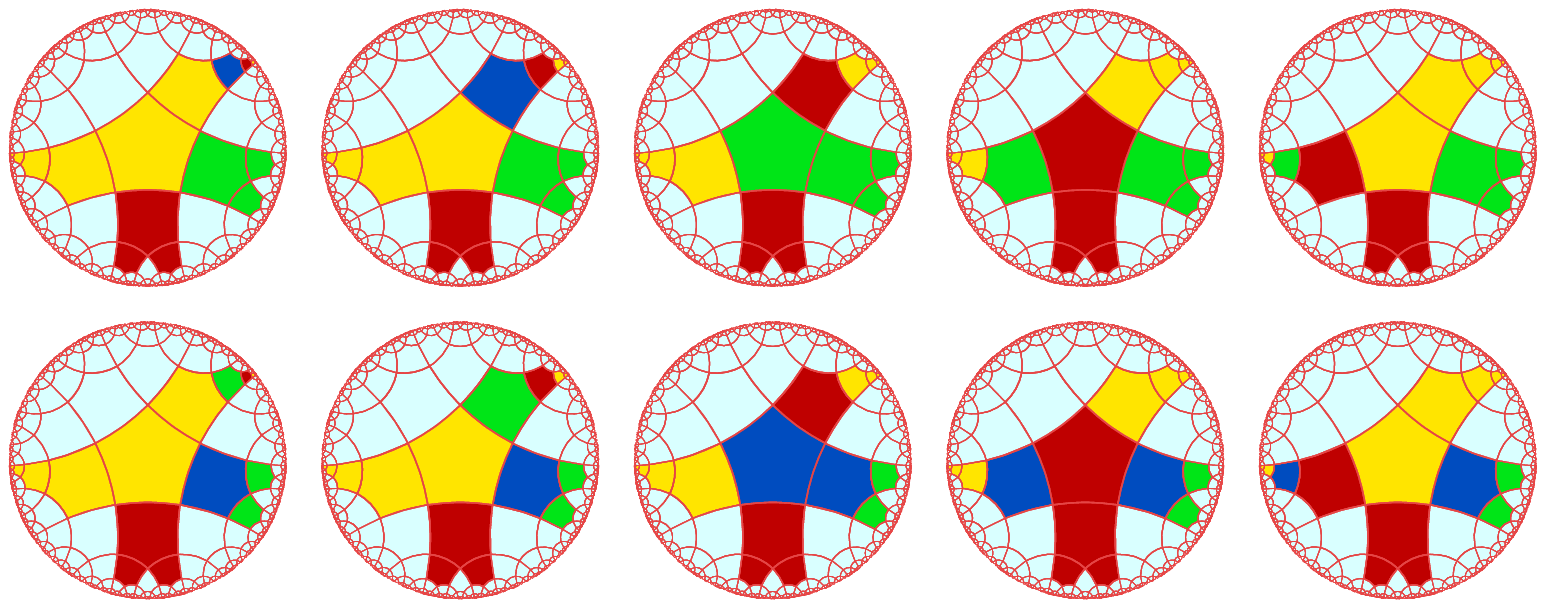}
\hfill}
\begin{fig}\label{f_cvcl}
\leurre
The motion of the locomotives through the converters.
\end{fig}
}
\vskip 10pt
On Figure~\ref{f_m_flt}, we can see the behaviour of a filter: the filter let go the locomotive with
the same colour while it kills the locomotive with an opposite colour. The figure illustrates all
cases: the top two lines illustrate the case when the filter let the locomotive go. The third line
illustrates the case when the filter kills the locomotive: on that line, to left the case of the
\BB-filter and to right, that of the \GG-one. The bottom line of the figure illustrates the change
of colour of a filter: to left, the \BB-filter becomes \GG, while to right, the \GG-filter 
becomes \BB. In both cases, the cell at 1(1) can see the front of the \BB-locomotive arriving along
the track 1($i$) with $i$ in \hbox{$\{$6,18,7,3$\}$} as far as the front of the locomotive 
eventually reaches the cell 1(7) which can be seen from the cell 1(1) although it is a partial 
neighbour  of that latter cell.
\vskip 10pt
\vtop{
\ligne{\hfill
\includegraphics[scale=0.5]{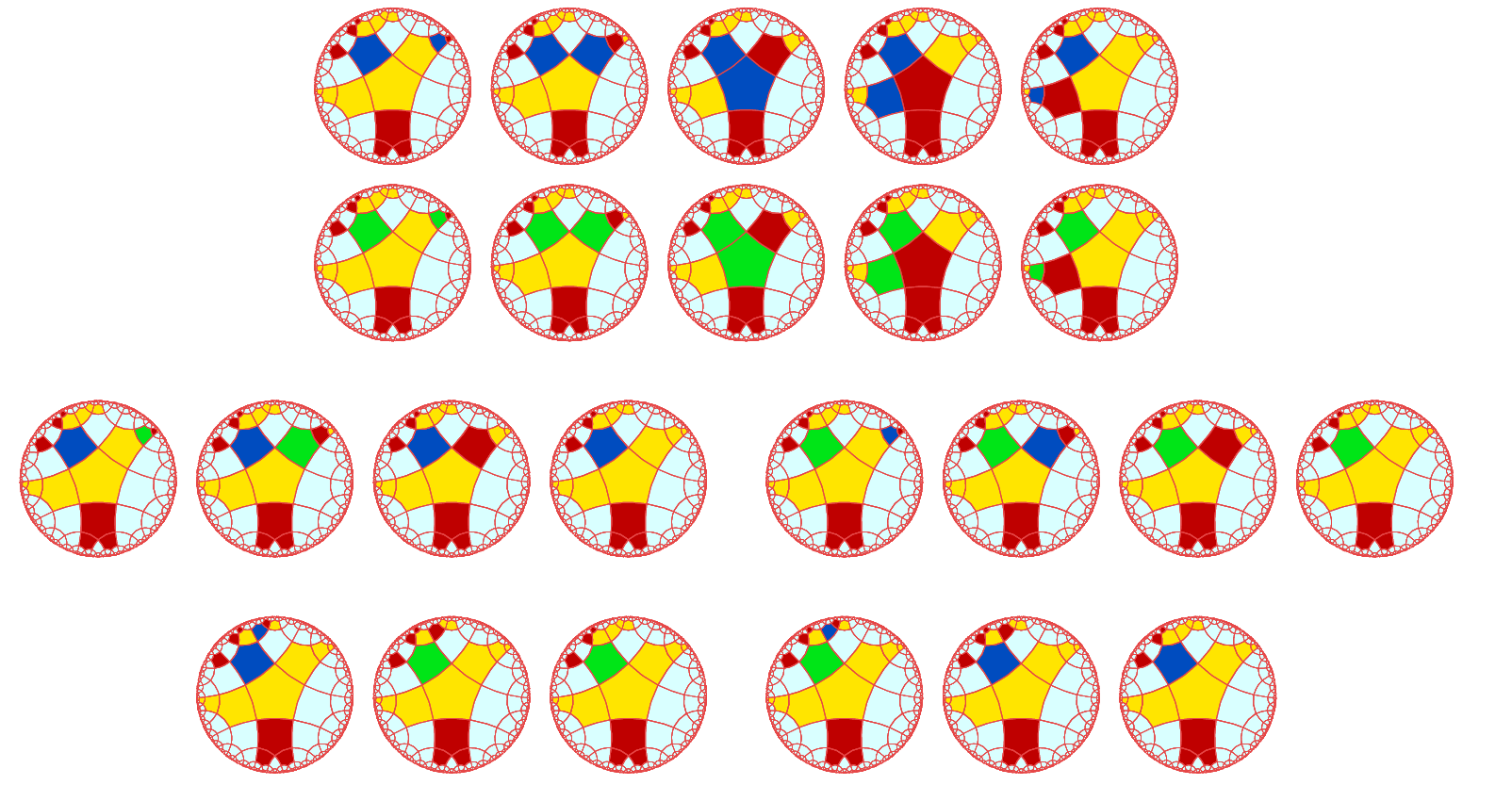}
\hfill}
\begin{fig}\label{f_m_flt}
\leurre
The motion of the locomotives through the filters. First row: the \BB-filter let the 
\BB-locomotive go. Second row, the \GG-filter let the \GG-locomotive go. Third row, to left the
\BB-filter kills a \GG-locomotive; to right, the \GG-filter kills a \BB-locomotive.
Fourth row: changing the colour of the filter: to left, from \BB- to \GG-; to right, from \GG- 
to \BB-. 
\end{fig}
}
\vskip 10pt
We can now turn to the passive fixed switch. As already indicated, we have four possible 
configurations illustrated by Figure~\ref{f_i_fix}. Each configuration is defined by the track which
leaves the switch: it may be a vertical track or a horizontal one. In the first case the locomotive
may come from above, in the second one, it may come from below. When we deal with a horizontal
track, the locomotive may come from the left, or it may come from the right.

From the rotation invariance of the rules, the configuration of Figure~\ref{f_i_fix} may be turned
by an angle which is a multiple of ${2\pi}\over 5$. Accordingly it is also true for the pictures
of Figure~\ref{f_m_fix}.

\vskip 10pt
\vtop{
\ligne{\hfill
\includegraphics[scale=0.475]{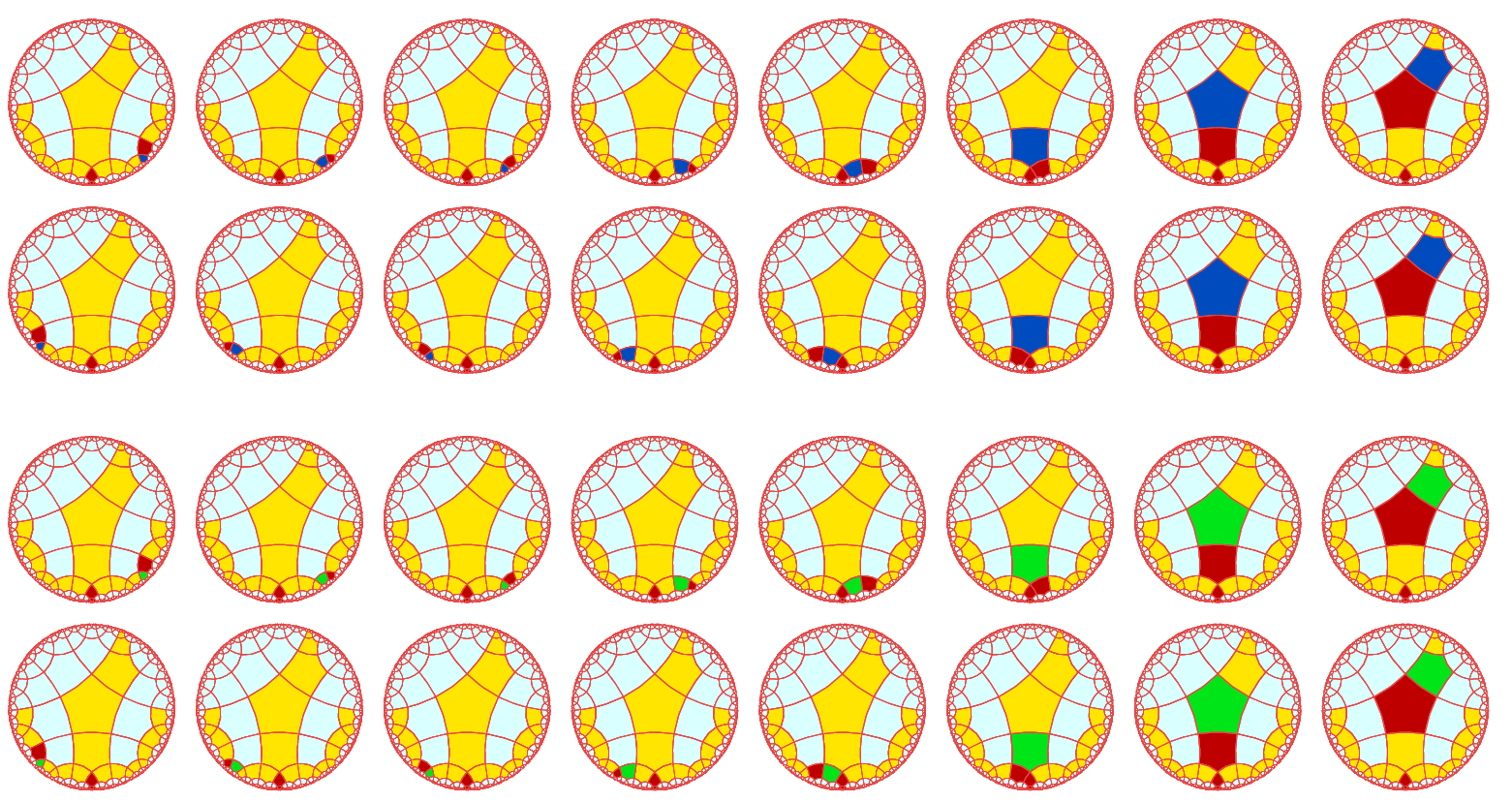}
\hfill}
\begin{fig}\label{f_m_fix}
\leurre
The \BB- and \GG-locomotives passively crossing a fixed switch: first row, a \BB-locomotive arriving
from the right. Second row, a \BB-locomotive is arriving from the left. Third row: a \GG-locomotive
is arriving from the right. Fourth row: a \GG-locomotive is arriving from the left.
\end{fig}
}

Figure~\ref{f_m_fix} 
illustrates the crossing of a passive fixed switch by a \BB-locomotive and by a \GG-one.

\subsubsection{The switches}\label{sbbswitch}

   In order to better understand the schemes later presented, it is important to indicate the
global frame in which the simulating computation holds.

   It holds within a {\bf grid} which we define as constituted by the vertical tracks and the
horizontal ones. However, up to now, vertical tracks and horizontal ones  were not clearly defined.
It is time to do that right now.

Figure~\ref{f_grille} illustrates a part of a grid in the hyperbolic plane. The vertical tracks are
represented by mauve arcs in the figure while horizontal ones are represented by blue arcs. 
However, the tracks are not lines but sequences of tiles in the pentagrid. We can identify tracks
by lines in the figure considering that our window encloses a large area of the hyperbolic plane.

Let us describe more exactly what our tracks consist of. Figure~\ref{penta} shows us five sectors 
around the central tile, that which we call tile~0. Let us fix cell~0 once for all and once for all 
too, let us fix the neighbour one of the tile~0. Our computation takes place in sector~4. In that 
sector our vertical tracks are segments of a branch of the tree rooted at the cell~(4,1) and the
horizontal tracks are pieces of levels in that tree. We remind the reader that a level of the tree
in a sector is the set of tiles which are at the same distance from the head of the sector.
However, we do not take any branch as support of vertical tracks: a vertical passes through white
tiles and for such a tile, its neighbours in the track are neighbour~1 and neighbour~4.
A horizontal track lies in a level. For a white tile of a horizontal track, its neighbours in the
track are neighbour~2 and neighbour~7. For a black tile of a horizontal track, its neighbours in the
track are neighbour~3 and neighbour~7. In order to have the freedom of organisation we can deduce
from Figure~\ref{f_grille}, it is needed to assume that the closest level for a horizontal track
in the circuit we shall describe is at least 10: there are 17711 tiles on level~10 as far as
$f_{21}= 17711$. Indeed, the number of nodes at level~$n$ is $f_{2n+1}$ where $f$ is the Fibonacci
function with initial conditions $f_0=f_1=1$, see for instance \cite{mmbook1,mmJUCS} for proofs of
that property of the levels in a sector. As each node produces one black node exactly among its
sons, there are $f_{2n-1}$ black nodes on the level~$n$, so that there are $f_{2n}$ white nodes on
that level. Accordingly, 10946 vertical tracks can start from level~10. For the simulation of
a universal register machine, it is much more than what is required as will soon be seen. It should 
be noted that at level~20, the distance between two consecutive vertical tracks raised at level~10
is more than 17711. The previous figures show us that a distance of 10 levels is a small distance
between the extremal tiles of a vertical track.
\vskip 10pt
\vtop{
\ligne{\hfill
\includegraphics[scale=0.6]{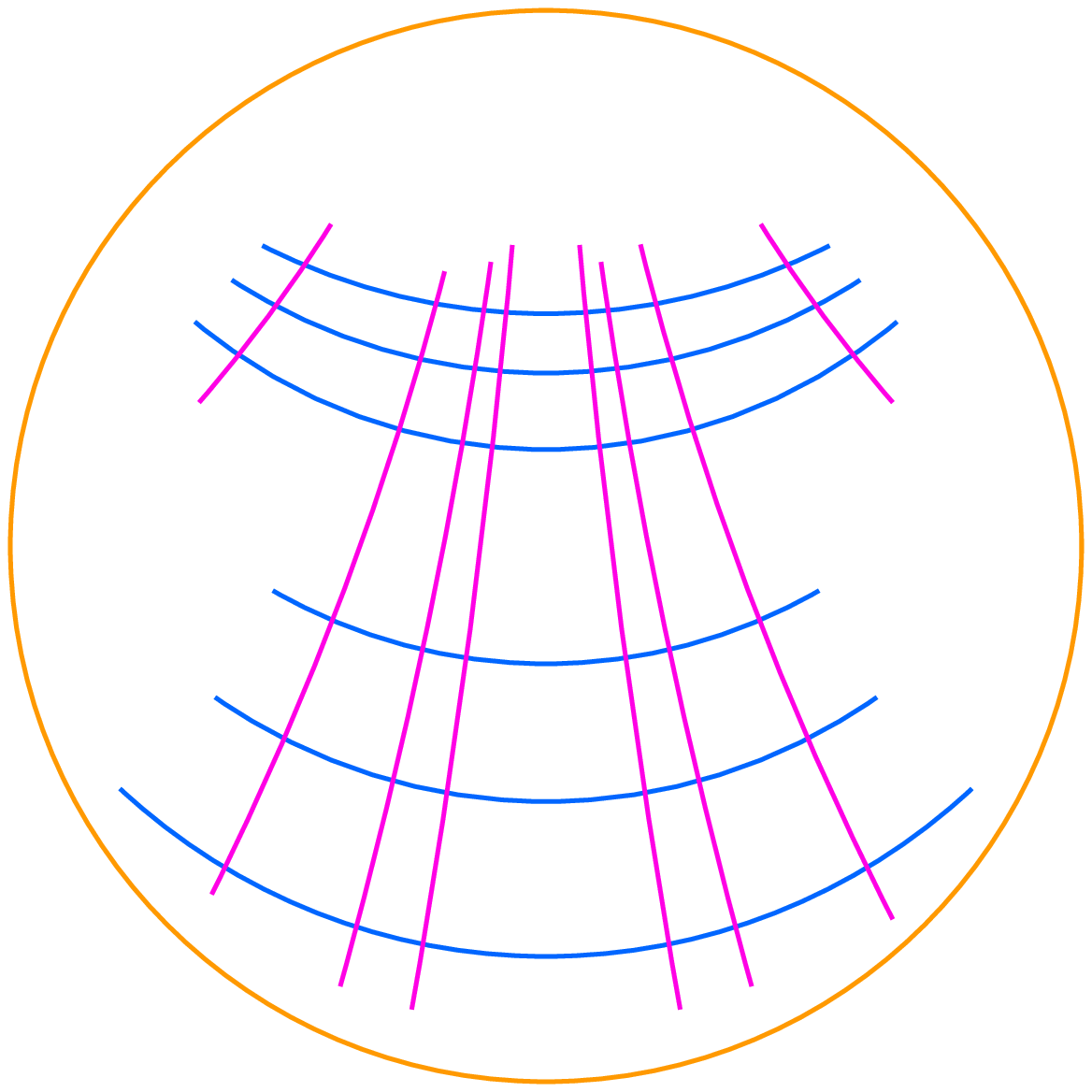}
\hfill}
\begin{fig}\label{f_grille}
\leurre
Representation of a grid in the hyperbolic plane. 
\end{fig}
}

   We postpone the implementation of the register machine to Subsection\ref{sbregs} which describes
how registers are implemented in our simulation.
\vskip 5pt
   Presently, we deal with the implementation of the switches. The present sub-subsection deals 
with the flip-flop switch and with the memory one.

A flip-flop changes the selection after an active crossing and there is never a passive crossing 
through a flip-flop switch. In a memory switch, its passive part defines the selection of
its active part. Accordingly, we need two kinds of structures: a structure which let
the locomotive go, the other one which kills it. In the flip-flop, the change is
made by the crossing of the switch, in the memory switch the change is triggered by
the crossing of the passive switch. Basically, it is the same constraint. 
Accordingly, the controlling structure in the flip-flop and in the memory switch should be 
{\it programmable}, a point which we explain further.

Figure~\ref{f_flip_flop} illustrates the implementation of the flip-flop. 
Figure~\ref{f_memo} illustrates both parts of the memory switch: the active and the passive ones.

In the flip-flop,  a locomotive arrives to~$A$. There, a {\bf fork} sends two 
locomotives of the same colour: one towards~$L$, the other towards~$R$. The locomotive which is 
sent to~$R$ meets a filter which is sitting there. If the filter has the colour of the locomotive, 
that latter goes on. Otherwise, the locomotive is destroyed. Symmetrically, another filter is 
sitting at~$L$. Now, the filter at~$R$ and that at~$L$ have opposite colours. Accordingly, the 
locomotive goes on further on the side of the switch where the filter has the same colour. That 
property defines the selected track: it is the leaving track on which the filter has the colour of 
the locomotive arriving at~$C$. When the locomotive goes on along the selected track, at some 
distance after the filter, it meets a fork: one branch of the fork let the locomotive go on along 
the selected track, the other sends the locomotive to~$B$. There, a fixed switch is sitting and 
any locomotive arriving there passively crosses the switch. From~$B$, the locomotive arrives to a 
fork sitting at ~$C$ which sends a copy of it to the filter at~$L$ and another copy to the filter 
at~$R$. Those copies change the filter to the opposite colour, so that now, the selected track of 
the switch is changed which is conformal to the definition of a flip-flop switch. Note that the 
change of the colour in a filter is performed by a \BB-locomotive so that a changer is possibly 
put on the track leaving $S$ and before the arrival to the fork leading to~$R$ and to~$L$.

\vskip 10pt
\vtop{
\ligne{\hfill
\includegraphics[scale=0.4]{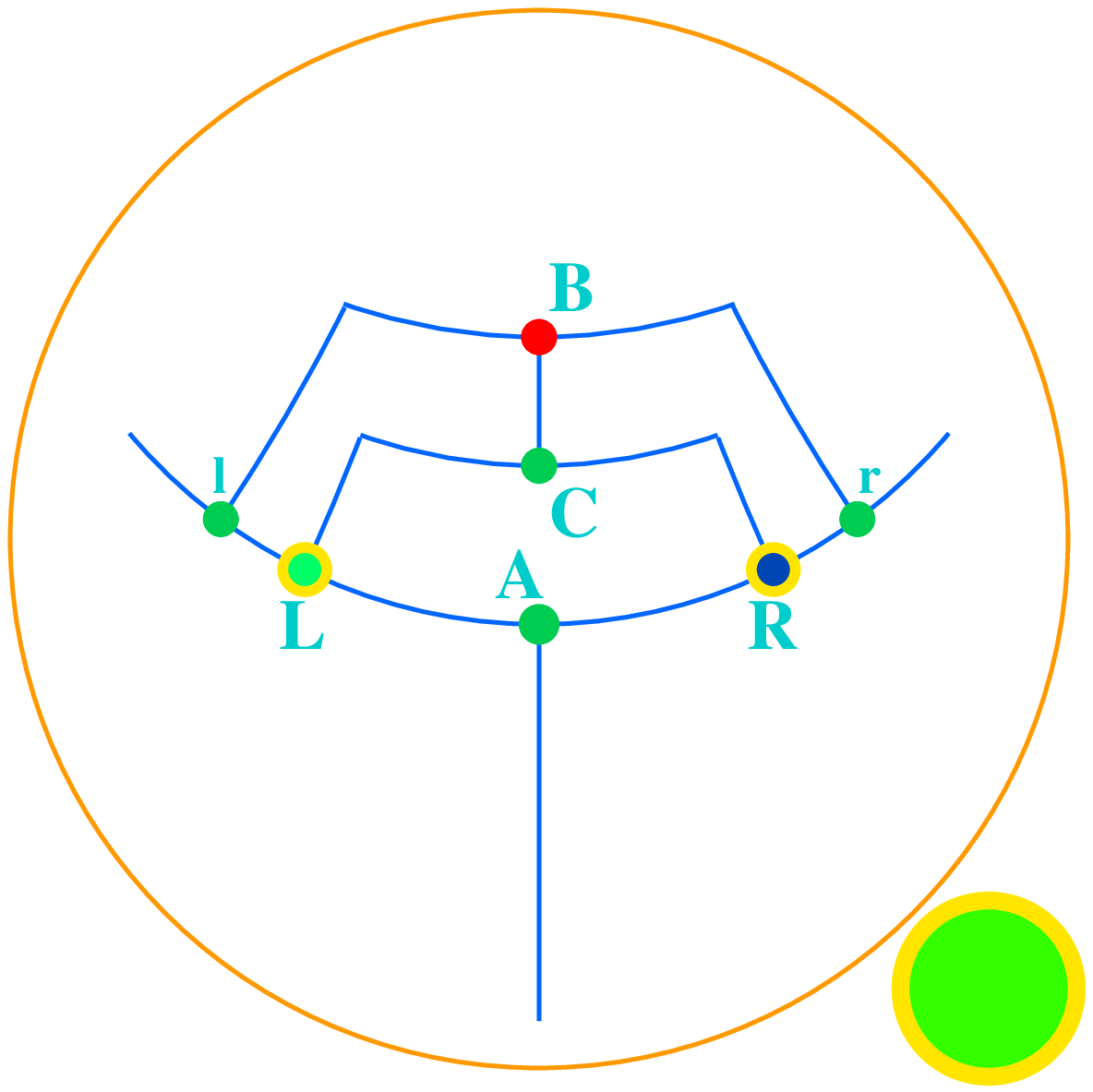}
\hfill}
\begin{fig}\label{f_flip_flop}
\leurre
The flip-flop and its symbolic representation. 
\end{fig}
}
\vskip 10pt

We assume that the arriving track at the flip-flop and the leaving tracks lie on different kind of
tracks: if the arriving track is vertical, horizontal, the leaving tracks are horizontal, vertical
respectively. When a locomotive crosses the switch, we may always assume that the change at~$L$ 
and~$R$ is performed before a new visit of the locomotive at the switch. We may also assume that
a given flip-flop switch is always crossed by a locomotive of the same colour: either always \BB-,
or always \GG-.

The active memory looks like the flip-flop but it is simpler as far as in its working when the 
locomotive arrives at~$A$ only the arcs from~$A$ to~$L$ and from~$L$ to~$R$ are concerned. We may 
assume that a passive crossing occurs when the locomotive is not involved in the
active switch.

\vskip 10pt
\vtop{
\ligne{\hfill
\includegraphics[scale=0.4]{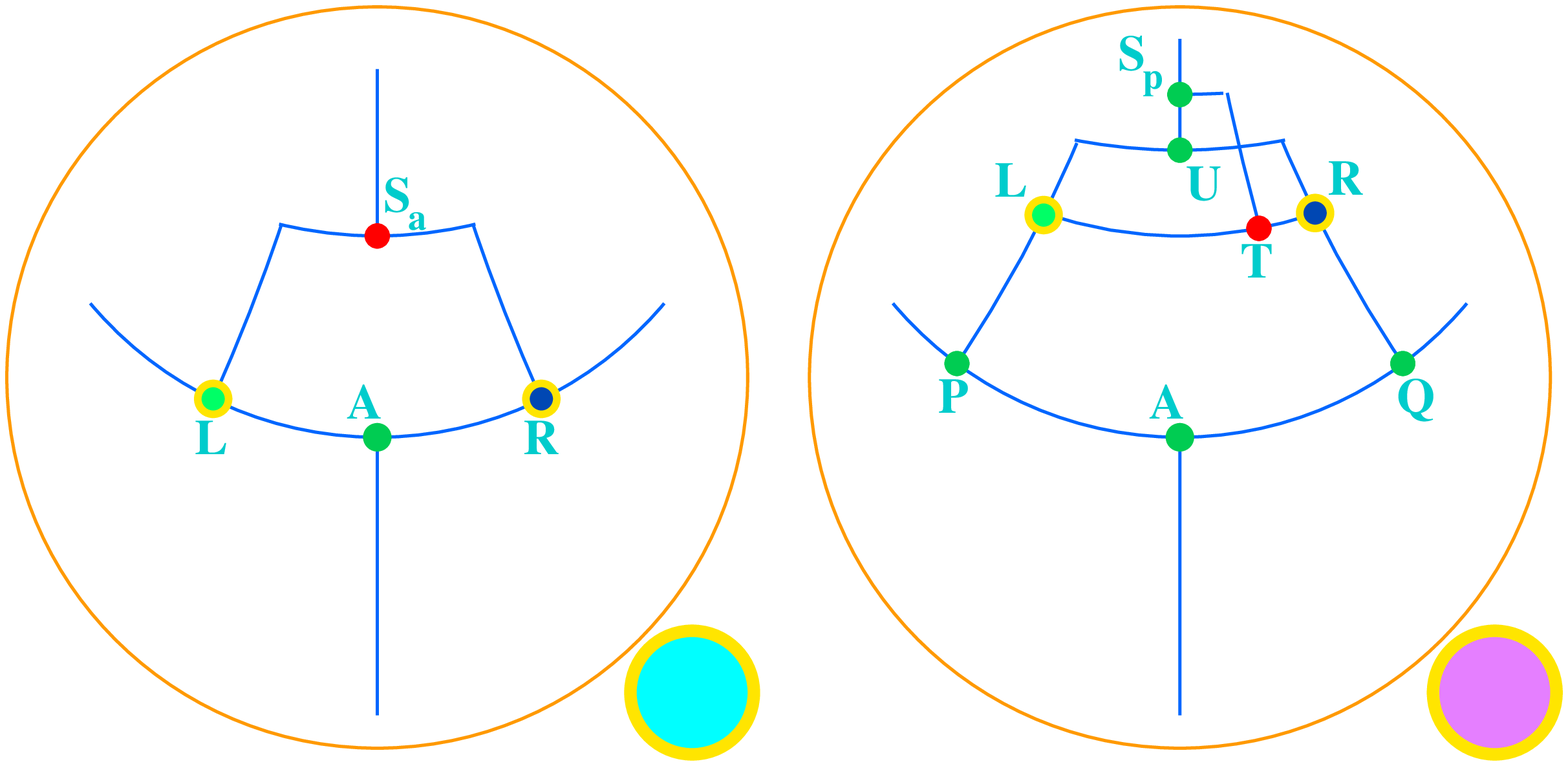}
\hfill}
\begin{fig}\label{f_memo}
\leurre
Memory switch: to left, the active switch; to right the passive one. Each one has a specific
symbolic representation.
\end{fig}
}

   The structure of the passive switch is very different. The locomotive arrives 
either at~$P$ or at~$Q$. Assume that it arrives at~$P$. The case of an arrival at~$Q$ 
is symmetrically dealt with. A fork at~$P$ sends a locomotive to~$A$, where a fixed switch 
let the locomotive leave the switch. The other locomotive is sent to~$L$. There, if the 
filter in the active part let the locomotive go, the filter at~$L$ stops the 
locomotive, so that the memory switch does not change the selection. If the filter
of the active part stops the locomotive, the filter of the passive switch let the 
locomotive go. That locomotive goes further to~$S_p$ via $T$ where a fixed switch is sitting. 
At~$S_p$, a fork sends the locomotive to~$U$ where an other fork sends a copy of the locomotive to 
the fork~$S_a$ of the active switch and according to the scheme of the active switch that action 
will exchange the role of its filters. The second locomotive sent by the fork at~$S_p$ goes to~$U$ 
where another fork duplicates that locomotive sending one copy to~$L$ and the other to~$R$, both
copies exchanging the roles of the corresponding filters. Accordingly, both parts
of the memory switch operates on the needed way. Here too, we have to make sure that
the arrival at~$L$ and~$R$ of the locomotives sent from~$T$ occurs later at the
permissive controller than the locomotive arriving from~$P$ or from~$Q$. The picture
of Figure~\ref{f_memo} ensures us that it is easy to fulfill that condition, possibly
appending tracks in order to introduce a possibly required delay.

    Note that the separation of the two parts of the memory switch allows to use one passive part
for changing several active parts in a same way, a situation we shall soon encounter.

    At last and not the least, the switches we have just implemented can be accessed in different
ways by the tracks: the possibility of free crossings allow us to satisfy the very structure of 
those implementations.

\subsubsection{The one-bit memory}\label{sbbunit}

   We are now ready to implement the one-bit memory in our setting.
The implementation is illustrated by Figure~\ref{fhypunit} which reminds us the
Euclidean implementation in its left-hand side part.
We adopt the same notations as in the Euclidean picture in order to facilitate the
comparison for the reader.
\vskip -5pt
\vtop{
\ligne{\hfill
\raise 40pt\hbox{\includegraphics[scale=0.6]{elem_gb.ps}}
\hskip-10pt
\includegraphics[scale=0.45]{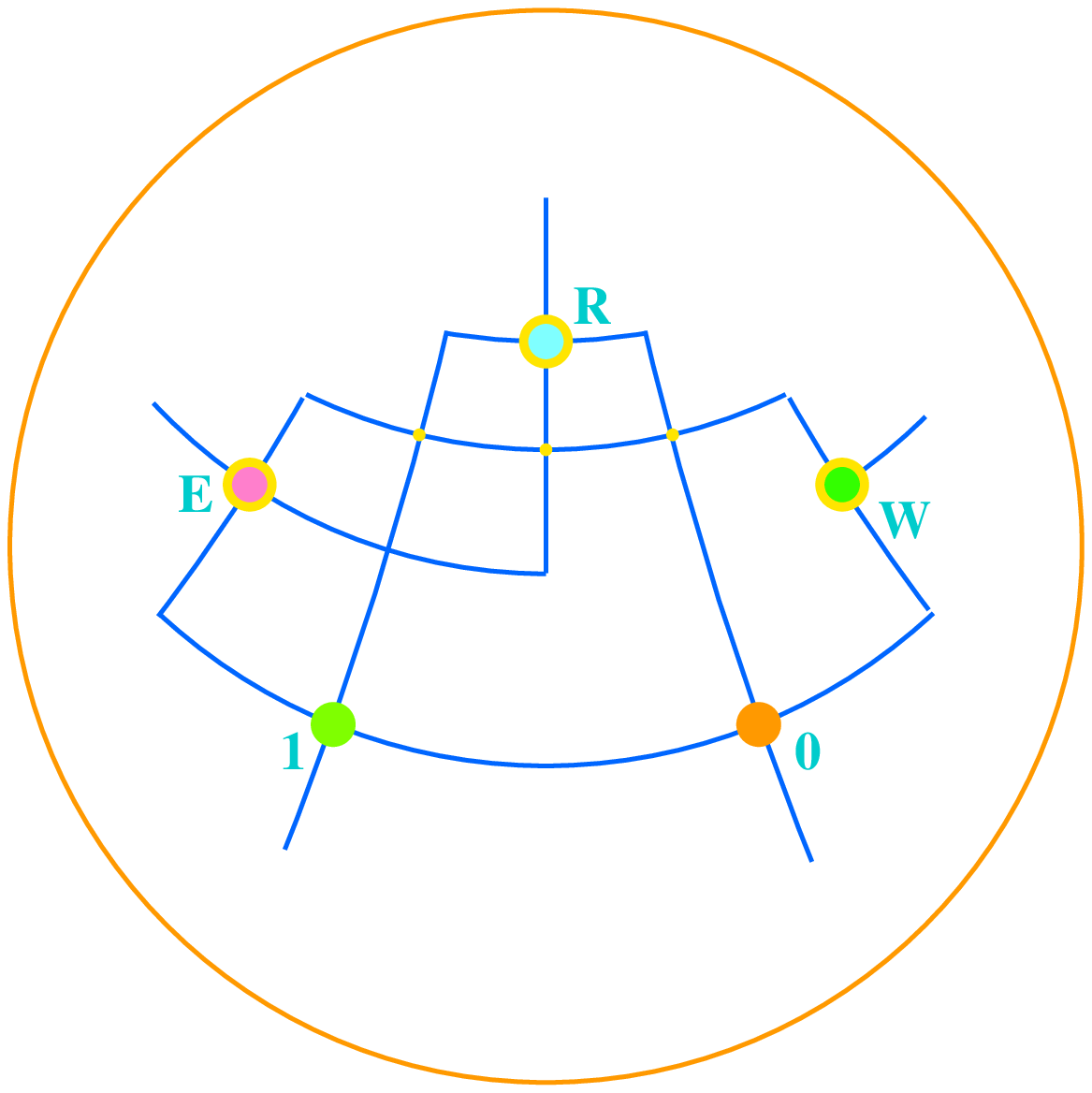}
\hfill}
\vspace{-30pt}
\begin{fig}\label{fhypunit}
\leurre
To left, the scheme of Figure~{\rm{\ref{basicelem}}}, to right, its implementation
in our setting. In the hyperbolic picture, the colours at~$W$, $R$ and~$E$, namely
yellow, light green and light purple respectively represent a flip-flop switch, an
active memory switch and a passive memory switch. The yellow points with an {\tt x}
letter indicate crossings.
\end{fig}
}

\def\zz{{\bf\tt 0}}
\def\uu{{\bf\tt 1}}
Here again, a \BB-locomotive is involved. If it arrives at the disc labelled with~$R$ it 
exits through the gate~\zz{} or through the gate~\uu{} according to the track selected by the 
active memory switch sitting at~$E$. 

If the locomotive arrives at $W$, the flip-flop sitting there sends the locomotive to~$E$ either 
through track~0 or track~1, depending on what is selected by the switch at~$R$. If the track 
is~$i$, then it arrives to~$E$ through its track $1$$-$$i$. Now, from what we have seen on
Figure~\ref{f_memo}, the switch at~$E$ sends a locomotive to~$R$ making the active memory
switch sitting there select the values which are compatible with those at~$E$. But from~$E$, 
another \BB-locomotive is sent out of the switch through the gate~$E$. Note that the memory 
switch of the left-hand side picture is split into two parts on the right-hand side figure. There, 
the active part sits at~$R$ while the passive one sits at~$E$. Note that there is a path 
joining~$E$ to~$R$.

\def\DDI{\hbox{$\mathbb {DD}$$_{\mathbb I}$}}
\def\DDD{\hbox{$\mathbb {DD}$$_{\mathbb D}$}}
\def\DDT{\hbox{$\mathbb {DD}$$_{\mathbb T}$}}
\def\EE{{\bf\tt E}}
As far as we shall later use on-bit memories, we shall use the names \RR, \EE, \WW, \zz{} and 
\uu{} to label the gates as in Figure~\ref{fhypunit}. Here two, we assume that the one-bit memories
used later will be implemented as indicated in Figure~\ref{fhypunit}: free crossings allow us
to make that assumption.

\subsubsection{A register and its discriminating structures}\label{sbbregdisc}

Let us illustrate the implementation of the structures used by the registers by a toy example
given by Figure~\ref{f_jouet}.

\vskip 10pt
\vtop{
\ligne{\hfill
\includegraphics[scale=0.65]{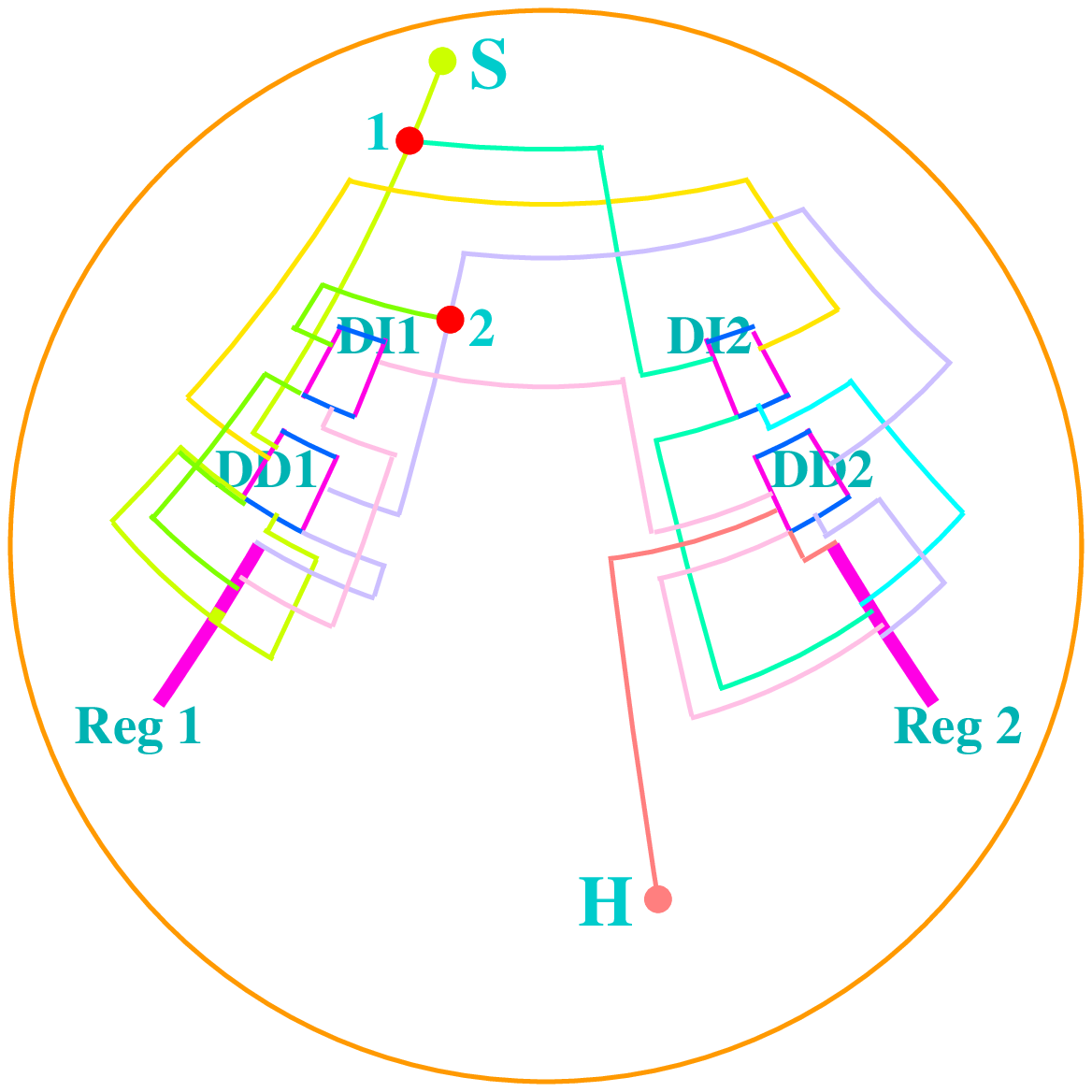}
\hfill}
\begin{fig}\label{f_jouet}
\leurre
A toy example. The content $x$ of~$R_1$ is transformed into $\lfloor {x\over 2}\rfloor$ stored in
$R_1$ too. Initially, the content of $R_2$ is~$0$.
\end{fig}
}	

\def\reg #1 {\hbox{$\mathcal R$(#1)}}
\def\MM{{\tt M}}
In Figure~\ref{f_jouet}, we have the implementation of a program simulating a two-registered 
machine which computes the half of the initial value contained in $R_1$. At the beginning, $R_2$ is
empty and it is used as an auxiliary structure. The illustrated program of the figure can be
written as follows:
\vskip 5pt
\ligne{\hfill
\vtop{\leftskip 0pt\parindent 0pt\hsize=120pt\tt
\ligne{1\hskip 10pt dec R1,2\hfill}
\ligne{\ \hskip 10pt dec R1,2\hfill}
\ligne{\ \hskip 10pt inc R2\hfill}
\ligne{\ \hskip 10pt jmp 1\hfill}
}
\hskip 20pt
\vtop{\leftskip 0pt\parindent 0pt\hsize=120pt\tt
\ligne{2\hskip 10pt dec R2,3\hfill}
\ligne{\ \hskip 10pt inc R1\hfill}
\ligne{\ \hskip 10pt jmp 2\hfill}
\ligne{3\hskip 10pt halt\hfill}
}
\hfill}
\vskip 5pt
Figure~\ref{f_reg} illustrates the implementation of a register in our setting.
As can be seen on the figure, we have a segment of line enclosed in a path surrounding it. 

   The value of the register is the number of \WW-tiles along that line from tile~0{} in the 
left-hand side of Figure~\ref{f_reg} up to tile~0 of the right-hand side part, both tiles included. 
Denoting the register by $\mathcal R$, we denote the \WW-cells representing its content~$n$ by 
\reg{m} {} with $0\leq m<n$. By construction, for any $n$, we have that \reg{n} {} is \YY.
In particular, when the register is empty, whose value is~0 by definition, \reg{0} {} is \YY.

   The register is the occasion to remind the reader the fact we have two kinds of locomotives.
The \BB-locomotive is devoted to increment the register while the \GG-one is devoted to
decrement it. 
The locomotive returning from the register has the same colour as it had when entering the register.
However, once it performed its operation, the locomotive does not remember which instruction of 
the program dictated it to execute that operation. Accordingly, we need two new structures outside 
a register: a structure to remember which instruction required to increment the register, the 
\DDI-structure, see~\ref{memo_incr}.a; and a structure to remember which instruction required to 
decrement the register, the \DDD-structure, see~\ref{memo_decr}.b. Note that the colour of the 
locomotive is enough to characterise which operation the locomotive did perform. Both structures 
are required for each register. Those structures too make use of one-bit memories.

\vskip 10pt
\vtop{
\ligne{\hfill
\includegraphics[scale=1.3]{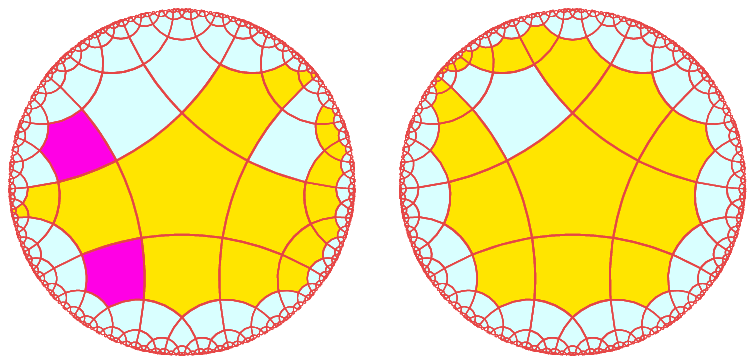}
\hfill}
\vspace{-5pt}
\begin{fig}\label{f_reg}
\leurre
To left, the beginning of a register when $n=0$; to right, its end when $n>0$. Both ends are 
marked as far as they play an important role.
\end{fig}
}

\label{memo_incr}
{\bf \ref{memo_incr}.a\ Remembering the incrementing instruction}

   As can be guessed from Figure~\ref{f_reg}, when the locomotive goes back from the register,
as far as we have two types of locomotive only, the locomotive does not remember from which point
of the program it was issued. In order to go back to that point in order to define the next 
instruction, the instruction must be marked in some way near the register itself. The \DDI-structure
is devoted to play that role. There is a \DDI-structure for each register $\mathcal R$. The 
structure contains as many units as there are instructions in the program to specifically 
increment $\mathcal R$. From each instruction $I$ incrementing $\mathcal R$, a path goes from the 
place of $I$ in the program to the unit of the \DDI{} attached to $\mathcal R$ associated by that
path to that unit. The unit itself contains a one-bit memory, see Figure~\ref{f_uDDI}.

At the initial configuration, in all units of each \DDI-structure, the one-bit memory contains~\zz.
When the locomotive starts its way to perform an incrementing instruction~$I$ over the register
$\mathcal R$, the track run by the locomotive arrives to the \WW-gate of the unit of the \DDI{}
attached to $\mathcal R$ which is associated with~$I$. As far as the locomotive arrives to a 
\WW-gate it rewrites the \zz-bit of the one-bit memory to \uu. Then, the locomotive exit the memory
through its \EE-gate. It meets a flip-flop at~$A$ which, form the initial configuration, sends it
towards $\mathcal R$. After crossing that flip-flop switch, the selected track is now a track 
which goes back to the program, to the instruction which has to be executed after $I$ has completed 
its operation on $\mathcal R$. 
At that time, there is a single unit in that \DDI{} whose one-bit memory contains \uu.

\vtop{
\ligne{\hfill
\includegraphics[scale=0.5]{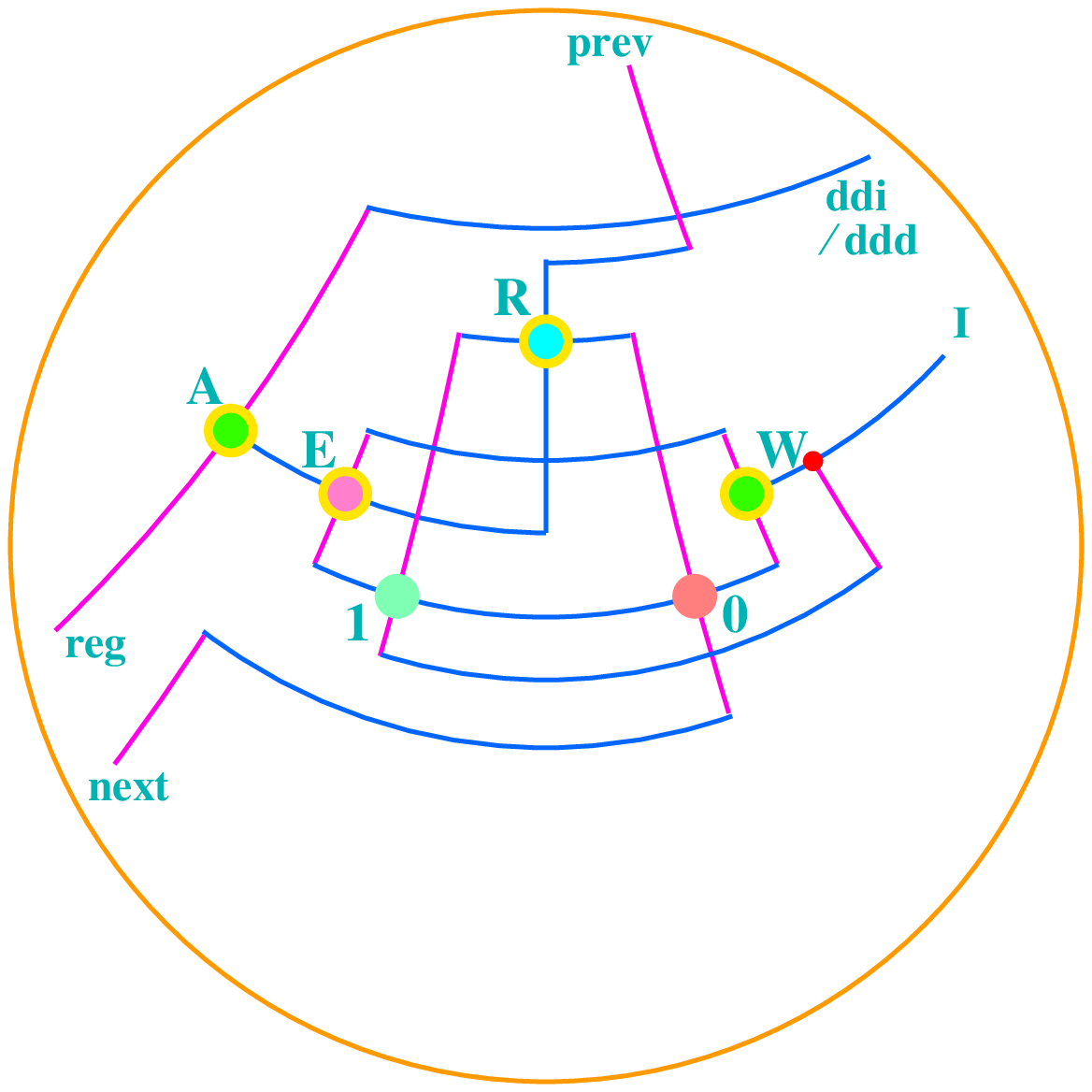}
\hfill}
\begin{fig}\label{f_uDDI}
\leurre
The unit of \DDI{} devoted to an incrementing instruction. The figure does not mark the
crossings which are clear from the picture itself.
\end{fig}
}

When the locomotive comes back from $\mathcal R$, it is a \BB-locomotive and a fork combined with 
appropriate filters of opposite colours sends that locomotive  to the \DDI{} of $\mathcal R$.

The locomotive visits each unit of the \DDI-structure until it meets the single one whose one-bit 
memory contains~\uu. To do that, the locomotive enters the memory of the unit through its \RR-gate.
If it reads~\zz, it is sent to the next unit. When it reads~\uu, it knows that the right unit is
reached. Leaving the one-bit memory through the \uu-gate, the locomotive is sent to the \WW-gate so
that it rewrites the content of the one-bit memory from~\uu{} to \zz. Leaving the memory through the\EE-gate, the locomotive again meets $A$ where the selected track of the flip-flop sends it on the
track leading to the right place of the \DDI{} or of the \DDD{} of the appropriate register. Once 
the flip-flop at $A$ is crossed, the selected track again becomes the track leading to $\mathcal R$.
Accordingly, when the locomotive leaves the \DDI-structure after preforming its operation, that 
\DDI{} recovered its initial configuration.

\label{memo_decr}
{\bf \ref{memo_decr}.b\ Remembering the decrementing instruction}

    If the locomotive has to decrement the register $\mathcal R$ it visits a similar structure
as \DDI, the \DDD-structure attache to $\mathcal R$. The principle of that structure is similar to 
that of \DDI. However a unit of \DDD{} is more complex than a unit of \DDI. The reason comes from
the difference between a decrementing instruction and an incrementing one. It is always possible to
increment a register. It is not possible to decrement an empty register as far as register machines
deal with natural numbers only. When a decrementing locomotive arrives at \reg{0} {} when 
$\mathcal R$ is empty, it leaves \reg{0} {} through a different track from which it leaves that
place when it had successfully decremented it.

   Accordingly, if the working of \DDD{} is similar to that of \DDI{} for an arriving locomotive 
from the program, it is not the case for a locomotive returning from the register. We leave the
case of an arriving locomotive from the program to the reader.

\vtop{
\ligne{\hfill
\includegraphics[scale=0.5]{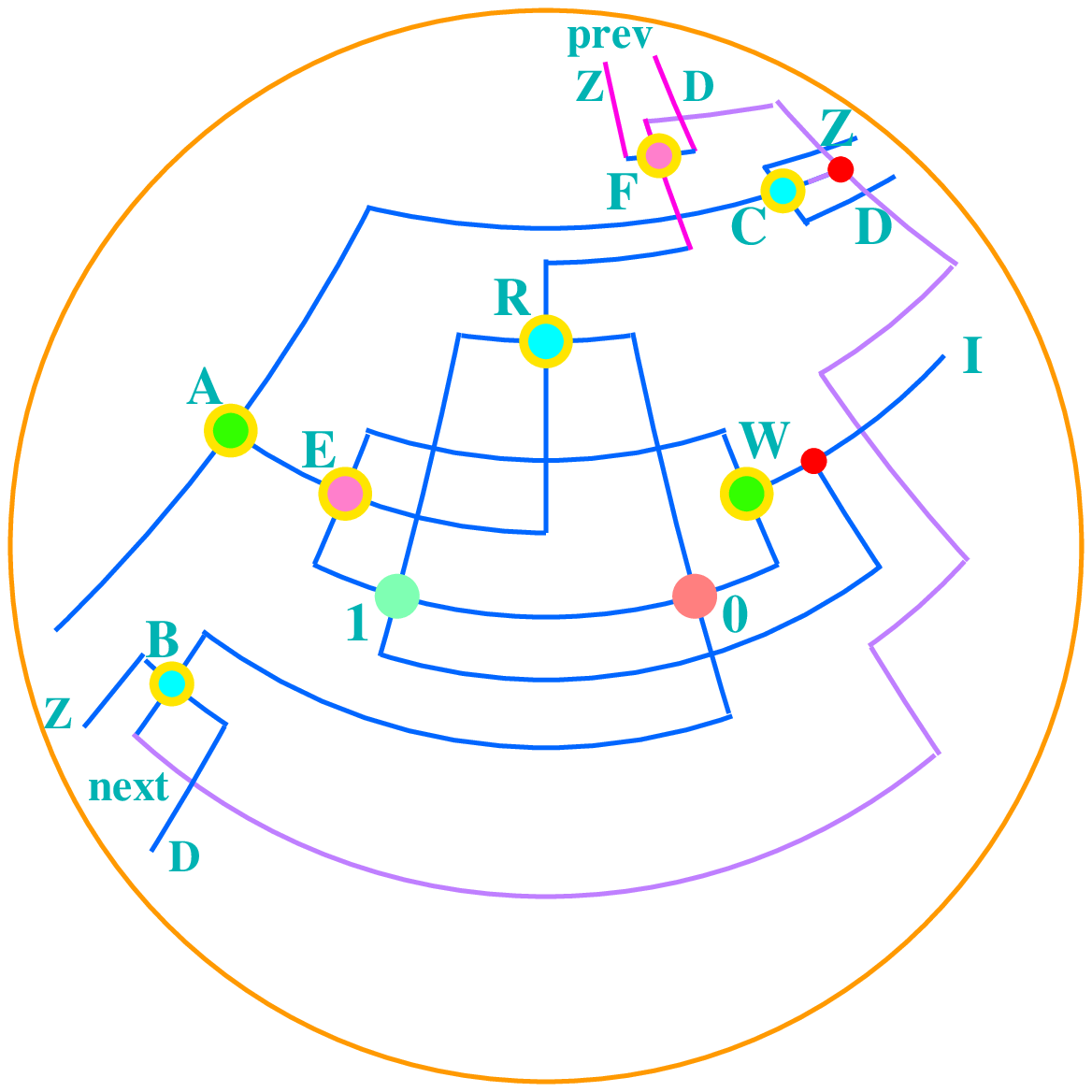}
\hfill}
\vspace{-5pt}
\begin{fig}\label{f_uDDD}
\leurre
The unit of \DDD{} devoted to a decrementing instruction.
\end{fig}
}

\def\ZZZ{{\bf\tt Z}}
\def\DD{{\bf\tt D}}
   Consider the case of a locomotive coming back from~$\mathcal R$. It comes from the \DD-track
after a successful operation or it comes through the \ZZZ-track because it could not decrement an 
empty $\mathcal R$. Both tracks behave in the same: they lead to the \RR-gate. If the locomotive
reads \zz, it goes to the next unit through the same kind of track as the one it used to arrive at 
the current unit. As can be seen, there are two possible exits for the locomotive entering the
one-bit through the \RR-gate when it comes from the register. But in both cases, the exit
splits again in two possibilities depending on which track the locomotive entered the unit.
In case of an exit to the next unit, the exit track is a \DD-, \ZZZ-track if it was respectively 
the case for the entering track. In case of an exit to the program, the exit track goes to the 
next instruction or to another one defined by the jump condition if the entering track was \DD- 
or \ZZZ-respectively. The distinction is known at the entry of the unit, see point $F$ in 
Figure~\ref{f_uDDD}. It is the reason why at $F$ a passive memory switch is sitting. As far as
that difference has to be known at~$B$ and at~$C$, there is at each of those points an active
memory switch connected to the passive one at~$F$. The connection to those active memory switches
is established by a track from~$F$ to a fork before reaching~$C$ which sends a copy to~$C$ and 
another copy to~$B$. Accordingly, the whole structure performed what is expected from it in all 
situations.

Figure~\ref{f_choix} illustrates the tracks followed by a locomotive returning from the register.
Its colour indicates which operation it performed. Coming from the register it arrives at $A$
where a fork is sitting. One branch has a \BB-filter at~$B$, meaning that the branch leads to \DDI.
The other branch has a \GG-filter at~$G$, so that the branch leads to \DDD. Accordingly, the
locomotive goes to the appropriate structure which, as already seen, sends it back to the right 
place in the program.
\vskip 10pt
\vtop{
\ligne{\hfill
\includegraphics[scale=0.4]{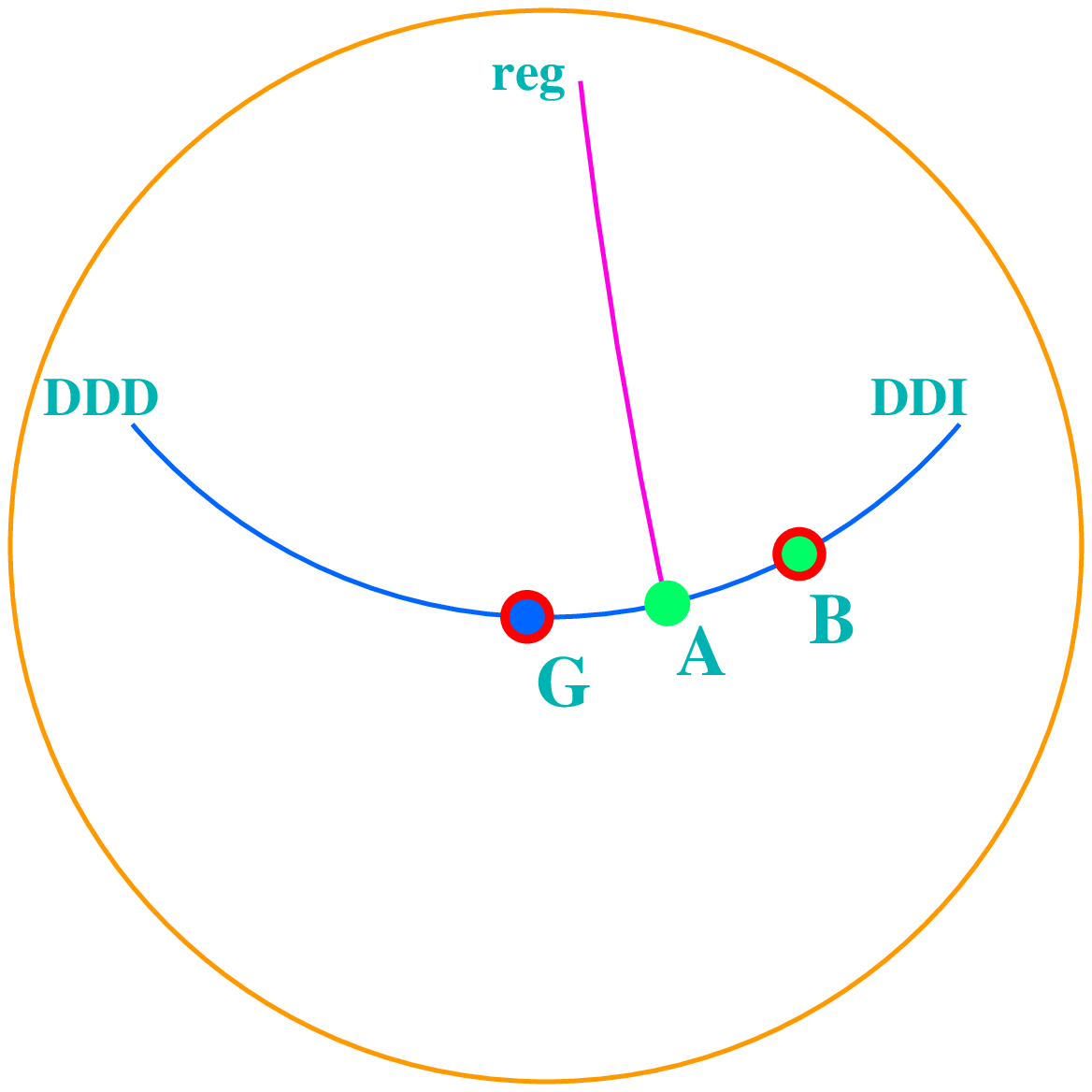}
\hfill}
\vspace{-5pt}
\begin{fig}\label{f_choix}
\leurre
The structure sending a locomotive returning from the register to the appropriate \DDI- or 
\DDD-structure. Note at $B$ and at $G$ the symbol for the filter: the forbidden colour is 
circumscribed by a red circle.
\end{fig}
}

\label{reg_op}
{\bf \ref{reg_op}.c\ Decrementing and incrementing the register}

The operation performed by a locomotive inside the register is illustrated by Figure~\ref{f_reg_op}.
The figure illustrates a standard situation for a non trivial value of the content: the \YY-tile is
at some distance from the beginning of the structure, depending on the initial content of the 
register which may be empty. On the leftmost picture of the figure, we can see the
\YY-cells 1(3), 1(4) and 1(5) around the \YY-cell at~0 which we call the {\bf hat} of the register.
In the rightmost picture of the figure we also see the hat around the \YY-cell at 1(1), it consists 
of the \YY-cells 1(2), 0 and 1(7). In the second picture from left, illustrating the incrementing 
operation, the hat consists of four \YY-cells around the \YY-cell at 1(4): 2(4), 3(4), 4(4) and 
2(5). 

Figure~\ref{f_reg_op} illustrates how the operation is performed. The figure is split into two
parts. To left, the register before the operation, to right, the register after it was performed.
In that right-hand side part, two possibilities, to left: incrementing; to right, decrementing
as mentioned.

\vtop{
\ligne{\hfill
\includegraphics[scale=0.8]{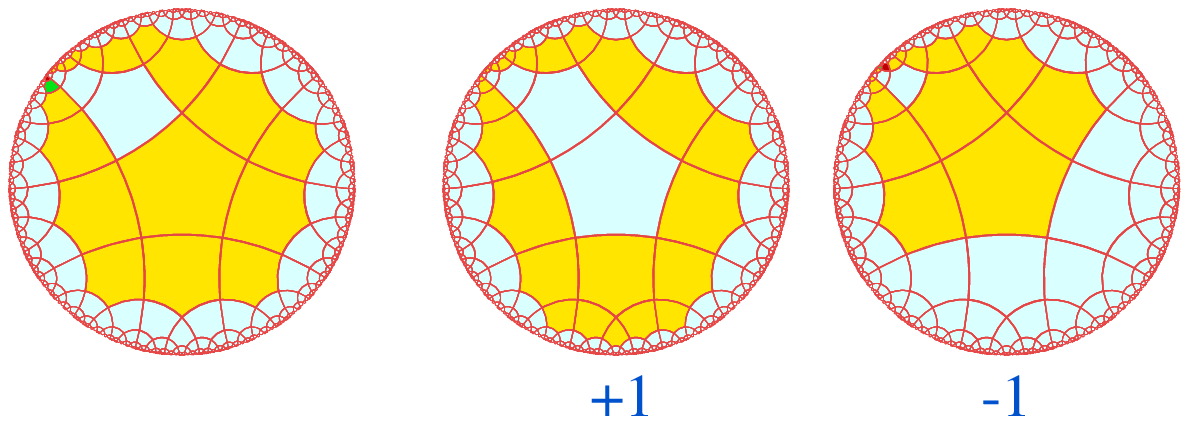}
\hfill}
\vspace{-10pt}
\begin{fig}\label{f_reg_op}
\leurre
To left, the register before the operation. To right, the configuration after the operation:
incrementing and, rightmost one, decrementing.
\end{fig}
}

The action illustrated by Figure~\ref{f_reg_op} has been tested by a computer program and can be 
checked in Figure~\ref{f_incdec} and in Table~\ref{t_rules}.
\vskip 10pt
\vtop{
\ligne{\hfill	
\includegraphics[scale=0.38]{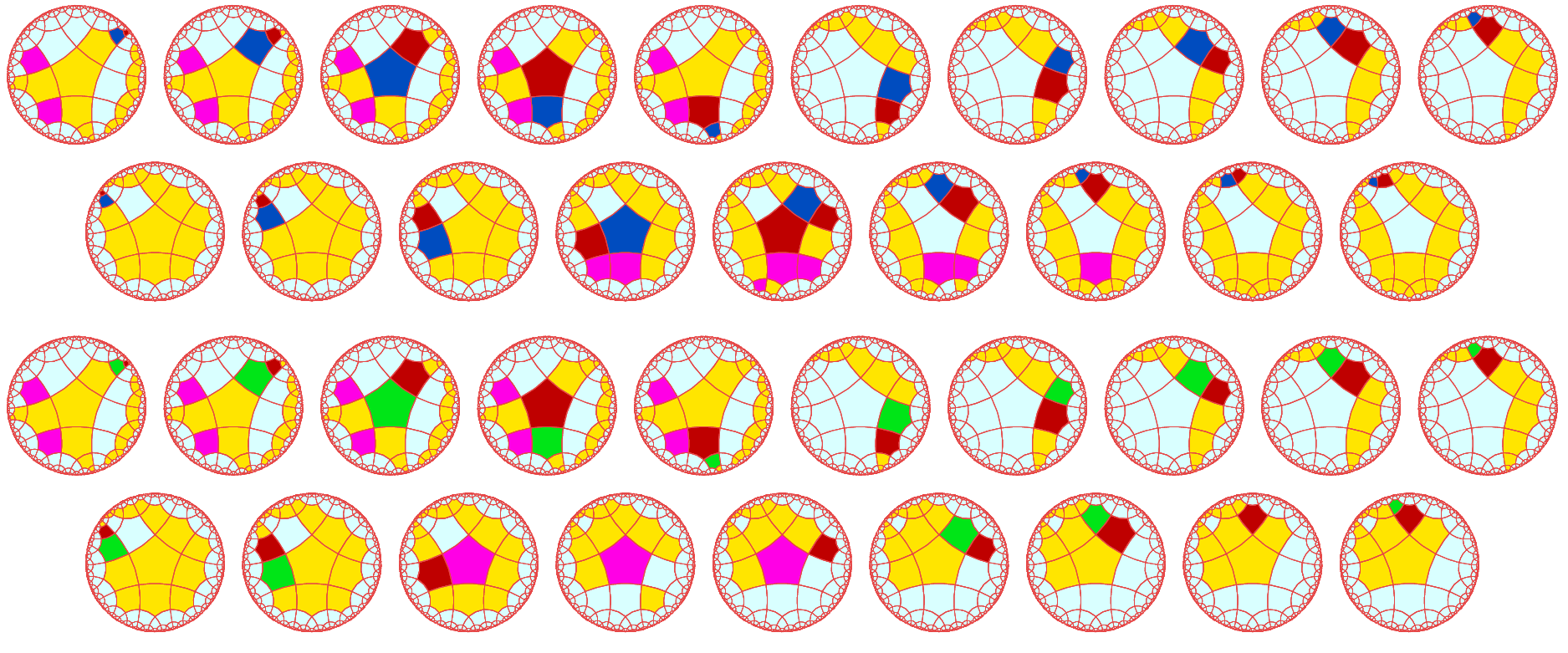}
\hfill}
\begin{fig}\label{f_incdec}
\leurre
Top, execution of an incrementing instruction when $n>0$ together with the motion of the 
locomotive through the beginning of the register and then the motion after the operation is
completed. Bottom, the similar executions for a decrementing instruction when $n>2$.
\end{fig}
}	

Figures illustrating the particular cases are given in Section~\ref{srules} to which we now turn.

\section{Rules}\label{srules}

   The definition of the rules will give us the opportunity to more explicitly study the behaviour
of the automaton.

   We start the section by introducing a formalism for writing rules and schemes of rules which 
will allow us to better present some rules. Then, in each further sub-sections,
we successively examine the cases studied in the sub-subsections of 
Subsection~\ref{newrailway}. 

\subsection{Fixing notations}\label{sbnot}

\def\schrule #1 #2 #3 {\hbox{\hbox{\ftt{#1}-\hbox{\ftt#2}:\hbox{\ftt#3}\hfill}}
}
We fix the rules in the following
format: \schrule c {s$_1$s$_2$s$_3$s$_4$s$_5$s$_6$s$_7$s$_8$s$_9$s$_{10}$} n {} where {\tt c} 
denotes the current state of the cell $\sigma$ to which the rule applies provided that $s_i$ 
with \hbox{$i\in\{1..10\}$} is the 
state of the neighbour $i$ of $\sigma$ and when the rule applies, the new state of $\sigma$ is 
{\tt n}. Of course, {\tt c}, s$_i$ and {\tt n} take their value in
\hbox{\WW, \YY, \BB, \RR, \GG, \MM}, numbered in that order. 
Note that neighbours $i$ with $i$ in \hbox{$\{1..5\}$} are full ones while neighbours $i$ when $i$
is in \hbox{$\{6..10\}$} are partial ones.
From now on, as far as we speak about the cellular automaton, we shall say cell in
place of tiles. As an example, the central tile of a window will become the central
cell of the window. The reason is that a cell is supported by a tile but it also contains
the finite automaton, the same for all cells, which rules the change of cell at each tip
of the discrete clock. All those conditions are just an application of the definition of
a cellular automaton. 

The first letter is the current state of the cell and the
last one is the new state obtained from the current state of the cell and the current 
states of its neighbours: if we order the neighbours according to the letters from~1 up
to~10 from left to right in the word delimited between two commas, the $i^{\rm th}$
letter gives the state of the corresponding $i^{\rm th}$ neighbour of the cell. That latter word 
is the {\bf neighbourhood word}. 

An observation must be made about rotation invariance. A rotation which leaves a tile globally 
unchanged operates a permutation on the sates of its full neighbours. Clearly, the same permutation
operates on the partial neighbours. The new rule resulting after such a permutation is called a
{\bf rotated form} of the rule. With that new definition of a rotated form, we can define the 
{\bf minimal form} of a rule: it is the lexicographically smallest one among the five rotated 
forms of the rule.

\subsection{Motion of a locomotive on the tracks}

   As already mentioned, we consider the tracks as combinations of segments of vertical lines
and arcs on levels of a tree, see Figure~\ref{f_grille} for instance. In order to establish the 
rules and to test them, we constructed two figures: the 'S'-track and the 'Z'-one, see
Figures~\ref{f_voie_ess} and \ref{f_voie_zed}.

Take, as an example, the motion rules on a vertical track for a blue locomotive displayed in
$(vT)$.
The number to right of a rule is the time measured from the application of the rule
\schrule Y {YWWYWWWWWW} Y {} applied to an idle configuration. The number to left of the rule is
its number in Table~\ref{t_rules}. Similar rules hold for a \GG-locomotive: in the rules, each
occurrence of \BB{} is replaced by \GG. Of course, rule~89 also applies to a \GG-locomotive
when the rear \RR{} of the locomotive is seen from the \YY-cell just left by the locomotive.

\vskip 5pt
\ligne{\hfill
$\vcenter{
\hbox to 250pt{\hfill
\vtop{
\ligne{\hfill\hbox to 71pt{\hfill$\downarrow$\hfill}\hfill
\hbox to 71pt{\hfill$\uparrow$\hfill}\hfill}
\ligne{\hfill
\vtop{\leftskip 0pt\parindent 0pt\hsize=81pt
\ligne{\hfill cell 0\hfill}
\ligne{5 \schrule {Y} {YWWYWWWWWW} {Y} $0,5$\hfill}
\ligne{63 \schrule {Y} {BWWYWWWWWW} {B} $1$\hfill}
\ligne{67 \schrule {B} {RWWYWWWWWW} {R} $2$\hfill}
\ligne{76 \schrule {R} {YWWBWWWWWW} {Y} $3$\hfill}
\ligne{89 \schrule {Y} {YWWRWWWWWW} {Y} $4$\hfill}
\ligne{\hfill cell 1(1)\hfill}
\ligne{6 \schrule {Y} {YWYWWWWYWW} {Y} $0,6$\hfill}
\ligne{57 \schrule {Y} {YWYWWWWBWW} {Y} $1$\hfill}
\ligne{60 \schrule {Y} {YWBWWWWRWW} {B} $2$\hfill}
\ligne{25 \schrule {B} {WRWWYWYWWW} {R} $3$\hfill}
\ligne{41 \schrule {R} {WYWWBWYWWW} {Y} $4$\hfill}
\ligne{49 \schrule {Y} {WYWWRWYWWW} {Y} $5$\hfill}
}
\hskip 20pt
\vtop{\leftskip 0pt\parindent 0pt\hsize=81pt
\ligne{\hfill cell 0\hfill}
\ligne{6 \schrule {Y} {YWWYWWWWWW} {Y} $0,5$\hfill}
\ligne{63 \schrule {Y} {YWWBWWWWWW} {B} $1$\hfill}
\ligne{147 \schrule {B} {YWWRWWWWWW} {R} $2$\hfill}
\ligne{152 \schrule {R} {BWWYWWWWWW} {Y} $3$\hfill}
\ligne{157 \schrule {Y} {RWWYWWWWWW} {Y} $4$\hfill}
\ligne{\hfill cell 4(1)\hfill}
\ligne{6 \schrule {Y} {YWYWWWWYWW} {Y} $0,6$\hfill}
\ligne{57 \schrule {Y} {YWYWWWWBWW} {Y} $1$\hfill}
\ligne{60 \schrule {Y} {YWBWWWWRWW} {B} $2$\hfill}
\ligne{25 \schrule {B} {WRWWYWYWWW} {R} $3$\hfill}
\ligne{41 \schrule {R} {WYWWBWYWWW} {Y} $4$\hfill}
\ligne{49 \schrule {Y} {WYWWRWYWWW} {Y} $5$\hfill}
}
\hfill}
}\hfill}}$\hfill$(vT)$\hskip15pt}
\vskip 5pt
In $(vT)$, the number(s) after an instruction is the time at which the rule is applied, 0 being the
initial time for the table.

Note that, in $(vT)$ instead of rule~25 one would expect the rule \schrule {B} {YWRWWWWYWW} {R} to
be applied. However, that latter rule is a rotated form of rule~25 as can easily be checked. Note 
that in rule~25, the track joins neighbours~2 and~5 as required for a white tile which is the case
for the support of the cell 4(1).

Say that a rule is {\bf conservative} if and only if the new state is the same as the current one
and also if the states of the neighbours are also unchanged when the rules for the neighbours are 
applied. If the state of the cell only is unchanged, we speak of a {\bf witness rule}. As an 
example, rule~75, namely \schrule {W} {WWYWWWBWWW} {W} is a witness rule as far as the \BB-cell
in the neighbourhood of the current cell to which the rule applies belong to a blue locomotive.

The rules dealing with the motion of a locomotive on tracks are given in Table~\ref{t_rules} from
rule~7 until rule~377. However, several rules among those ones also apply to the fork, to the
fixed switch, to the other auxiliary structures and to many cells of the registers too.

Note that among the mentioned rules, many of them deal with arcs of a circle. Together with rules
applying to a segment of line, it requires additional rules. They have the form
\schrule {Y} {WYWWYYYWWW} {Y} . If \FF{} stands for \BB{} or \GG, the front of a locomotive,
we also have \schrule {Y} {WYWWYYFWWW} {Y}, rule~37, rule~175 for a \BB-, \GG-locomotive
respectively.
As motion rules, we have the rules \schrule {Y} {WFWWYYRWWW} {F}, \schrule {F} {WRWWYYYWWW} {R}, 
the rule \schrule {R} {WYWWFYYWWW} {Y}, the rule \schrule {Y} {WYWWRFYWWW} {Y} and the rule
\schrule {Y} {WYWWYRYWWW} {Y} respectively 
rules~45, 52,
56, 59 and 62 for a \BB-locomotive, and rules , 181, 186, 188, 191  and 62 again for a 
\GG-locomotive. 
We also have the rules \schrule {Y} {WYWWYYWWWW} {Y}, \schrule {Y} {WFWWYYWWW} {F}, 
\schrule {F} {WRWWYYWWWW} {R},
\schrule {R} {WYWWFYWWWW} {Y}, \schrule {Y} {WYWWRFWWWW} {Y} and \schrule {Y} {WYWWYRWWWW} {Y}. We 
leave as an exercise to the reader to identify them in Table~\ref{t_rules}. 
We refer the reader to Figure~\ref{f_voie_ess} illustrating those motions where the rules can be 
checked. We also leave as an exercise to the reader to identify the motion rules on a 'Z'-track.

\subsection{Fixed switch and fork}\label{sbfxfk}

We start with the fixed switch, considering its passive version only as already mentioned.

We have four cases: a blue locomotive coming from the left-hand, right-hand side branch and 
the similar two sub-cases with a green locomotive.
We refer the reader to Figures~\ref{f_i_fix} and  \ref{f_auxil} to follow the application of the 
rules given by Table~\ref{tfx} where the notations are those introduced in Subsection~\ref{sbnot}. 

In the table, the rules are given for both locomotives: in the considered rule, replace the
occurrences of \FF, all of them by \BB{} or all of them by \GG. 

For other configurations of the fixed switch, the reader is referred to Figure~\ref{f_m_fix}
and to Table~\ref{t_rules}.

Note the symmetry of the neighbourhood words for the rules dealing with the transformation of
the current cell: the position of \FF{} in that word depends from which side of the switch the
locomotive is arriving.

In the case of a fork, rule 380, namely \schrule {Y} {YWWYYWWYRY} {Y} gives the neighbourhood of the
\YY-cell where all tracks meet. That cell is a cell of the track and it is at the same time a
good witness of what happens:

That \YY-cell is a good witness of the action of the fork: rules 488 and 518, namely 
\schrule {R} {YWBBWWWYRY} {W} and \schrule {R} {YWGGWWWYRY} {Y} indicate that the cell can see 
both fronts of the locomotives being created from the arriving one and rules 495 and 525, 
namely \schrule {Y} {YWRRWWWBYB} {Y} and \schrule {Y} {YWRRWWWGYG} {Y} show us that the cell
can see both rears and both fronts of the leaving locomotives. And last, rule 502, namely
\schrule {Y} {YWYYWWWRRR} {Y}, show us that the cell can see the rear of the leaving locomotive,
whichever its colour.

\def\fts#1{{\small #1}}
\vtop{\leftskip 0pt\parindent 0pt
\begin{tab}\label{tfx}
\leurre
Rules for the motion of a \BB-locomotive through the cell $1(3)$ in
a fixed switch with an upward vertical exit track. To left, to right, when the locomotive arrives 
from the left, from the right respectively. The index at the number indicates at which time(s) the
rule is applied, $0$ being the initial time for the table.
\end{tab}
\ligne{\hfill
\vtop{\leftskip 0pt\parindent 0pt\hsize=93pt
\ligne{1(3), $\bullet$\hskip 0.5pt$\rightarrow$, \BB-, \GG- \hfill}
\ligne{\hfill\schrule Y {YWYYWWWYRY} Y  \fts{380}$_{0,6}$ \hfill}
\ligne{\hfill\schrule Y {YWYYWWWFRY} Y  \fts{395,438}$_1$ \hfill}
\ligne{\hfill\schrule Y {YWFYWWWRRY} F  \fts{397,440}$_2$ \hfill}
\ligne{\hfill\schrule F {YWRYWWWYRY} R  \fts{403,446}$_3$ \hfill}
\ligne{\hfill\schrule R {FWYYWWWYRY} Y  \fts{412,454}$_4$ \hfill}
\ligne{\hfill\schrule Y {RWYYWWWYRY} Y  \fts{419}$_5$ \hfill}
}
\hfill
\vtop{\leftskip 0pt\parindent 0pt\hsize=93pt
\ligne{1(3), $\leftarrow$\hskip 0.5pt$\bullet$, \BB-,\GG-\hfill} 
\ligne{\hfill\schrule Y {YWYYWWWYRY} Y  \fts{380}$_{0,6}$ \hfill}
\ligne{\hfill\schrule Y {YWYYWWWYRF} Y  \fts{423,458}$_1$ \hfill}
\ligne{\hfill\schrule Y {YWYFWWWYRR} F  \fts{425,460}$_2$ \hfill}
\ligne{\hfill\schrule F {YWYRWWWYRY} R  \fts{430,465}$_3$ \hfill}
\ligne{\hfill\schrule R {FWYYWWWYRY} Y  \fts{412,454}$_4$ \hfill}
\ligne{\hfill\schrule Y {RWYYWWWYRY} Y  \fts{419}$_5$ \hfill}
}
\hfill}
\par}
\vskip 10pt

\ligne{\hfill
$\vcenter{\hbox{
\vtop{\leftskip 0pt\parindent 0pt\hsize=93pt
\ligne{1(3): \FF{} is either \BB{} or \GG\hfill}
\ligne{\hfill\schrule Y {YWYYWWWYRY} Y  \fts{380}$_{0,6}$ \hfill}
\ligne{\hfill\schrule Y {FWYYWWWYRY} F  \fts{481,512}$_1$ \hfill}
\ligne{\hfill\schrule F {RWYYWWWYRY} R  \fts{483,514}$_2$ \hfill}
\ligne{\hfill\schrule R {YWFFWWWYRY} Y  \fts{488,518}$_3$ \hfill}
\ligne{\hfill\schrule Y {YWRRWWWFRF} Y  \fts{495,525}$_4$ \hfill}
\ligne{\hfill\schrule Y {YWYYWWWRRR} Y  \fts{502}$_5$ \hfill}
}
}}$
\hfill$(Fk)$\hskip 15pt}
\vskip 10pt
Here too, the index at a number of a rule indicates the time of its application.

\def\HH{{\tt H}}
\subsection{Converters and filters}\label{schflts}

In the case the converter, the structure contains a \BB- or a \GG-cell. It transforms a locomotive
of the opposite colour into a locomotive of its colour. Looking at Figure~\ref{f_auxil}, it is
considered that the locomotive arrives from the right. Accordingly, the first cell of the track 
seeing the \BB- or the \GG-cell is cell 1(5). However, we shall follow the move of a locomotive
from the cell 1(4) where the colour of the changer is sitting. It is a perfect place to witness
the transformation:
\vskip 10pt
\ligne{\hfill  
$\vcenter{\hbox to 280pt{
\vtop{\leftskip 0pt\parindent 0pt\hsize=93pt
\ligne{1(4):$\{$\FF,\HH$\}$ = $\{$\BB,\GG$\}$\hfill}
\ligne{\schrule F {YWGGWYRWWW} F \fts{536,572}$_{0,4}$\hfill}
\ligne{\schrule F {YWGGWHRWWW} F \fts{548,583}$_1$\hfill}
\ligne{\schrule F {FWGGWRRWWW} F \fts{555,590}$_2$\hfill}
\ligne{\schrule F {RWGGWYRWWW} F \fts{563,598}$_3$\hfill}
}
\hskip 20pt
\vtop{\leftskip 0pt\parindent 0pt\hsize=93pt
\ligne{0:$\{$\FF,\HH$\}$ = $\{$\BB,\GG$\}$\hfill}
\ligne{\schrule Y {WYRFYWWWWW} Y \fts{532,570}$_{0,5}$\hfill}
\ligne{\schrule Y {WYRFHWWWWW} F \fts{547,582}$_1$\hfill}
\ligne{\schrule F {WYRFRWWWWW} R \fts{551,586}$_2$\hfill}
\ligne{\schrule R {WFRFYWWWWW} Y \fts{559,594}$_3$\hfill}
\ligne{\schrule Y {WRRFYWWWWW} Y \fts{566,601}$_4$\hfill}
}
}}$
\hfill$(Cv)$\hskip 15pt}
\vskip 10pt
The table also provides us with the rules applied to cell~0.
Note, for instance, that rules 536 and 572 have the same neighbourhood word but a different current
state. Note the use of the first rule of the table after the application of the last one was 
completed. 

From the use of the filters which was done in Subsection~\ref{newrailway}, we know that it was
required that the colour of those structures could be changed: in other words they must be 
programmable. Looking again at Figure~\ref{f_auxil}, we can see four \YY-cells at the top of the 
filters. Those four \YY-cells can be seen as the
end of a track arriving at the filter. We require the filter to be changed by a \BB-locomotive.
This time, we take cell~0 as a witness of the motion in the filter working and we take the cell 
1(1) to illustrate the change of the filtering colour, see Figure~\ref{f_auxil}.
\vskip 10pt
\ligne{\hfill
$\vcenter{
\hbox to 280pt{\hfill
\vtop{
\ligne{\hfill
\vtop{\leftskip 0pt\parindent 0pt\hsize=83pt
\ligne{0: same colour\hfill}
\ligne{\schrule Y {FYRWYWWWWW} Y \fts{604,653}\hfill}
\ligne{\schrule Y {FYRWFWWWWW} F \fts{616,663}\hfill}
\ligne{\schrule F {FYRWRWWWWW} R \fts{621,667}\hfill}
\ligne{\schrule R {FFRWYWWWWW} Y \fts{629,675}\hfill}
\ligne{\schrule Y {FRRWYWWWWW} Y \fts{637,682}\hfill}
}
\hfill
\vtop{\leftskip 0pt\parindent 0pt\hsize=83pt
\ligne{0: opposite colours\hfill}
\ligne{\schrule Y {FYRWYWWWWW} Y \fts{604,653}\hfill}
\ligne{\schrule Y {FYRWHWWWWW} Y \fts{643,687}\hfill}
\ligne{\schrule Y {FYRWRWWWWW} Y \fts{648,691}\hfill}
}
\hfill}
\vskip 5pt
\ligne{\hfill
\vtop{\leftskip 0pt\parindent 0pt\hsize=80pt
\ligne{1(1),\hfill}
\ligne{change colour:\hfill}
\ligne{\schrule F {YWYWWYYYRR} F \fts{605,654}\hfill}
\ligne{\schrule F {YWYWWYYFRR} H \fts{695,704}\hfill}
\ligne{\schrule H {YWYWWYYRRR} H \fts{700,708}\hfill}
\ligne{\schrule H {YWYWWYYYRR} H \fts{654,605}\hfill}
}
\hfill}
}}}$
\hfill$(Flt)$\hskip 15pt}
\vskip 10pt
In the table, \FF{} and \HH{} mean opposite colours where \BB{} and \GG{} are opposite of one
another. Note that rules 700 and 708 have the same neighbourhood word and different current states.
The same observation holds for rules 654 and 605. Note that after the change of colour, the 
conservative rule for the cell~1(1) turns from rule~605,654 to rule~654,605 when the filter is
\BB,\GG{} respectively.

\subsection{Registers}~\label{sbregs}

   From Sub subsection~\ref{sbbregdisc}, we know that when the register is empty,
\reg{0} {} is \YY. From Figure~\ref{f_reg}, we know that the transformation from \reg{n} {} to
\reg{n+1} {} or \reg{n-1} {} entails much work. First, the \YY-cell moves from \reg{n} {} to
\reg{n+1} {} or to \reg{n-1} {} which, in the latter case, assumes that $n>0$. It means also that 
the previous hat is either destroyed or modified, and that a new hat is built from scratch or from
already non empty cells.

   The case when $n=0$ requires a special configuration for \reg{0} {} in order the decrementing
operation should not be processed. Instead the \GG-locomotive should go along a specific track
as mentioned in Subsection~\ref{sbbregdisc}. Now, when a locomotive goes to \reg{n} , when $n>1$ in 
order to perform an operation, it must cross \reg{0} . Accordingly, there are two possible tracks
followed by a locomotive leaving \reg{0} , depending on whether it could not perform the operation
or it could do it as far as it has to increment $\mathcal R$ or it has to operate on \reg{n} {} for
$n>0$. The left-hand side part of Figure~\ref{f_reg} is repeated in Figure~\ref{f_debreg} in order
to better see the situation around \reg{0} . The two tracks which can be followed by a locomotive
leaving \reg{0} are separate from cell~0: one track goes on to 1(3), the other goes on to 1(2). 
In order to differentiate the configurations, cell 1(2) has two \MM-neighbours: 2(2), shared 
with 0 and 2(3), shared with 1(3). 
\vskip 5pt
\vtop{
\vspace{-20pt}
\ligne{\hfill
\includegraphics[scale=2]{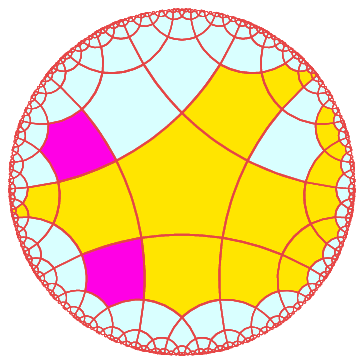}
\hfill}
\begin{fig}\label{f_debreg}
\leurre
	The idle configuration of \reg{$0$} {}. Note that the cell~$1(4)$ is \WW{} when $n>0$ and 
that it is \YY{} as here, when $n=0$. 
\end{fig}
}

Figure~\ref{f_incdec} shows us the motions of the locomotive during an incrementing operation and 
during a decrementing one, assuming that $n>1$ for those operations.

The rules for incrementing the register are rules~535 to~684 and also
rules~771 to~774. The rules of decrementing the register are rules~685 to~770. Those rules are
gathered as follows. For the incrementing operation, rules~535 up to~608 deal with the case
when $n>1$. Rules~609 to~639 manage the crossing of $\mathcal R$(0) and the progression of the
\BB-locomotive to $\mathcal R$($n$). Rules~640 to~684 deal with incrementing the empty register.
Rule~685 deals with the return of the \BB-locomotive from $\mathcal R$($n$+1) including
the crossing of $\mathcal R$(0). At last, rules~771 to~774 deal with the incrementation of the
register when $n=1$. 

For the decrementing operation, rules~685 to~714 manage the decrementing of the register when $n>1$.
Rules~715 to~734 deal with the progression of a \GG-locomotive from $\mathcal R$(0) up to
$\mathcal R$($n$) when $n>1$. Rules~735 to~737 deal with the return of the \GG-locomotive from
$\mathcal R$($n$$-$1), also crossing $\mathcal R$(0). Rules~738 to~764 manage the \GG-locomotive
when it faces an empty register. At last, rules~765 to~770 deal with the case when $n=1$ which
boils down to an empty register.

Figure~\ref{f_regoppart} illustrates the behaviour of the locomotive in particular cases.
First, when the register is empty, {\it i.e.} its value is 0: incrementing resets it to~1
while decrementing fails: the return track is different in that case. Then we have the case when the
value of the register is~1. Incrementing it sets it to~2 while decrementing it sets it to~0.
At last, we consider the case when the value of the register is 2 and the locomotive has to 
decrement it, setting it to~1.
\vskip 10pt
\vtop{
\ligne{\hfill
\includegraphics[scale=0.4]{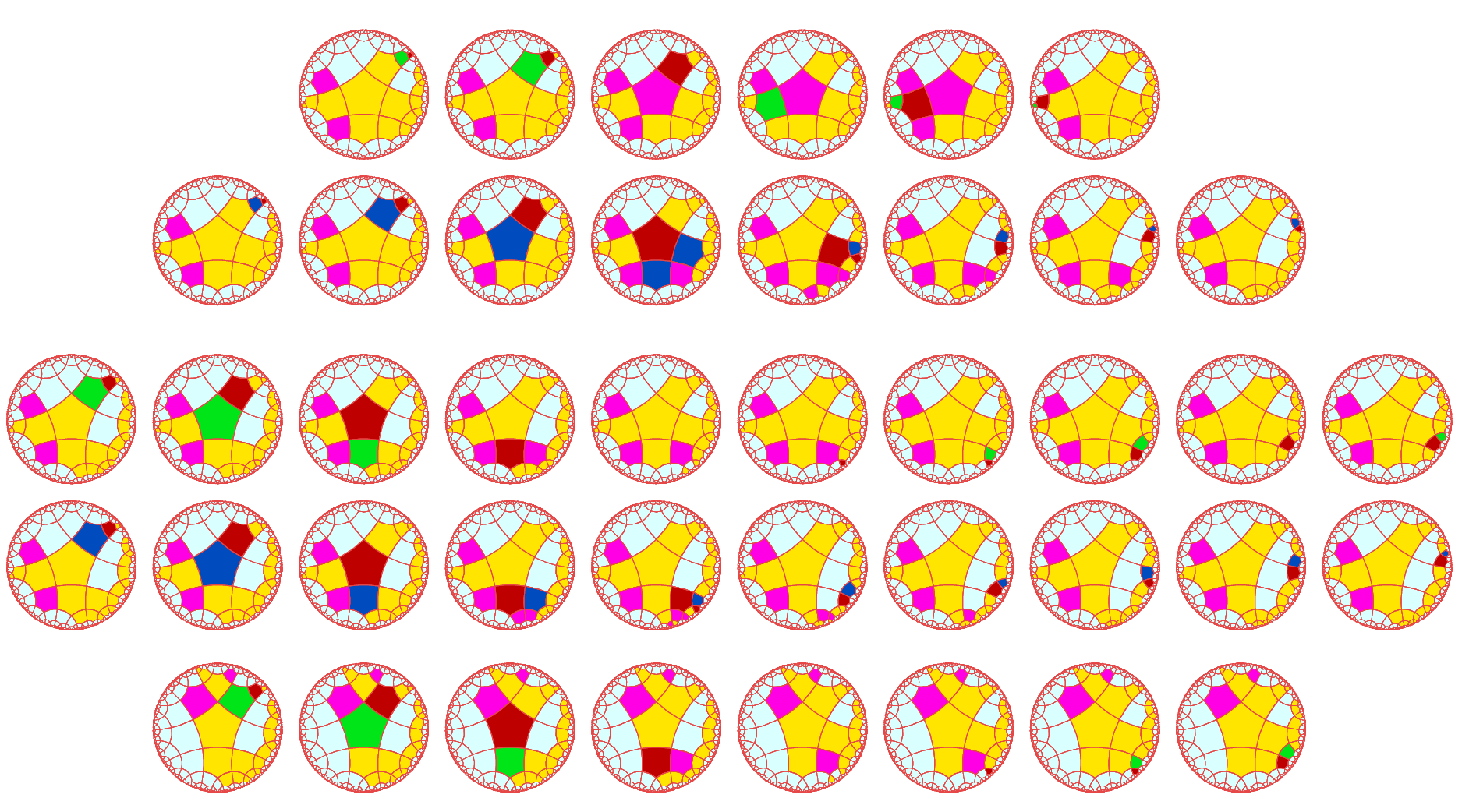}
\hfill}
\begin{fig}\label{f_regoppart}
\leurre
The particular cases:
Top two row: $n=0$. First, decrementing fails, the locomotive exits from another way.
motion of a locomotive when the register is empty. Second, incrementing from~$0$ to~$1$.
Next two rows, $n=1$. First, decrementing from $1$ to~$0$. Second, incrementing from~$1$ to~$2$.
Bottom row: $n=2$. Decrementing from~$2$ to~$1$.
\end{fig}
}
\vskip 10pt
The corresponding rules are rules~975 up to~1072 from Table~\ref{t_rules}. There are too many of
those rules in order to detail their working. Their correctness was checked by a computer program 
which also created the figures illustrating that working. The same observation holds for
Figure~\ref{f_regopsgene} which illustrates rules~712 up to~850. That figure also illustrate the 
motion of the locomotive for each operation: before performing it and then when it completed it.
That illustrates rules~851 up to 974 of Table~\ref{t_rules}.
\vskip 10pt
\vtop{
\ligne{\hfill
\includegraphics[scale=0.4]{move_opsgene.ps}
\hfill}
\begin{fig}\label{f_regopsgene}
\leurre
The case $n> 1$. Top row: incrementing. Bottom row, decrementing.
\end{fig}
}
\vskip 10pt
It is worth having a closer look at the transformation of the hat during the operations on
a register. See Figure~\ref{zoom_ops} and the instructions given at $(zIop)$ and at $(zDop)$.

\vskip 10pt
\vtop{
\ligne{\hfill
\includegraphics[scale=0.6]{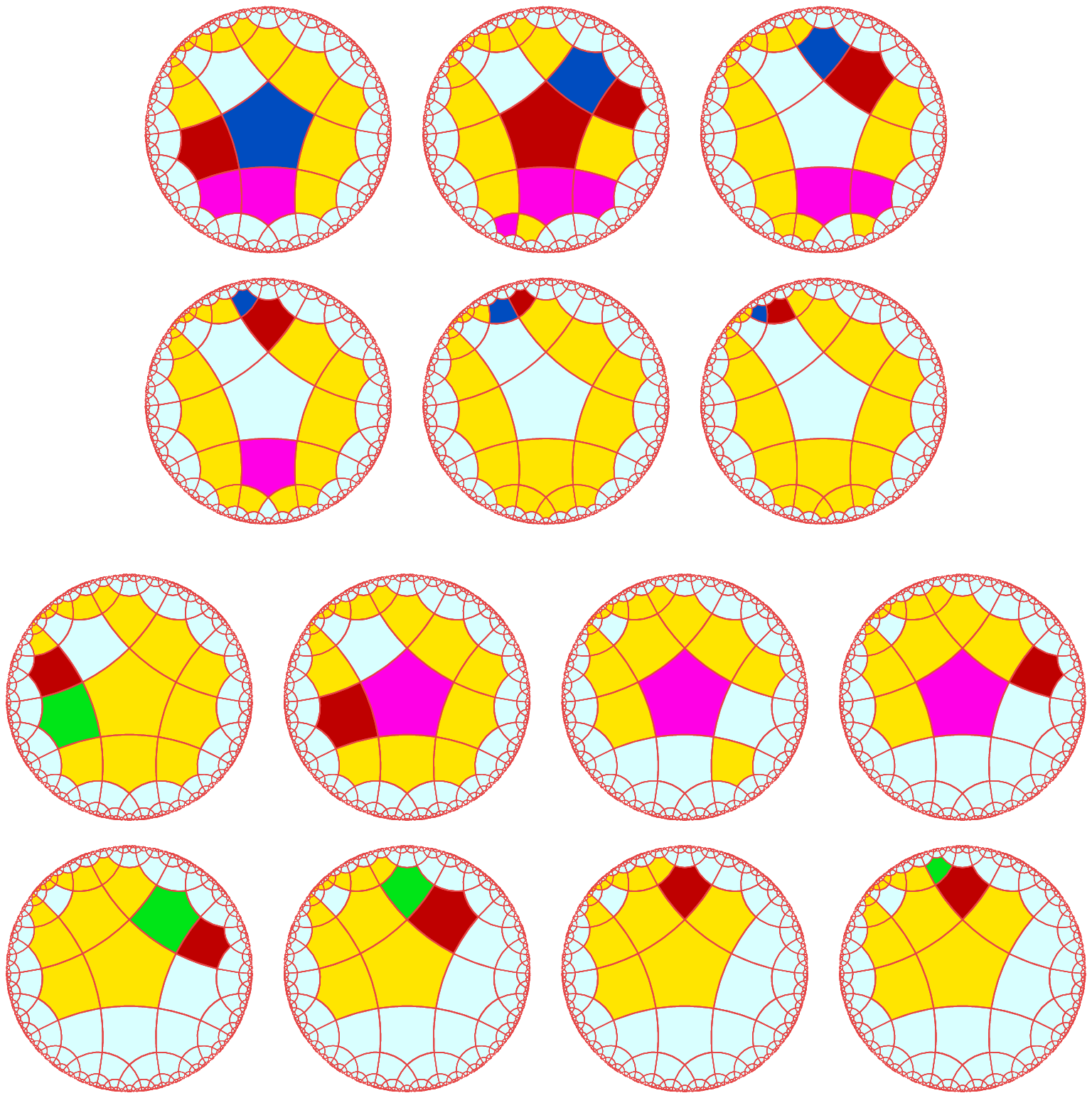}
\hfill}
\begin{fig}\label{zoom_ops}
\leurre
Closer look at the operations on a register when $n>1$, where $n$ is its content.
Top two rows: incrementing operation; bottom two rows: decrementing one. 
\end{fig}
}
\vskip 10pt

In lines under $(zIop$), we have the instructions applied to cell (3,1) and to cell (3,10). The first
cell is a good witness of the transformation touching its neighbours and itself by an incrementing
operation. Cell (3,10) belongs to the new hat and we can see how it is changed from \WW{} to \YY{} .
We can see how the new hat is created in 4 steps once cell~(3,1) is changed from~\YY{} to~\MM{} .
The cell (3,1) becomes the new marker of the end of the register after it has been incremented.
\vskip 10pt
\ligne{\hfill
$\vcenter{\leftskip 0pt\parindent 0pt\hsize=250pt
\ligne{\hfill
\vtop{\leftskip 0pt\parindent 0pt\hsize=85pt
\ligne{(3,1):\hfill}
\ligne{\schrule Y {YYWWYYYWWW} Y \fts{718}\hfill}
\ligne{\schrule Y {YYWWYYBWWW} M \fts{730}\hfill}
\ligne{\schrule M {BMWWYYRWWW} M \fts{735}\hfill}
\ligne{\schrule M {RYYWMYYMWW} M \fts{750}\hfill}
\ligne{\schrule M {WYYWMYYYWY} M \fts{766}\hfill}
\ligne{\schrule M {WYYYYYYYWY} Y \fts{777}\hfill}
\ligne{\schrule Y {WYYYYYYYYY} Y \fts{712}\hfill}
}
\hfill
\vtop{\leftskip 0pt\parindent 0pt\hsize=85pt
\ligne{(3,10):\hfill}
\ligne{\schrule W {WWWWWWYWWW} W \fts{15}\hfill}
\ligne{\schrule W {WWWWWWMWWW} W \fts{742}\hfill}
\ligne{\schrule W {YWWWWWMWWW} W \fts{755}\hfill}
\ligne{\schrule W {YYWWWWMWWW} Y \fts{780}\hfill}
\ligne{\schrule Y {YYWWWWYWWW} Y \fts{719}\hfill}
}
\hfill}}$
\hfill$(zIop)$\hskip 15pt}
\vskip 10pt
In the lines of $(zDop)$, we have the instructions which are applied to a register with value $n$
greater than~1 to decrement it. Cell~0 is a good witness of the transformation: its \YY-neighbours
become \WW{} for 5 of them. Also, the cell (3,1) is also an interesting witness of the transformation.
For instance, rule~755 witnesses the fact that cell~0 was changed from~\WW{} to~\YY{} : indeed,
cell~0 is the neighbour~1 of the cell (3,1).
\vskip 10pt
\ligne{\hfill
$\vcenter{\leftskip 0pt\parindent 0pt\hsize=250pt
\ligne{\hfill
\vtop{\leftskip 0pt\parindent 0pt\hsize=85pt
\ligne{0:\hfill}
\ligne{\schrule Y {WYYYYYYYYY} Y \fts{712}\hfill}
\ligne{\schrule Y {WYYYYYGYYY} Y \fts{796}\hfill}
\ligne{\schrule Y {WGYYYYRYYY} M \fts{798}\hfill}
\ligne{\schrule M {WRYYYYYYYY} M \fts{802}\hfill}
\ligne{\schrule M {YYWWYYYWYY} M \fts{809}\hfill}
\ligne{\schrule M {YYWWYYYWWR} Y \fts{817}\hfill}
\ligne{\schrule Y {YYWWGYYWWR} Y \fts{821}\hfill}
\ligne{\schrule Y {YYWWRGYWWW} Y \fts{826}\hfill}
\ligne{\schrule Y {YYWWYRYWWW} Y \fts{831}\hfill}
\ligne{\schrule Y {YYWWYYYWWW} Y \fts{718}\hfill}
}
\hfill
\vtop{\leftskip 0pt\parindent 0pt\hsize=85pt
\ligne{(3,1):\hfill}
\ligne{\schrule Y {YYWWYYYWWW} Y \fts{718}\hfill}
\ligne{\schrule Y {YYWWYYGWWW} Y \fts{800}\hfill}
\ligne{\schrule Y {MYWWYYRWWW} W \fts{806}\hfill}
\ligne{\schrule W {MYWWYWYWWW} W \fts{813}\hfill}
\ligne{\schrule W {MWWWWWYWWW} W \fts{754}\hfill}
\ligne{\schrule W {YWWWWWYWWW} W \fts{9}\hfill}
}
\hfill}}$
\hfill$(zDop)$\hskip 15pt}
\vskip 10pt
Table~\ref{t_rules} displays the whole set of rules used by the automaton. We use the same 
notation as what was done in the previous parts of the paper. We remind the reader the fact that 
in the table, the neighbouring word is the minimal lexicographic order among the possible 
equivalent forms of the rule obtained through a circular permutation. However, on 
Figure~\ref{f_incdec}, it is not difficult to check the rules indicated in the text. It is needed 
to see that in the text we considered rules applied to a few cells while the rules of 
Table~\ref{t_rules} were devised for all possible cells involved in the computation: cells directly impacted by the computation as well as witnessing cells.

With Table~\ref{t_rules}, with all the tables and all the figures from Subsection~\ref{newrailway}, 
the proof of Theorem~\ref{letheo} is completed. \hfill$\Box$

\section{Conclusion}

We can see that we have 1072 rules in the table, conservative, witness and motion rules. 
Note that the registers alone require 361 rules and 98 of those ones are required for
the special cases when the content of the register is 0, 1 or 2, that latter case in the case
of decrementing only. 

We remain with the open question whether it is possible to reduce the number of
states. 
\newpage
\newcommand{\tttregle}[4]
{%
\ligne{\hbox to 15pt{\hfill\ftt #1}
\hbox to 10pt{\hfill\ftt #2\hfill}
\hbox to 40pt{\hfill\ftt #3\hfill}
\hbox to 10pt{\hfill\ftt #4\hfill}
\hfill}
\vskip 0pt
}

\ligne{\hfill
\vtop{\leftskip 0pt\parindent 0pt\hsize=80pt
\ligne{\hfill Locomotives \hfill}
\ligne{\hfill on S-track\hfill}
\vskip-2pt
\ligne{\hrulefill}
\tttregle {  1} {W} {WWWWWWWWWW} {W} 
\tttregle {  2} {B} {WWWWWWWWWW} {B} 
\tttregle {  3} {R} {WWWWWWWWWW} {R} 
\tttregle {  4} {G} {WWWWWWWWWW} {G} 
\vskip 1pt
\ligne{\hfill blue locomotive\hfill}
\ligne{\hfill $\downarrow$ motion\hfill}
\vskip 1pt
\tttregle {  5} {Y} {YWWYWWWWWW} {Y} 
\tttregle {  6} {Y} {YWYWWWWYWW} {Y} 
\tttregle {  7} {W} {YWYYYYYBYY} {W} 
\tttregle {  8} {Y} {YYWWWWWWWW} {Y} 
\tttregle {  9} {W} {YWWWWWYWWW} {W} 
\tttregle { 10} {Y} {WBWWYYWWWW} {B} 
\tttregle { 11} {Y} {WYWWYYYWWW} {Y} 
\tttregle { 12} {W} {YWWWWWWWWW} {W} 
\tttregle { 13} {W} {WYWWWWYWWW} {W} 
\tttregle { 14} {W} {YWWWWWBWWW} {W} 
\tttregle { 15} {W} {WWWWWWYWWW} {W} 
\tttregle { 16} {W} {YWWWWWYWWY} {W} 
\tttregle { 17} {W} {YWYYWYWYYY} {W} 
\tttregle { 18} {W} {WWYYYYYWYY} {W} 
\tttregle { 19} {Y} {WYWWYYWWWW} {Y} 
\tttregle { 20} {Y} {WWWWYYWWWW} {Y} 
\tttregle { 21} {W} {YWYYYYYYYY} {W} 
\tttregle { 22} {W} {YWYBWYYWRY} {W} 
\tttregle { 23} {W} {WYWWWYYWWW} {W} 
\tttregle { 24} {Y} {WWWWRBWWWW} {Y} 
\tttregle { 25} {B} {WRWWYWYWWW} {R} 
\tttregle { 26} {W} {YWWWWRWWWW} {W} 
\tttregle { 27} {R} {BYWWWWWWWW} {Y} 
\tttregle { 28} {W} {BWWWWWRWWW} {W} 
\tttregle { 29} {W} {BWWWWYWWWW} {W} 
\tttregle { 30} {W} {RWWWWWYWWW} {W} 
\tttregle { 31} {W} {RWWWWWWWWW} {W} 
\tttregle { 32} {W} {WRWWWWBWWW} {W} 
\tttregle { 33} {W} {WWWWWWBWWW} {W} 
\tttregle { 34} {W} {WWWWWWRWWW} {W} 
\tttregle { 35} {W} {YWBYYYYRYY} {W} 
\tttregle { 36} {B} {WRWWYYWWWW} {R} 
\tttregle { 37} {Y} {WYWWYYBWWW} {Y} 
\tttregle { 38} {Y} {YBWWWWWWWW} {B} 
\tttregle { 39} {W} {YWYRWYYWYB} {W} 
\tttregle { 40} {Y} {WWWWYRWWWW} {Y} 
\tttregle { 41} {R} {WYWWBWYWWW} {Y} 
\tttregle { 42} {Y} {RYWWWWWWWW} {Y} 
}
\hfill
\vtop{\leftskip 0pt\parindent 0pt\hsize=80pt
\tttregle { 43} {W} {YWRYYYYYBY} {W} 
\tttregle { 44} {R} {WYWWBYWWWW} {Y} 
\tttregle { 45} {Y} {WBWWYYRWWW} {B} 
\tttregle { 46} {B} {YRWWWWWWWW} {R} 
\tttregle { 47} {W} {BWWWWWWWWW} {W} 
\tttregle { 48} {W} {YWYYWYYWYR} {W} 
\tttregle { 49} {Y} {WYWWRWYWWW} {Y} 
\tttregle { 50} {W} {YWYBYYYYRY} {W} 
\tttregle { 51} {Y} {WYWWRBWWWW} {Y} 
\tttregle { 52} {B} {WRWWYYYWWW} {R} 
\tttregle { 53} {W} {YWYYWYYWYY} {W} 
\tttregle { 54} {W} {YWYRYYYYYB} {W} 
\tttregle { 55} {Y} {WYWWYRWWWW} {Y} 
\tttregle { 56} {R} {WYWWBYYWWW} {Y} 
\tttregle { 57} {Y} {YWYWWWWBWW} {Y} 
\tttregle { 58} {W} {YWYYBYYYYR} {W} 
\tttregle { 59} {Y} {WYWWRBYWWW} {Y} 
\tttregle { 60} {Y} {YWBWWWWRWW} {B} 
\tttregle { 61} {W} {YWYYRBYYYY} {W} 
\tttregle { 62} {Y} {WYWWYRYWWW} {Y} 
\tttregle { 63} {Y} {BWWYWWWWWW} {B} 
\tttregle { 64} {W} {BWYYYRYYYY} {W} 
\tttregle { 65} {W} {YWWWWWBWWY} {W} 
\tttregle { 66} {W} {YWYYWBYWYY} {W} 
\tttregle { 67} {B} {RWWYWWWWWW} {R} 
\tttregle { 68} {W} {RWYYYYBYYY} {W} 
\tttregle { 69} {W} {BWWWWWRWWY} {W} 
\tttregle { 70} {W} {BWYYWYWYYY} {W} 
\tttregle { 71} {W} {WWYYYYBWYY} {W} 
\tttregle { 72} {Y} {BWYWWWWYWW} {B} 
\tttregle { 73} {W} {YWYYYYBYYY} {W} 
\tttregle { 74} {W} {BWYYWRYWYY} {W} 
\tttregle { 75} {W} {WYWWWYBWWW} {W} 
\tttregle { 76} {R} {YWWBWWWWWW} {Y} 
\tttregle { 77} {W} {YWYYYYRYYY} {W} 
\tttregle { 78} {W} {RWWWWWYWWY} {W} 
\tttregle { 79} {W} {RWYYWBWYYY} {W} 
\tttregle { 80} {W} {WWYYYYRWYY} {W} 
\tttregle { 81} {B} {RWYWWWWYWW} {R} 
\tttregle { 82} {W} {BWYYYYRYYY} {W} 
\tttregle { 83} {Y} {BYWWWWWWWW} {B} 
\tttregle { 84} {W} {BWWWWWYWWW} {W} 
\tttregle { 85} {Y} {YWYWWWBYWW} {Y} 
\tttregle { 86} {W} {WYWWWWBWWW} {W} 
\tttregle { 87} {W} {RWYYWYBWYY} {W} 
\tttregle { 88} {W} {WBWWWYRWWW} {W} 
}
\hfill
\vtop{\leftskip 0pt\parindent 0pt\hsize=80pt
\tttregle { 89} {Y} {YWWRWWWWWW} {Y} 
\tttregle { 90} {W} {YWYYWRWYYY} {W} 
\tttregle { 91} {R} {YWBWWWWYWW} {Y} 
\tttregle { 92} {W} {RWYYYBYYYY} {W} 
\tttregle { 93} {B} {RYWWWWWWWW} {R} 
\tttregle { 94} {W} {RWWWWWBWWW} {W} 
\tttregle { 95} {Y} {BWYWWWRYWW} {B} 
\tttregle { 96} {W} {WBWWWWRWWW} {W} 
\tttregle { 97} {W} {YWYYWYRWYY} {W} 
\tttregle { 98} {W} {WRWWWYYWWW} {W} 
\tttregle { 99} {Y} {YWRWWWWBWW} {Y} 
\tttregle {100} {W} {YWYYBRYYYY} {W} 
\tttregle {101} {R} {YBWWWWWWWW} {Y} 
\tttregle {102} {W} {YWWWWWRWWW} {W} 
\tttregle {103} {B} {RWYWWWYYWW} {R} 
\tttregle {104} {W} {WRWWWWYWWW} {W} 
\tttregle {105} {Y} {YWYWWWWRWW} {Y} 
\tttregle {106} {W} {YWYYRYYYYB} {W} 
\tttregle {107} {Y} {YRWWWWWWWW} {Y} 
\tttregle {108} {R} {YWBWWWYYWW} {Y} 
\tttregle {109} {W} {YWYBYYYYYR} {W} 
\tttregle {110} {Y} {WYWWYBWWWW} {Y} 
\tttregle {111} {Y} {YWRWWWYBWW} {Y} 
\tttregle {112} {W} {YWYRYYYYBY} {W} 
\tttregle {113} {Y} {WYWWBRWWWW} {B} 
\tttregle {114} {Y} {YWYWWWYRWW} {Y} 
\tttregle {115} {W} {YWYYWYWYYB} {W} 
\tttregle {116} {W} {YWBYYYYYRY} {W} 
\tttregle {117} {B} {WYWWRYWWWW} {R} 
\tttregle {118} {W} {YWYBWYWYYR} {W} 
\tttregle {119} {W} {YWRYYYYBYY} {W} 
\tttregle {120} {R} {WBWWYYWWWW} {Y} 
\tttregle {121} {W} {YWYRWYWYBY} {W} 
\tttregle {122} {W} {YWYYYYYRYY} {W} 
\tttregle {123} {Y} {WRWWYYWWWW} {Y} 
\tttregle {124} {W} {YWBYWYWYRY} {W} 
\tttregle {125} {W} {WWYYYBYWYY} {W} 
\tttregle {126} {W} {YWRYWYWBYY} {W} 
\tttregle {127} {W} {WWYYBRYWYY} {W} 
\tttregle {128} {W} {YWYYWYWRYY} {W} 
\tttregle {129} {W} {WWYYRYYWYB} {W} 
\tttregle {130} {W} {WWYBYYYWYR} {W} 
\tttregle {131} {Y} {WWWWYBWWWW} {Y} 
\vskip 1pt
\ligne{\hfill $\uparrow$ motion\hfill}
\vskip 1pt
\tttregle {132} {W} {WWYBYYYWRY} {W} 
\tttregle {133} {W} {WWYRYYYWYB} {W} 
}
\hfill}

\ligne{\hfill
\vtop{\leftskip 0pt\parindent 0pt\hsize=80pt
\tttregle {134} {W} {YWYYWYWBYY} {W} 
\tttregle {135} {W} {WWYYBYYWYR} {W} 
\tttregle {136} {W} {YWBYWYWRYY} {W} 
\tttregle {137} {W} {WWYYRBYWYY} {W} 
\tttregle {138} {W} {YWRYWYWYBY} {W} 
\tttregle {139} {W} {WWYYYRYWYY} {W} 
\tttregle {140} {W} {YWYBWYWYRY} {W} 
\tttregle {141} {W} {YWYRWYWYYB} {W} 
\tttregle {142} {W} {YWYYWYWYYR} {W} 
\tttregle {143} {Y} {YWWBWWWWWW} {B} 
\tttregle {144} {W} {YWYYWBWYYY} {W} 
\tttregle {145} {W} {YWYYWYBWYY} {W} 
\tttregle {146} {W} {WWBWWWYYWW} {W} 
\tttregle {147} {B} {YWWRWWWWWW} {R} 
\tttregle {148} {W} {BWWWWWYWWY} {W} 
\tttregle {149} {W} {BWYYWRWYYY} {W} 
\tttregle {150} {W} {BWYYWYRWYY} {W} 
\tttregle {151} {W} {WWRWWWYBWW} {W} 
\tttregle {152} {R} {BWWYWWWWWW} {Y} 
\tttregle {153} {W} {RWWWWWBWWY} {W} 
\tttregle {154} {W} {RWYYWYWYYY} {W} 
\tttregle {155} {W} {RWYYWBYWYY} {W} 
\tttregle {156} {W} {WWYWWWYRWW} {W} 
\tttregle {157} {Y} {RWWYWWWWWW} {Y} 
\tttregle {158} {W} {YWWWWWRWWY} {W} 
\tttregle {159} {W} {YWYYWRYWYY} {W} 
\tttregle {160} {W} {YWYYWYYWYB} {W} 
\tttregle {161} {W} {YWYBWYYWYR} {W} 
\vskip 1pt
\ligne{\hfill green locomotive\hfill}
\ligne{\hfill $\downarrow$ motion\hfill}
\vskip 1pt
\tttregle {162} {W} {YWYYYYYGYY} {W} 
\tttregle {163} {Y} {WGWWYYWWWW} {G} 
\tttregle {164} {W} {YWWWWWGWWW} {W} 
\tttregle {165} {W} {YWYGWYYWRY} {W} 
\tttregle {166} {Y} {WWWWRGWWWW} {Y} 
\tttregle {167} {G} {WRWWYWYWWW} {R} 
\tttregle {168} {R} {GYWWWWWWWW} {Y} 
\tttregle {169} {W} {GWWWWWRWWW} {W} 
\tttregle {170} {W} {GWWWWYWWWW} {W} 
\tttregle {171} {W} {WRWWWWGWWW} {W} 
\tttregle {172} {W} {WWWWWWGWWW} {W} 
\tttregle {173} {W} {YWGYYYYRYY} {W} 
\tttregle {174} {G} {WRWWYYWWWW} {R} 
\tttregle {175} {Y} {WYWWYYGWWW} {Y} 
\tttregle {176} {Y} {YGWWWWWWWW} {G} 
\tttregle {177} {W} {YWYRWYYWYG} {W} 
}
\hfill
\vtop{\leftskip 0pt\parindent 0pt\hsize=80pt
\tttregle {178} {R} {WYWWGWYWWW} {Y} 
\tttregle {179} {W} {YWRYYYYYGY} {W} 
\tttregle {180} {R} {WYWWGYWWWW} {Y} 
\tttregle {181} {Y} {WGWWYYRWWW} {G} 
\tttregle {182} {G} {YRWWWWWWWW} {R} 
\tttregle {183} {W} {GWWWWWWWWW} {W} 
\tttregle {184} {W} {YWYGYYYYRY} {W} 
\tttregle {185} {Y} {WYWWRGWWWW} {Y} 
\tttregle {186} {G} {WRWWYYYWWW} {R} 
\tttregle {187} {W} {YWYRYYYYYG} {W} 
\tttregle {188} {R} {WYWWGYYWWW} {Y} 
\tttregle {189} {Y} {YWYWWWWGWW} {Y} 
\tttregle {190} {W} {YWYYGYYYYR} {W} 
\tttregle {191} {Y} {WYWWRGYWWW} {Y} 
\tttregle {192} {Y} {YWGWWWWRWW} {G} 
\tttregle {193} {W} {YWYYRGYYYY} {W} 
\tttregle {194} {Y} {GWWYWWWWWW} {G} 
\tttregle {195} {W} {GWYYYRYYYY} {W} 
\tttregle {196} {W} {YWWWWWGWWY} {W} 
\tttregle {197} {W} {YWYYWGYWYY} {W} 
\tttregle {198} {G} {RWWYWWWWWW} {R} 
\tttregle {199} {W} {RWYYYYGYYY} {W} 
\tttregle {200} {W} {GWWWWWRWWY} {W} 
\tttregle {201} {W} {GWYYWYWYYY} {W} 
\tttregle {202} {W} {WWYYYYGWYY} {W} 
\tttregle {203} {Y} {GWYWWWWYWW} {G} 
\tttregle {204} {W} {YWYYYYGYYY} {W} 
\tttregle {205} {W} {GWYYWRYWYY} {W} 
\tttregle {206} {W} {WYWWWYGWWW} {W} 
\tttregle {207} {R} {YWWGWWWWWW} {Y} 
\tttregle {208} {W} {RWYYWGWYYY} {W} 
\tttregle {209} {G} {RWYWWWWYWW} {R} 
\tttregle {210} {W} {GWYYYYRYYY} {W} 
\tttregle {211} {Y} {GYWWWWWWWW} {G} 
\tttregle {212} {W} {GWWWWWYWWW} {W} 
\tttregle {213} {Y} {YWYWWWGYWW} {Y} 
\tttregle {214} {W} {WYWWWWGWWW} {W} 
\tttregle {215} {W} {RWYYWYGWYY} {W} 
\tttregle {216} {W} {WGWWWYRWWW} {W} 
\tttregle {217} {R} {YWGWWWWYWW} {Y} 
\tttregle {218} {W} {RWYYYGYYYY} {W} 
\tttregle {219} {G} {RYWWWWWWWW} {R} 
\tttregle {220} {W} {RWWWWWGWWW} {W} 
\tttregle {221} {Y} {GWYWWWRYWW} {G} 
\tttregle {222} {W} {WGWWWWRWWW} {W} 
\tttregle {223} {Y} {YWRWWWWGWW} {Y} 
}
\hfill
\vtop{\leftskip 0pt\parindent 0pt\hsize=80pt
\tttregle {224} {W} {YWYYGRYYYY} {W} 
\tttregle {225} {R} {YGWWWWWWWW} {Y} 
\tttregle {226} {G} {RWYWWWYYWW} {R} 
\tttregle {227} {W} {YWYYRYYYYG} {W} 
\tttregle {228} {R} {YWGWWWYYWW} {Y} 
\tttregle {229} {W} {YWYGYYYYYR} {W} 
\tttregle {230} {Y} {WYWWYGWWWW} {Y} 
\tttregle {231} {Y} {YWRWWWYGWW} {Y} 
\tttregle {232} {W} {YWYRYYYYGY} {W} 
\tttregle {233} {Y} {WYWWGRWWWW} {G} 
\tttregle {234} {W} {YWYYWYWYYG} {W} 
\tttregle {235} {W} {YWGYYYYYRY} {W} 
\tttregle {236} {G} {WYWWRYWWWW} {R} 
\tttregle {237} {W} {YWYGWYWYYR} {W} 
\tttregle {238} {W} {YWRYYYYGYY} {W} 
\tttregle {239} {R} {WGWWYYWWWW} {Y} 
\tttregle {240} {W} {YWYRWYWYGY} {W} 
\tttregle {241} {W} {YWGYWYWYRY} {W} 
\tttregle {242} {W} {WWYYYGYWYY} {W} 
\tttregle {243} {W} {YWRYWYWGYY} {W} 
\tttregle {244} {W} {WWYYGRYWYY} {W} 
\tttregle {245} {W} {WWYYRYYWYG} {W} 
\tttregle {246} {W} {WWYGYYYWYR} {W} 
\tttregle {247} {Y} {WWWWYGWWWW} {Y} 
\vskip 1pt
\ligne{\hfill $\uparrow$ motion\hfill}
\vskip 1pt
\tttregle {248} {W} {WWYGYYYWRY} {W} 
\tttregle {249} {W} {WWYRYYYWYG} {W} 
\tttregle {250} {W} {YWYYWYWGYY} {W} 
\tttregle {251} {W} {WWYYGYYWYR} {W} 
\tttregle {252} {W} {YWGYWYWRYY} {W} 
\tttregle {253} {W} {WWYYRGYWYY} {W} 
\tttregle {254} {W} {YWRYWYWYGY} {W} 
\tttregle {255} {W} {YWYGWYWYRY} {W} 
\tttregle {256} {W} {YWYRWYWYYG} {W} 
\tttregle {257} {Y} {YWWGWWWWWW} {G} 
\tttregle {258} {W} {YWYYWGWYYY} {W} 
\tttregle {259} {W} {YWYYWYGWYY} {W} 
\tttregle {260} {W} {WWGWWWYYWW} {W} 
\tttregle {261} {G} {YWWRWWWWWW} {R} 
\tttregle {262} {W} {GWWWWWYWWY} {W} 
\tttregle {263} {W} {GWYYWRWYYY} {W} 
\tttregle {264} {W} {GWYYWYRWYY} {W} 
\tttregle {265} {W} {WWRWWWYGWW} {W} 
\tttregle {266} {R} {GWWYWWWWWW} {Y} 
\tttregle {267} {W} {RWWWWWGWWY} {W} 
\tttregle {268} {W} {RWYYWGYWYY} {W} 
}
\hfill}

\ligne{\hfill
\vtop{\leftskip 0pt\parindent 0pt\hsize=80pt
\tttregle {269} {W} {YWYYWYYWYG} {W} 
\tttregle {270} {W} {YWYGWYYWYR} {W} 
\vskip -3pt
\ligne{\hrulefill}
\ligne{\hfill Locomotives \hfill}
\ligne{\hfill on Z-track\hfill}
\vskip -3pt
\ligne{\hrulefill}
\vskip -2pt
\ligne{\hfill blue locomotive\hfill}
\ligne{\hfill $\downarrow$ motion\hfill}
\vskip 1pt
\tttregle {271} {W} {YWBYWYYYRW} {W} 
\tttregle {272} {Y} {WRWWWWBWWW} {Y} 
\tttregle {273} {W} {YWWWWWYYWW} {W} 
\tttregle {274} {W} {YWYYWWYYYY} {W} 
\tttregle {275} {W} {YWWWWYWYWW} {W} 
\tttregle {276} {W} {WWYYYWYYYY} {W} 
\tttregle {277} {Y} {WWYWWWWYWW} {Y} 
\tttregle {278} {W} {YWRYWYYBYW} {W} 
\tttregle {279} {Y} {WYWWWWRWWW} {Y} 
\tttregle {280} {W} {YWYYWYYRYW} {W} 
\tttregle {281} {W} {YWYYWYYYYW} {W} 
\tttregle {282} {W} {YWYYWYBYYW} {W} 
\tttregle {283} {W} {YWWWWBWYWW} {W} 
\tttregle {284} {W} {BWYYWYRYYW} {W} 
\tttregle {285} {W} {YWWWWWBYWW} {W} 
\tttregle {286} {W} {BWYYWWYYYY} {W} 
\tttregle {287} {W} {BWWWWRWYWW} {W} 
\tttregle {288} {W} {WWYYYWBYYY} {W} 
\tttregle {289} {W} {RWYYWBYYYW} {W} 
\tttregle {290} {W} {BWWWWWRYWW} {W} 
\tttregle {291} {W} {RWYYWWBYYY} {W} 
\tttregle {292} {W} {RWWWWYWYWW} {W} 
\tttregle {293} {W} {WWYYYWRYYY} {W} 
\tttregle {294} {W} {YWYYWRYYYW} {W} 
\tttregle {295} {W} {RWWWWWYYWW} {W} 
\tttregle {296} {W} {YWYYWWRYYY} {W} 
\tttregle {297} {W} {YWYYWWYBYY} {W} 
\tttregle {298} {W} {YWBYWWYRYY} {W} 
\tttregle {299} {W} {YWRYWWYYBY} {W} 
\tttregle {300} {W} {YWYBWWYYRY} {W} 
\tttregle {301} {W} {WWYYYWYBYY} {W} 
\tttregle {302} {W} {YWYRWWYYYB} {W} 
\tttregle {303} {W} {WWBYYWYRYY} {W} 
\tttregle {304} {W} {YWYYWWYYYR} {W} 
\tttregle {305} {W} {WWRYYWYYBY} {W} 
\tttregle {306} {W} {WWYBYWYYRY} {W} 
\tttregle {307} {Y} {WWYWWWWBWW} {Y} 
\vskip 1pt
\ligne{\hfill $\uparrow$ motion\hfill}
\vskip 1pt
\tttregle {308} {W} {WWYBYWYYYR} {W} 
}
\hfill
\vtop{\leftskip 0pt\parindent 0pt\hsize=80pt
\tttregle {309} {W} {WWYRYWYYBY} {W} 
\tttregle {310} {W} {YWYYWWYYYB} {W} 
\tttregle {311} {W} {WWBYYWYYRY} {W} 
\tttregle {312} {W} {YWYBWWYYYR} {W} 
\tttregle {313} {W} {WWRYYWYBYY} {W} 
\tttregle {314} {W} {YWYRWWYYBY} {W} 
\tttregle {315} {W} {WWYYYWYRYY} {W} 
\tttregle {316} {W} {YWBYWWYYRY} {W} 
\tttregle {317} {W} {YWRYWWYBYY} {W} 
\tttregle {318} {W} {YWYYWWYRYY} {W} 
\tttregle {319} {W} {YWYYWBYYYW} {W} 
\tttregle {320} {W} {BWWWWWYYWW} {W} 
\tttregle {321} {W} {YWYYWWBYYY} {W} 
\tttregle {322} {W} {BWYYWRYYYW} {W} 
\tttregle {323} {W} {RWWWWWBYWW} {W} 
\tttregle {324} {W} {BWYYWWRYYY} {W} 
\tttregle {325} {W} {BWWWWYWYWW} {W} 
\tttregle {326} {W} {RWYYWYBYYW} {W} 
\tttregle {327} {W} {YWWWWWRYWW} {W} 
\tttregle {328} {W} {RWYYWWYYYY} {W} 
\tttregle {329} {W} {RWWWWBWYWW} {W} 
\tttregle {330} {W} {YWYYWYRYYW} {W} 
\tttregle {331} {W} {YWWWWRWYWW} {W} 
\tttregle {332} {W} {YWYYWYYBYW} {W} 
\tttregle {333} {W} {YWBYWYYRYW} {W} 
\vskip 1pt
\ligne{\hfill green locomotive\hfill}
\ligne{\hfill $\downarrow$ motion\hfill}
\vskip 1pt
\tttregle {334} {W} {YWGYWYYYRW} {W} 
\tttregle {335} {Y} {WRWWWWGWWW} {Y} 
\tttregle {336} {W} {YWRYWYYGYW} {W} 
\tttregle {337} {W} {YWYYWYGYYW} {W} 
\tttregle {338} {W} {YWWWWGWYWW} {W} 
\tttregle {339} {W} {GWYYWYRYYW} {W} 
\tttregle {340} {W} {YWWWWWGYWW} {W} 
\tttregle {341} {W} {GWYYWWYYYY} {W} 
\tttregle {342} {W} {GWWWWRWYWW} {W} 
\tttregle {343} {W} {WWYYYWGYYY} {W} 
\tttregle {344} {W} {RWYYWGYYYW} {W} 
\tttregle {345} {W} {GWWWWWRYWW} {W} 
\tttregle {346} {W} {RWYYWWGYYY} {W} 
\tttregle {347} {W} {YWYYWWYGYY} {W} 
\tttregle {348} {W} {YWGYWWYRYY} {W} 
\tttregle {349} {W} {YWRYWWYYGY} {W} 
\tttregle {350} {W} {YWYGWWYYRY} {W} 
\tttregle {351} {W} {WWYYYWYGYY} {W} 
\tttregle {352} {W} {YWYRWWYYYG} {W} 
}
\hfill
\vtop{\leftskip 0pt\parindent 0pt\hsize=80pt
\tttregle {353} {W} {WWGYYWYRYY} {W} 
\tttregle {354} {W} {WWRYYWYYGY} {W} 
\tttregle {355} {W} {WWYGYWYYRY} {W} 
\tttregle {356} {Y} {WWYWWWWGWW} {Y} 
\ligne{\hfill $\uparrow$ motion\hfill}
\tttregle {357} {W} {WWYGYWYYYR} {W} 
\tttregle {358} {W} {WWYRYWYYGY} {W} 
\tttregle {359} {W} {YWYYWWYYYG} {W} 
\tttregle {360} {W} {WWGYYWYYRY} {W} 
\tttregle {361} {W} {YWYGWWYYYR} {W} 
\tttregle {362} {W} {WWRYYWYGYY} {W} 
\tttregle {363} {W} {YWYRWWYYGY} {W} 
\tttregle {364} {W} {YWGYWWYYRY} {W} 
\tttregle {365} {W} {YWRYWWYGYY} {W} 
\tttregle {366} {W} {YWYYWGYYYW} {W} 
\tttregle {367} {W} {GWWWWWYYWW} {W} 
\tttregle {368} {W} {YWYYWWGYYY} {W} 
\tttregle {369} {W} {GWYYWRYYYW} {W} 
\tttregle {370} {W} {RWWWWWGYWW} {W} 
\tttregle {371} {W} {GWYYWWRYYY} {W} 
\tttregle {372} {W} {GWWWWYWYWW} {W} 
\tttregle {373} {W} {RWYYWYGYYW} {W} 
\tttregle {374} {W} {RWWWWGWYWW} {W} 
\tttregle {375} {W} {YWYYWYYGYW} {W} 
\tttregle {376} {W} {YWGYWYYRYW} {W} 
\vskip -3pt
\ligne{\hrulefill}
\ligne{\hfill passive fixed switch\hfill}
\vskip -4pt
\ligne{\hrulefill}
\ligne{\hfill blue locomotive\hfill}
\ligne{\hfill $\bullet$\hskip 0.5pt$\rightarrow$\hfill}
\vskip 1pt
\tttregle {377} {W} {WYWWWWYYWW} {W} 
\tttregle {378} {W} {YWYBWYWWRY} {W} 
\tttregle {379} {W} {WWWWWYYWWW} {W} 
\tttregle {380} {Y} {YWYYWWWYRY} {Y} 
\tttregle {381} {Y} {YYWWRYWWWW} {Y} 
\tttregle {382} {Y} {YRWWYWYWWW} {Y} 
\tttregle {383} {R} {YYWRWWYWRR} {R} 
\tttregle {384} {W} {RWWWRRYWWW} {W} 
\tttregle {385} {R} {RRWWRWWWWW} {R} 
\tttregle {386} {W} {WRRWWWYRWW} {W} 
\tttregle {387} {R} {RWWWWWRWWW} {R} 
\tttregle {388} {W} {RWWWWWRWWW} {W} 
\tttregle {389} {W} {RWWWWRWWWW} {W} 
\tttregle {390} {R} {WRWWWWRWWW} {R} 
\tttregle {391} {Y} {YWWWWWWWWW} {Y} 
\tttregle {392} {W} {YWYRWYWWYB} {W} 
\tttregle {393} {W} {YWYYWYWWYR} {W} 
}
\hfill}

\ligne{\hfill
\vtop{\leftskip 0pt\parindent 0pt\hsize=80pt
\tttregle {394} {W} {YWYYWYWWYY} {W} 
\tttregle {395} {Y} {YWYYWWWBRY} {Y} 
\tttregle {396} {Y} {YBWWRYWWWW} {B} 
\tttregle {397} {Y} {YWBYWWWRRY} {B} 
\tttregle {398} {B} {YRWWRYWWWW} {R} 
\tttregle {399} {Y} {YRWWYWBWWW} {Y} 
\tttregle {400} {R} {YBWRWWYWRR} {R} 
\tttregle {401} {W} {RWWWRRBWWW} {W} 
\tttregle {402} {W} {YWYYWBWWYY} {W} 
\tttregle {403} {B} {YWRYWWWYRY} {R} 
\tttregle {404} {R} {BYWWRYWWWW} {Y} 
\tttregle {405} {Y} {BRWWYWRWWW} {Y} 
\tttregle {406} {R} {YRWRWWBWRR} {R} 
\tttregle {407} {W} {RWWWRRRWWW} {W} 
\tttregle {408} {W} {WBYYYYYYYY} {W} 
\tttregle {409} {W} {WYWWWWBYWW} {W} 
\tttregle {410} {W} {BWYYWRWWYY} {W} 
\tttregle {411} {W} {WWWWWYBWWW} {W} 
\tttregle {412} {R} {BWYYWWWYRY} {Y} 
\tttregle {413} {Y} {RYWWRYWWWW} {Y} 
\tttregle {414} {Y} {RRWWYWYWWW} {Y} 
\tttregle {415} {R} {YYWRWWRWRR} {R} 
\tttregle {416} {W} {WBWWWWRYWW} {W} 
\tttregle {417} {W} {RWYYWYWWYY} {W} 
\tttregle {418} {W} {WWWWWYRWWW} {W} 
\tttregle {419} {Y} {RWYYWWWYRY} {Y} 
\tttregle {420} {W} {WRWWWWYBWW} {W} 
\tttregle {421} {W} {WYWWWWYRWW} {W} 
\tttregle {422} {Y} {BWWWWWWWWW} {Y} 
\vskip 1pt
\ligne{\hfill $\leftarrow$\hskip 0.5pt$\bullet$\hfill}
\vskip 1pt
\tttregle {423} {Y} {YWYYWWWYRB} {Y} 
\tttregle {424} {Y} {YRWWBWYWWW} {B} 
\tttregle {425} {Y} {YWYBWWWYRR} {B} 
\tttregle {426} {Y} {YYWWRBWWWW} {Y} 
\tttregle {427} {B} {YRWWRWYWWW} {R} 
\tttregle {428} {R} {BYWRWWYWRR} {R} 
\tttregle {429} {W} {WRRWWWBRWW} {W} 
\tttregle {430} {B} {YWYRWWWYRY} {R} 
\tttregle {431} {W} {BWYYYYYYYY} {W} 
\tttregle {432} {Y} {BYWWRRWWWW} {Y} 
\tttregle {433} {R} {BRWWYWYWWW} {Y} 
\tttregle {434} {R} {RYWRWWBWRR} {R} 
\tttregle {435} {W} {WRRWWWRRWW} {W} 
\vskip 1pt
\ligne{\hfill green locomotive\hfill}
\ligne{\hfill $\bullet$\hskip 0.5pt$\rightarrow$\hfill}
\vskip 1pt
\tttregle {436} {W} {YWYGWYWWRY} {W} 
}
\hfill
\vtop{\leftskip 0pt\parindent 0pt\hsize=80pt
\tttregle {437} {W} {YWYRWYWWYG} {W} 
\tttregle {438} {Y} {YWYYWWWGRY} {Y} 
\tttregle {439} {Y} {YGWWRYWWWW} {G} 
\tttregle {440} {Y} {YWGYWWWRRY} {G} 
\tttregle {441} {G} {YRWWRYWWWW} {R} 
\tttregle {442} {Y} {YRWWYWGWWW} {Y} 
\tttregle {443} {R} {YGWRWWYWRR} {R} 
\tttregle {444} {W} {RWWWRRGWWW} {W} 
\tttregle {445} {W} {YWYYWGWWYY} {W} 
\tttregle {446} {G} {YWRYWWWYRY} {R} 
\tttregle {447} {R} {GYWWRYWWWW} {Y} 
\tttregle {448} {Y} {GRWWYWRWWW} {Y} 
\tttregle {449} {R} {YRWRWWGWRR} {R} 
\tttregle {450} {W} {WGYYYYYYYY} {W} 
\tttregle {451} {W} {WYWWWWGYWW} {W} 
\tttregle {452} {W} {GWYYWRWWYY} {W} 
\tttregle {453} {W} {WWWWWYGWWW} {W} 
\tttregle {454} {R} {GWYYWWWYRY} {Y} 
\tttregle {455} {W} {WGWWWWRYWW} {W} 
\tttregle {456} {W} {WRWWWWYGWW} {W} 
\tttregle {457} {Y} {GWWWWWWWWW} {Y} 
\vskip 1pt
\ligne{\hfill $\leftarrow$\hskip 0.5pt$\bullet$\hfill}
\vskip 1pt
\tttregle {458} {Y} {YWYYWWWYRG} {Y} 
\tttregle {459} {Y} {YRWWGWYWWW} {G} 
\tttregle {460} {Y} {YWYGWWWYRR} {G} 
\tttregle {461} {Y} {YYWWRGWWWW} {Y} 
\tttregle {462} {G} {YRWWRWYWWW} {R} 
\tttregle {463} {R} {GYWRWWYWRR} {R} 
\tttregle {464} {W} {WRRWWWGRWW} {W} 
\tttregle {465} {G} {YWYRWWWYRY} {R} 
\tttregle {466} {W} {GWYYYYYYYY} {W} 
\tttregle {467} {Y} {GYWWRRWWWW} {Y} 
\tttregle {468} {R} {GRWWYWYWWW} {Y} 
\tttregle {469} {R} {RYWRWWGWRR} {R} 
\vskip -3pt
\ligne{\hrulefill}
\ligne{\hfill fork\hfill}
\vskip -4pt
\ligne{\hrulefill}
\vskip -1pt
\ligne{\hfill blue locomotive\hfill}
\vskip -1pt
\tttregle {470} {W} {WYWWWWYBWW} {W} 
\tttregle {471} {W} {WBWWWWYRWW} {W} 
\tttregle {472} {W} {WRWWWWBYWW} {W} 
\tttregle {473} {W} {WYWWWWRYWW} {W} 
\tttregle {474} {Y} {YYWRRYWWWW} {Y} 
\tttregle {475} {Y} {YRRWYWYWWW} {Y} 
\tttregle {476} {R} {YWWWWRWWWW} {R} 
\tttregle {477} {R} {YYWWWRYRWW} {R} 
\tttregle {478} {R} {YWWWWWRWWW} {R} 
}
\hfill
\vtop{\leftskip 0pt\parindent 0pt\hsize=80pt
\tttregle {479} {W} {RRWWWWYWWW} {W} 
\tttregle {480} {W} {BWYYWYWWYY} {W} 
\tttregle {481} {Y} {BWYYWWWYRY} {B} 
\tttregle {482} {W} {RWYYWBWWYY} {W} 
\tttregle {483} {B} {RWYYWWWYRY} {R} 
\tttregle {484} {Y} {BYWRRYWWWW} {B} 
\tttregle {485} {Y} {BRRWYWYWWW} {B} 
\tttregle {486} {R} {YYWWWRBRWW} {R} 
\tttregle {487} {W} {YWYYWRWWYY} {W} 
\tttregle {488} {R} {YWBBWWWYRY} {Y} 
\tttregle {489} {B} {RYWRRBWWWW} {R} 
\tttregle {490} {B} {RRRWYWBWWW} {R} 
\tttregle {491} {R} {BWWWWRWWWW} {R} 
\tttregle {492} {R} {BBWWWRRRWW} {R} 
\tttregle {493} {R} {BWWWWWRWWW} {R} 
\tttregle {494} {W} {RRWWWWBWWW} {W} 
\tttregle {495} {Y} {YWRRWWWBRB} {Y} 
\tttregle {496} {R} {YBWRRRWWWW} {Y} 
\tttregle {497} {R} {YRRWBWRWWW} {Y} 
\tttregle {498} {W} {RWWWWRBWWW} {W} 
\tttregle {499} {R} {RRWWWRYRWW} {R} 
\tttregle {500} {W} {RWWWWBRWWW} {W} 
\tttregle {501} {W} {RRWWWWRWWW} {W} 
\tttregle {502} {Y} {YWYYWWWRRR} {Y} 
\tttregle {503} {Y} {YRWRRYWWWW} {Y} 
\tttregle {504} {Y} {YRRWRWYWWW} {Y} 
\tttregle {505} {W} {YWWWWRRWWW} {W} 
\tttregle {506} {W} {YWYYWYWWYB} {W} 
\tttregle {507} {W} {YWYBWYWWYR} {W} 
\vskip 1pt
\ligne{\hfill green locomotive\hfill}
\vskip 1pt
\tttregle {508} {W} {WYWWWWYGWW} {W} 
\tttregle {509} {W} {WGWWWWYRWW} {W} 
\tttregle {510} {W} {WRWWWWGYWW} {W} 
\tttregle {511} {W} {GWYYWYWWYY} {W} 
\tttregle {512} {Y} {GWYYWWWYRY} {G} 
\tttregle {513} {W} {RWYYWGWWYY} {W} 
\tttregle {514} {G} {RWYYWWWYRY} {R} 
\tttregle {515} {Y} {GYWRRYWWWW} {G} 
\tttregle {516} {Y} {GRRWYWYWWW} {G} 
\tttregle {517} {R} {YYWWWRGRWW} {R} 
\tttregle {518} {R} {YWGGWWWYRY} {Y} 
\tttregle {519} {G} {RYWRRGWWWW} {R} 
\tttregle {520} {G} {RRRWYWGWWW} {R} 
\tttregle {521} {R} {GWWWWRWWWW} {R} 
\tttregle {522} {R} {GGWWWRRRWW} {R} 
\tttregle {523} {R} {GWWWWWRWWW} {R} 
\tttregle {524} {W} {RRWWWWGWWW} {W} 
}
\hfill}

\ligne{\hfill
\vtop{\leftskip 0pt\parindent 0pt\hsize=80pt
\tttregle {525} {Y} {YWRRWWWGRG} {Y} 
\tttregle {526} {R} {YGWRRRWWWW} {Y} 
\tttregle {527} {R} {YRRWGWRWWW} {Y} 
\tttregle {528} {W} {RWWWWRGWWW} {W} 
\tttregle {529} {W} {RWWWWGRWWW} {W} 
\tttregle {530} {W} {YWYYWYWWYG} {W} 
\tttregle {531} {W} {YWYGWYWWYR} {W}
\ligne{\hrulefill}
\vskip -3pt 
\ligne{\hfill converters\hfill}
\vskip -3pt 
\ligne{\hrulefill}
\ligne{\hfill from blue to green\hfill}
\vskip -2pt 
\tttregle {532} {Y} {WYRGYWWWWW} {Y} 
\tttregle {533} {Y} {YWYWWRWWWW} {Y} 
\tttregle {534} {R} {YWRRWGYWWW} {R} 
\tttregle {535} {W} {RYWWWRYWWW} {W} 
\tttregle {536} {G} {YWGGWYRWWW} {G} 
\tttregle {537} {W} {GRWWWGYRWW} {W} 
\tttregle {538} {G} {GWWWWGWWWW} {G} 
\tttregle {539} {G} {GWWWWWGWWW} {G} 
\tttregle {540} {W} {GWWWWWGWWW} {W} 
\tttregle {541} {W} {GGWWWWGWWW} {W} 
\tttregle {542} {W} {YGWWWYYGWW} {W} 
\tttregle {543} {W} {WGWWWWGWWW} {W} 
\tttregle {544} {Y} {RWWWWWWWWW} {Y} 
\tttregle {545} {Y} {YWBWWWGWWW} {B} 
\tttregle {546} {W} {YGWWWBYGWW} {W} 
\tttregle {547} {Y} {WYRGBWWWWW} {G} 
\tttregle {548} {G} {YWGGWBRWWW} {G} 
\tttregle {549} {B} {YWRWWWGWWW} {R} 
\tttregle {550} {W} {BGWWWRYGWW} {W} 
\tttregle {551} {G} {WYRGRWWWWW} {R} 
\tttregle {552} {Y} {GWYWWRWWWW} {G} 
\tttregle {553} {R} {GWRRWGYWWW} {R} 
\tttregle {554} {W} {RYWWWRGWWW} {W} 
\tttregle {555} {G} {GWGGWRRWWW} {G} 
\tttregle {556} {W} {GRWWWGGRWW} {W} 
\tttregle {557} {R} {GWYWWWGWWW} {Y} 
\tttregle {558} {W} {RGWWWYGGWW} {W} 
\tttregle {559} {R} {WGRGYWWWWW} {Y} 
\tttregle {560} {G} {RWYWWRWWWW} {R} 
\tttregle {561} {R} {RWRRWGGWWW} {R} 
\tttregle {562} {W} {RGWWWRRWWW} {W} 
\tttregle {563} {G} {RWGGWYRWWW} {G} 
\tttregle {564} {W} {GRWWWGRRWW} {W} 
\tttregle {565} {W} {YGWWWYRGWW} {W} 
\tttregle {566} {Y} {WRRGYWWWWW} {Y} 
\tttregle {567} {R} {YWGWWRWWWW} {Y} 
}
\hfill
\vtop{\leftskip 0pt\parindent 0pt\hsize=80pt
\tttregle {568} {R} {YWRRWGRWWW} {R} 
\tttregle {569} {W} {RRWWWRYWWW} {W} 
\vskip 1pt 
\ligne{\hfill from green to blue\hfill}
\vskip 1pt 
\tttregle {570} {Y} {WYRBYWWWWW} {Y} 
\tttregle {571} {R} {YWRRWBYWWW} {R} 
\tttregle {572} {B} {YWGGWYRWWW} {B} 
\tttregle {573} {W} {BRWWWGYRWW} {W} 
\tttregle {574} {G} {BWWWWGWWWW} {G} 
\tttregle {575} {G} {BWWWWWGWWW} {G} 
\tttregle {576} {W} {GWWWWWBWWW} {W} 
\tttregle {577} {W} {GGWWWWBWWW} {W} 
\tttregle {578} {W} {YBWWWYYGWW} {W} 
\tttregle {579} {W} {WGWWWWBWWW} {W} 
\tttregle {580} {Y} {YWGWWWBWWW} {G} 
\tttregle {581} {W} {YBWWWGYGWW} {W} 
\tttregle {582} {Y} {WYRBGWWWWW} {B} 
\tttregle {583} {B} {YWGGWGRWWW} {B} 
\tttregle {584} {G} {YWRWWWBWWW} {R} 
\tttregle {585} {W} {GBWWWRYGWW} {W} 
\tttregle {586} {B} {WYRBRWWWWW} {R} 
\tttregle {587} {Y} {BWYWWRWWWW} {B} 
\tttregle {588} {R} {BWRRWBYWWW} {R} 
\tttregle {589} {W} {RYWWWRBWWW} {W} 
\tttregle {590} {B} {BWGGWRRWWW} {B} 
\tttregle {591} {W} {BRWWWGBRWW} {W} 
\tttregle {592} {R} {BWYWWWBWWW} {Y} 
\tttregle {593} {W} {RBWWWYBGWW} {W} 
\tttregle {594} {R} {WBRBYWWWWW} {Y} 
\tttregle {595} {B} {RWYWWRWWWW} {R} 
\tttregle {596} {R} {RWRRWBBWWW} {R} 
\tttregle {597} {W} {RBWWWRRWWW} {W} 
\tttregle {598} {B} {RWGGWYRWWW} {B} 
\tttregle {599} {W} {BRWWWGRRWW} {W} 
\tttregle {600} {W} {YBWWWYRGWW} {W} 
\tttregle {601} {Y} {WRRBYWWWWW} {Y} 
\tttregle {602} {R} {YWBWWRWWWW} {Y} 
\tttregle {603} {R} {YWRRWBRWWW} {R}
\vskip -4pt 
\ligne{\hrulefill}
\ligne{\hfill filters\hfill}
\vskip -4pt 
\ligne{\hrulefill}
\ligne{\hfill blue filter\hfill}
\vskip -2pt 
\ligne{\hfill blue locomotive\hfill}
\vskip -2pt 
\tttregle {604} {Y} {BYRWYWWWWW} {Y} 
\tttregle {605} {B} {YWYWWYYYRR} {B} 
\tttregle {606} {W} {BYWYYYYWWY} {W} 
\tttregle {607} {W} {BRWWRWYWWW} {W} 
\tttregle {608} {R} {WYWWWWBRWW} {R} 
}
\hfill
\vtop{\leftskip 0pt\parindent 0pt\hsize=80pt
\tttregle {609} {Y} {YWYWWRBWWW} {Y} 
\tttregle {610} {W} {YBRWWYYWWW} {W} 
\tttregle {611} {R} {WWWWWWBWWW} {R} 
\tttregle {612} {R} {YWRRWWYWWW} {R} 
\tttregle {613} {W} {WRWWWWYRWW} {W} 
\tttregle {614} {Y} {YWYWWBWWWW} {Y} 
\tttregle {615} {Y} {YWBWWBWWWW} {B} 
\tttregle {616} {Y} {BYRWBWWWWW} {B} 
\tttregle {617} {B} {YWYWWYBYRR} {B} 
\tttregle {618} {W} {BBWYYYYWWY} {W} 
\tttregle {619} {W} {YWWWWBRWWW} {W} 
\tttregle {620} {B} {YWRWWBWWWW} {R} 
\tttregle {621} {B} {BYRWRWWWWW} {R} 
\tttregle {622} {B} {BWYWWYRYRR} {B} 
\tttregle {623} {W} {BRWYYYBWWY} {W} 
\tttregle {624} {Y} {BWYWWRBWWW} {B} 
\tttregle {625} {W} {YBRWWYBWWW} {W} 
\tttregle {626} {R} {BWRRWWYWWW} {R} 
\tttregle {627} {W} {BWWWWRRWWW} {W} 
\tttregle {628} {R} {BWYWWBWWWW} {Y} 
\tttregle {629} {R} {BBRWYWWWWW} {Y} 
\tttregle {630} {B} {RWYWWBYYRR} {B} 
\tttregle {631} {W} {BYWYYYRWWY} {W} 
\tttregle {632} {B} {RWYWWRBWWW} {R} 
\tttregle {633} {W} {BBRWWYRWWW} {W} 
\tttregle {634} {R} {RWRRWWBWWW} {R} 
\tttregle {635} {W} {WRWWWWRRWW} {W} 
\tttregle {636} {Y} {RWYWWBWWWW} {Y} 
\tttregle {637} {Y} {BRRWYWWWWW} {Y} 
\tttregle {638} {B} {YWYWWRYYRR} {B} 
\tttregle {639} {R} {YWBWWRBWWW} {Y} 
\tttregle {640} {W} {RBRWWBYWWW} {W} 
\tttregle {641} {R} {YWRRWWRWWW} {R} 
\vskip 1pt 
\ligne{\hfill green locomotive\hfill}
\vskip 1pt 
\tttregle {642} {Y} {YWGWWBWWWW} {G} 
\tttregle {643} {Y} {BYRWGWWWWW} {Y} 
\tttregle {644} {B} {YWYWWYGYRR} {B} 
\tttregle {645} {W} {BGWYYYYWWY} {W} 
\tttregle {646} {W} {YWWWWGRWWW} {W} 
\tttregle {647} {G} {YWRWWBWWWW} {R} 
\tttregle {648} {Y} {BYRWRWWWWW} {Y} 
\tttregle {649} {B} {YWYWWYRYRR} {B} 
\tttregle {650} {W} {BRWYYYYWWY} {W} 
\tttregle {651} {R} {YWYWWBWWWW} {Y} 
\tttregle {652} {W} {RWWWWYYWWW} {W} 
}
\hfill}

\ligne{\hfill
\vtop{\leftskip 0pt\parindent 0pt\hsize=80pt
\ligne{\hfill green filter\hfill}
\vskip 1pt 
\ligne{\hfill green locomotive\hfill}
\vskip 1pt 
\tttregle {653} {Y} {GYRWYWWWWW} {Y} 
\tttregle {654} {G} {YWYWWYYYRR} {G} 
\tttregle {655} {W} {GYWYYYYWWY} {W} 
\tttregle {656} {W} {GRWWRWYWWW} {W} 
\tttregle {657} {R} {WYWWWWGRWW} {R} 
\tttregle {658} {Y} {YWYWWRGWWW} {Y} 
\tttregle {659} {W} {YGRWWYYWWW} {W} 
\tttregle {660} {R} {WWWWWWGWWW} {R} 
\tttregle {661} {Y} {YWYWWGWWWW} {Y} 
\tttregle {662} {Y} {YWGWWGWWWW} {G} 
\tttregle {663} {Y} {GYRWGWWWWW} {G} 
\tttregle {664} {G} {YWYWWYGYRR} {G} 
\tttregle {665} {W} {GGWYYYYWWY} {W} 
\tttregle {666} {G} {YWRWWGWWWW} {R} 
\tttregle {667} {G} {GYRWRWWWWW} {R} 
\tttregle {668} {G} {GWYWWYRYRR} {G} 
\tttregle {669} {W} {GRWYYYGWWY} {W} 
\tttregle {670} {Y} {GWYWWRGWWW} {G} 
\tttregle {671} {W} {YGRWWYGWWW} {W} 
\tttregle {672} {R} {GWRRWWYWWW} {R} 
\tttregle {673} {W} {GWWWWRRWWW} {W} 
\tttregle {674} {R} {GWYWWGWWWW} {Y} 
\tttregle {675} {R} {GGRWYWWWWW} {Y} 
\tttregle {676} {G} {RWYWWGYYRR} {G} 
\tttregle {677} {W} {GYWYYYRWWY} {W} 
\tttregle {678} {G} {RWYWWRGWWW} {R} 
\tttregle {679} {W} {GGRWWYRWWW} {W} 
\tttregle {680} {R} {RWRRWWGWWW} {R} 
\tttregle {681} {Y} {RWYWWGWWWW} {Y} 
\tttregle {682} {Y} {GRRWYWWWWW} {Y} 
\tttregle {683} {G} {YWYWWRYYRR} {G} 
\tttregle {684} {R} {YWGWWRGWWW} {Y} 
\tttregle {685} {W} {RGRWWGYWWW} {W} 
\vskip 1pt 
\ligne{\hfill blue locomotive\hfill}
\vskip 1pt 
\tttregle {686} {Y} {YWBWWGWWWW} {B} 
\tttregle {687} {Y} {GYRWBWWWWW} {Y} 
\tttregle {688} {G} {YWYWWYBYRR} {G} 
\tttregle {689} {W} {GBWYYYYWWY} {W} 
\tttregle {690} {B} {YWRWWGWWWW} {R} 
\tttregle {691} {Y} {GYRWRWWWWW} {Y} 
\tttregle {692} {G} {YWYWWYRYRR} {G} 
\tttregle {693} {W} {GRWYYYYWWY} {W} 
\tttregle {694} {R} {YWYWWGWWWW} {Y} 
}
\hfill
\vtop{\leftskip 0pt\parindent 0pt\hsize=80pt
\ligne{\hfill changing colour\hfill}
\vskip 1pt 
\ligne{\hfill from blue to green\hfill}
\vskip 1pt 
\tttregle {695} {B} {YWYWWYYBRR} {G} 
\tttregle {696} {W} {BYWYBYYWWR} {W} 
\tttregle {697} {Y} {BBWRRWWWWW} {Y} 
\tttregle {698} {B} {YWRWWWBYWW} {R} 
\tttregle {699} {W} {YWWWWRBWWW} {W} 
\tttregle {700} {G} {YWYWWYYRRR} {G} 
\tttregle {701} {W} {GYWYRYYWWY} {W} 
\tttregle {702} {Y} {GRWRRWWWWW} {Y} 
\tttregle {703} {R} {YWYWWWGYWW} {Y} 
\vskip 1pt 
\ligne{\hfill from green to blue\hfill}
\vskip 1pt 
\tttregle {704} {G} {YWYWWYYBRR} {B} 
\tttregle {705} {W} {GYWYBYYWWR} {W} 
\tttregle {706} {Y} {GBWRRWWWWW} {Y} 
\tttregle {707} {B} {YWRWWWGYWW} {R} 
\tttregle {708} {B} {YWYWWYYRRR} {B} 
\tttregle {709} {W} {BYWYRYYWWY} {W} 
\tttregle {710} {Y} {BRWRRWWWWW} {Y} 
\tttregle {711} {R} {YWYWWWBYWW} {Y} 
\vskip -4pt 
\ligne{\hrulefill}
\ligne{\hfill registers\hfill}
\vskip -4pt 
\ligne{\hrulefill}
\ligne{\hfill general case\hfill}
\vskip -2pt 
\ligne{\hfill incrementing\hfill}
\vskip -1pt 
\tttregle {712} {Y} {WYYYYYYYYY} {Y} 
\tttregle {713} {W} {WYYWYYYYYB} {W} 
\tttregle {714} {W} {WYYWBYYYYR} {W} 
\tttregle {715} {W} {WYWWRBYWWW} {W} 
\tttregle {716} {Y} {YYWWYYWWWW} {Y} 
\tttregle {717} {R} {BWWWWWWWWW} {Y} 
\tttregle {718} {Y} {YYWWYYYWWW} {Y} 
\tttregle {719} {Y} {YYWWWWYWWW} {Y} 
\tttregle {720} {Y} {YYWWYWYWWW} {Y} 
\tttregle {721} {W} {WYYWRBYYYY} {W} 
\tttregle {722} {W} {WYWWYRYWWW} {W} 
\tttregle {723} {Y} {WYYYYYBYYY} {Y} 
\tttregle {724} {W} {WYYWYRYYYY} {W} 
\tttregle {725} {W} {WYWWYYYWWW} {W} 
\tttregle {726} {Y} {YBWWYYWWWW} {B} 
\tttregle {727} {Y} {WBYYYYRYYY} {B} 
\tttregle {728} {W} {WYYWYYYYYY} {W} 
\tttregle {729} {B} {YRWWYYWWWW} {R} 
\tttregle {730} {Y} {YYWWYYBWWW} {M} 
\tttregle {731} {Y} {YBWWWWYWWW} {M} 
\tttregle {732} {B} {WRMYYYYMYY} {R} 
\tttregle {733} {R} {BYWWMMWWWW} {Y} 
}
\hfill
\vtop{\leftskip 0pt\parindent 0pt\hsize=80pt
\tttregle {734} {W} {RWWWWMWWWW} {W} 
\tttregle {735} {M} {BMWWYYRWWW} {M} 
\tttregle {736} {M} {MRWWWWBWWW} {Y} 
\tttregle {737} {W} {MWWWWWMWWW} {Y} 
\tttregle {738} {W} {MWWWWYWWWW} {W} 
\tttregle {739} {W} {MWWWWWRWWW} {W} 
\tttregle {740} {W} {MWWWWWWWWW} {W} 
\tttregle {741} {W} {WMWWWWMWWW} {M} 
\tttregle {742} {W} {WWWWWWMWWW} {W} 
\tttregle {743} {Y} {BYWWYYMWWW} {Y} 
\tttregle {744} {Y} {YMWWWWBWWW} {M} 
\tttregle {745} {W} {YWWWWWMWWW} {W} 
\tttregle {746} {Y} {BYWWYWYWWW} {B} 
\tttregle {747} {Y} {YYWWWWBWWW} {R} 
\tttregle {748} {R} {WYMYBYYYMR} {W} 
\tttregle {749} {Y} {RYWWYMWWWW} {Y} 
\tttregle {750} {M} {RYYWMYYMWW} {M} 
\tttregle {751} {Y} {MYWWMYRWWW} {Y} 
\tttregle {752} {Y} {MMWWWWYWWW} {Y} 
\tttregle {753} {W} {MWWWWMMWWW} {M} 
\tttregle {754} {W} {MWWWWWYWWW} {W} 
\tttregle {755} {W} {YWWWWMWWWW} {W} 
\tttregle {756} {M} {YYWWWWMWWW} {Y} 
\tttregle {757} {Y} {RMWWRBMWWW} {Y} 
\tttregle {758} {M} {YMWWWWRWWW} {M} 
\tttregle {759} {W} {BWWWWWMWWW} {W} 
\tttregle {760} {W} {WMWWWWBWWW} {W} 
\tttregle {761} {B} {RRWWYWYWWW} {R} 
\tttregle {762} {R} {BYWWWWRWWW} {Y} 
\tttregle {763} {W} {WYMYRBYYMY} {W} 
\tttregle {764} {W} {WBYWYYRYYY} {W} 
\tttregle {765} {Y} {WYWWYMWWWW} {Y} 
\tttregle {766} {M} {WYYWMYYYWY} {M} 
\tttregle {767} {Y} {MYWWYYWWWW} {Y} 
\tttregle {768} {Y} {MYWWWWYWWW} {Y} 
\tttregle {769} {W} {MWWWYMYWWW} {Y} 
\tttregle {770} {W} {MWWWWMYWWW} {W} 
\tttregle {771} {Y} {YYWWWWMWWW} {Y} 
\tttregle {772} {Y} {WMWWYRMWWW} {Y} 
\tttregle {773} {M} {YMYWWWWWWW} {Y} 
\tttregle {774} {Y} {MWWWWWMWWW} {Y} 
\tttregle {775} {W} {WYMYYRYYYY} {W} 
\tttregle {776} {W} {WRYWYYYBYY} {W} 
\tttregle {777} {M} {WYYYYYYYWY} {Y} 
\tttregle {778} {Y} {MYWWWYYWWW} {Y} 
\tttregle {779} {Y} {MWWWYYYWWW} {Y} 
}
\hfill}

\ligne{\hfill
\vtop{\leftskip 0pt\parindent 0pt\hsize=80pt
\tttregle {780} {W} {YYWWWWMWWW} {Y} 
\tttregle {781} {Y} {WYWWYYMWWW} {Y} 
\tttregle {782} {Y} {YMYWWWWYWW} {Y} 
\tttregle {783} {W} {WYBWYYYRYY} {W} 
\tttregle {784} {W} {WYYWYYBYYY} {W} 
\tttregle {785} {W} {WYRWYYYYBY} {W} 
\tttregle {786} {W} {WYYWYYYYRY} {W} 
\tttregle {787} {W} {WYWWYYBWWW} {W} 
\tttregle {788} {W} {WYRWYYYYYY} {W} 
\tttregle {789} {R} {WYWWYWYWWW} {Y} 
\tttregle {790} {W} {WYWWYYRWWW} {W} 
\vskip 1pt 
\ligne{\hfill decrementing\hfill}
\vskip 1pt 
\tttregle {791} {W} {WYYWYYYYYG} {W} 
\tttregle {792} {W} {WYYWGYYYYR} {W} 
\tttregle {793} {W} {WYWWRGYWWW} {W} 
\tttregle {794} {R} {GWWWWWWWWW} {Y} 
\tttregle {795} {W} {WYYWRGYYYY} {W} 
\tttregle {796} {Y} {WYYYYYGYYY} {Y} 
\tttregle {797} {Y} {YGWWYYWWWW} {G} 
\tttregle {798} {Y} {WGYYYYRYYY} {M} 
\tttregle {799} {G} {YRWWYYWWWW} {R} 
\tttregle {800} {Y} {YYWWYYGWWW} {Y} 
\tttregle {801} {Y} {YGWWWWYWWW} {Y} 
\tttregle {802} {M} {WRYYYYYYYY} {M} 
\tttregle {803} {W} {MYYWYRYYYY} {Y} 
\tttregle {804} {R} {MYWWYYWWWW} {Y} 
\tttregle {805} {Y} {RWYWWWMWWW} {Y} 
\tttregle {806} {Y} {MYWWYYRWWW} {W} 
\tttregle {807} {Y} {YRWWWWMWWW} {W} 
\tttregle {808} {Y} {MYWWYYYWWW} {W} 
\tttregle {809} {M} {YYWWYYYWYY} {M} 
\tttregle {810} {Y} {MYYWYYYYYY} {Y} 
\tttregle {811} {Y} {YYWWYYMWWW} {Y} 
\tttregle {812} {Y} {YYYWWWMWWW} {Y} 
\tttregle {813} {W} {MWWWYWYWWW} {W} 
\tttregle {814} {W} {MYWWYYWWWW} {W} 
\tttregle {815} {Y} {WWWWWWMWWW} {W} 
\tttregle {816} {Y} {YWWWWWMWWW} {R} 
\tttregle {817} {M} {YYWWYYYWWR} {Y} 
\tttregle {818} {W} {MWWWRYWWWW} {W} 
\tttregle {819} {Y} {MRWWYYWWWW} {G} 
\tttregle {820} {R} {YWWWWWMWWW} {R} 
\tttregle {821} {Y} {YYWWGYYWWR} {Y} 
\tttregle {822} {Y} {YYYWYYGYYY} {Y} 
\tttregle {823} {Y} {YGWWYYYWWW} {G} 
}
\hfill
\vtop{\leftskip 0pt\parindent 0pt\hsize=80pt
\tttregle {824} {W} {YWWWRGWWWW} {W} 
\tttregle {825} {R} {GWWWWWYWWW} {W} 
\tttregle {826} {Y} {YYWWRGYWWW} {Y} 
\tttregle {827} {Y} {YGYWYYRYYY} {Y} 
\tttregle {828} {G} {YRWWYYYWWW} {R} 
\tttregle {829} {Y} {YYWWYWGWWW} {Y} 
\tttregle {830} {R} {YWWWGYWWWW} {Y} 
\tttregle {831} {Y} {YYWWYRYWWW} {Y} 
\tttregle {832} {Y} {YRYWYYYYYY} {Y} 
\tttregle {833} {R} {YYWWYYYWWW} {R} 
\tttregle {834} {Y} {YYWWYWRWWW} {Y} 
\tttregle {835} {Y} {YRWWWWYWWW} {G} 
\tttregle {836} {Y} {YWWWRYWWWW} {Y} 
\tttregle {837} {Y} {YRYWYYYGYY} {Y} 
\tttregle {838} {R} {YYWWGYYWWW} {Y} 
\tttregle {839} {Y} {YGWWYWRWWW} {G} 
\tttregle {840} {G} {YRWWWWYWWW} {R} 
\tttregle {841} {Y} {YYGWYYYRYY} {Y} 
\tttregle {842} {G} {YRWWYWYWWW} {R} 
\tttregle {843} {W} {YYYWYYGYYY} {W} 
\tttregle {844} {Y} {WGWWYYYWWW} {G} 
\tttregle {845} {Y} {YYRWYYYYGY} {Y} 
\tttregle {846} {R} {YYWWGWYWWW} {Y} 
\tttregle {847} {Y} {YYYWYYYYRY} {Y} 
\tttregle {848} {Y} {YYWWRWYWWW} {Y} 
\tttregle {849} {W} {WYYWYYGYYY} {W} 
\tttregle {850} {W} {WGYWYYRYYY} {W} 
\vskip 1pt 
\ligne{\hfill $\downarrow$ in the register\hfill}
\vskip 1pt 
\ligne{\hfill blue locomotive\hfill}
\vskip 1pt 
\tttregle {851} {Y} {WYYWYWMMWW} {Y} 
\tttregle {852} {W} {YWWWMYYWWW} {W} 
\tttregle {853} {Y} {YMYWMYWWWW} {Y} 
\tttregle {854} {M} {YWWWWYYWWW} {M} 
\tttregle {855} {Y} {YWWYWWMWWW} {Y} 
\tttregle {856} {W} {YWWWWMYWWW} {W} 
\tttregle {857} {W} {YMWWWWYWWW} {W} 
\tttregle {858} {Y} {YMWYWWYWWW} {Y} 
\tttregle {859} {M} {YYWWWWYWWW} {M} 
\tttregle {860} {W} {YWWWWYMWWW} {W} 
\tttregle {861} {Y} {YWWYWWWWYW} {Y} 
\tttregle {862} {W} {YWWWYYWWWW} {W} 
\tttregle {863} {Y} {WYYWBWMMWW} {B} 
\tttregle {864} {W} {YWWWMYBWWW} {W} 
\tttregle {865} {B} {WYYWRWMMWW} {R} 
\tttregle {866} {W} {BWWWMYRWWW} {W} 
}
\hfill
\vtop{\leftskip 0pt\parindent 0pt\hsize=80pt
\tttregle {867} {Y} {BMYWMYWWWW} {Y} 
\tttregle {868} {M} {YWWWWYBWWW} {M} 
\tttregle {869} {Y} {BMWYWWYWWW} {B} 
\tttregle {870} {M} {YYWWWWBWWW} {M} 
\tttregle {871} {R} {WYBWYWMMWW} {Y} 
\tttregle {872} {W} {RWWWMYYWWW} {W} 
\tttregle {873} {Y} {RMYWMBWWWW} {Y} 
\tttregle {874} {M} {YWWWWYRWWW} {M} 
\tttregle {875} {B} {RMWYWWYWWW} {R} 
\tttregle {876} {M} {BYWWWWRWWW} {M} 
\tttregle {877} {W} {BWWWWYMWWW} {W} 
\tttregle {878} {Y} {WYRWYWMMWW} {Y} 
\tttregle {879} {Y} {YMYWMRWWWW} {Y} 
\tttregle {880} {R} {YMWBWWYWWW} {Y} 
\tttregle {881} {M} {RYWWWWYWWW} {M} 
\tttregle {882} {W} {RWWWWBMWWW} {W} 
\tttregle {883} {W} {WMWWWWRWWW} {W} 
\tttregle {884} {Y} {BWWYWWWWYW} {B} 
\tttregle {885} {Y} {YMWRWWYWWW} {Y} 
\tttregle {886} {W} {YWWWWRMWWW} {W} 
\tttregle {887} {B} {RWWYWWWWYW} {R} 
\tttregle {888} {W} {BWWWYYWWWW} {W} 
\tttregle {889} {R} {YWWBWWWWYW} {Y} 
\tttregle {890} {W} {RWWWYBWWWW} {W} 
\tttregle {891} {Y} {BWWWWWRWWW} {Y} 
\tttregle {892} {Y} {YWWRWWWWYW} {Y} 
\tttregle {893} {W} {YWWWYRWWWW} {W} 
\tttregle {894} {R} {YYWWWWWWWW} {Y} 
\tttregle {895} {Y} {RWWWWWYWWW} {Y} 
\ligne{\hfill green locomotive\hfill}
\tttregle {896} {Y} {WYYWGWMMWW} {G} 
\tttregle {897} {W} {YWWWMYGWWW} {W} 
\tttregle {898} {G} {WYYWRWMMWW} {R} 
\tttregle {899} {W} {GWWWMYRWWW} {W} 
\tttregle {900} {Y} {GMYWMYWWWW} {Y} 
\tttregle {901} {M} {YWWWWYGWWW} {M} 
\tttregle {902} {Y} {GMWYWWYWWW} {G} 
\tttregle {903} {M} {YYWWWWGWWW} {M} 
\tttregle {904} {R} {WYGWYWMMWW} {Y} 
\tttregle {905} {Y} {RMYWMGWWWW} {Y} 
\tttregle {906} {G} {RMWYWWYWWW} {R} 
\tttregle {907} {M} {GYWWWWRWWW} {M} 
\tttregle {908} {W} {GWWWWYMWWW} {W} 
\tttregle {909} {W} {WMWWWWGWWW} {W} 
\tttregle {910} {R} {YMWGWWYWWW} {Y} 
}
\hfill}

\ligne{\hfill
\vtop{\leftskip 0pt\parindent 0pt\hsize=80pt
\tttregle {911} {W} {RWWWWGMWWW} {W} 
\tttregle {912} {Y} {GWWYWWWWYW} {G} 
\tttregle {913} {G} {RWWYWWWWYW} {R} 
\tttregle {914} {W} {GWWWYYWWWW} {W} 
\tttregle {915} {R} {YWWGWWWWYW} {Y} 
\tttregle {916} {W} {RWWWYGWWWW} {W} 
\tttregle {917} {Y} {GWWWWWRWWW} {Y} 
\vskip 1pt 
\ligne{\hfill $\uparrow$ in the register\hfill}
\vskip 1pt 
\ligne{\hfill blue locomotive\hfill}
\vskip 1pt 
\tttregle {918} {W} {WWWYYYWWYY} {W} 
\tttregle {919} {W} {WWWWYWYWWW} {W} 
\tttregle {920} {W} {WYYWWWYYWW} {W} 
\tttregle {921} {Y} {WYYWWWYWWW} {Y} 
\tttregle {922} {Y} {YMYYWWYWYW} {Y} 
\tttregle {923} {M} {YYWYYYYWYY} {M} 
\tttregle {924} {Y} {YYWYWWYWWW} {Y} 
\tttregle {925} {W} {MWWWYYYWWW} {W} 
\tttregle {926} {W} {WYWWYYYYWW} {W} 
\tttregle {927} {W} {YWYWWYYYYY} {W} 
\tttregle {928} {W} {YWYWWBYYWW} {W} 
\tttregle {929} {Y} {WWWWWWWWWW} {Y} 
\tttregle {930} {W} {WWWYYYWWBY} {W} 
\tttregle {931} {W} {BWYWWRYYWW} {W} 
\tttregle {932} {W} {WWWBYYWWRY} {W} 
\tttregle {933} {W} {RWYWWYBYWW} {W} 
\tttregle {934} {W} {WWWRYYWWYB} {W} 
\tttregle {935} {W} {YWYWWYRYWW} {W} 
\tttregle {936} {W} {WWWYBYWWYR} {W} 
\tttregle {937} {W} {WYWWYYBYWW} {W} 
\tttregle {938} {W} {WWWYRBWWYY} {W} 
\tttregle {939} {W} {YWYWWYYYWW} {W} 
\tttregle {940} {W} {WBWWYYRYWW} {W} 
\tttregle {941} {W} {WWWYYRWWYY} {W} 
\tttregle {942} {W} {WRWWYYYBWW} {W} 
\tttregle {943} {W} {WYYWWWBYWW} {W} 
\tttregle {944} {W} {WYWWYYYRWW} {W} 
\tttregle {945} {W} {WBYWWWRYWW} {W} 
\tttregle {946} {Y} {WYYWWWBWWW} {Y} 
\tttregle {947} {Y} {YBWWWYWWWW} {B} 
\tttregle {948} {W} {WRYWWWYBWW} {W} 
\tttregle {949} {Y} {WBYWWWRWWW} {Y} 
\tttregle {950} {B} {YRWWWYWWWW} {R} 
\tttregle {951} {W} {WYYWWWYRWW} {W} 
\tttregle {952} {Y} {WYWWRYWWWW} {Y} 
\tttregle {953} {Y} {WRYWWWYWWW} {Y} 
\tttregle {954} {R} {YYWWWYWWWW} {Y} 
}
\hfill
\vtop{\leftskip 0pt\parindent 0pt\hsize=80pt
\tttregle {955} {W} {YRWWWWYWWW} {W} 
\tttregle {956} {W} {WYRWWWYYYW} {W} 
\vskip 1pt 
\ligne{\hfill green locomotive\hfill}
\vskip 1pt 
\tttregle {957} {W} {YWYWWGYYWW} {W} 
\tttregle {958} {W} {WWWYYYWWGY} {W} 
\tttregle {959} {W} {GWYWWRYYWW} {W} 
\tttregle {960} {W} {WWWGYYWWRY} {W} 
\tttregle {961} {W} {RWYWWYGYWW} {W} 
\tttregle {962} {W} {WWWRYYWWYG} {W} 
\tttregle {963} {W} {WWWYGYWWYR} {W} 
\tttregle {964} {W} {WYWWYYGYWW} {W} 
\tttregle {965} {W} {WWWYRGWWYY} {W} 
\tttregle {966} {W} {WGWWYYRYWW} {W} 
\tttregle {967} {W} {WRWWYYYGWW} {W} 
\tttregle {968} {W} {WYYWWWGYWW} {W} 
\tttregle {969} {W} {WGYWWWRYWW} {W} 
\tttregle {970} {Y} {WYYWWWGWWW} {Y} 
\tttregle {971} {Y} {YGWWWYWWWW} {G} 
\tttregle {972} {W} {WRYWWWYGWW} {W} 
\tttregle {973} {Y} {WGYWWWRWWW} {Y} 
\tttregle {974} {G} {YRWWWYWWWW} {R} 
\vskip -4pt
\ligne{\hrulefill}
\ligne{\hfill particular cases\hfill}
\vskip -4pt 
\ligne{\hrulefill}
\ligne{\hfill $n=0$\hfill}
\vskip -1pt 
\ligne{\hfill incrementing\hfill}
\vskip -1pt 
\tttregle {975} {Y} {WYYYYWMMYW} {Y} 
\tttregle {976} {Y} {YMWWYYYWWW} {Y} 
\tttregle {977} {Y} {WYYYBWMMYW} {B} 
\tttregle {978} {B} {WYYYRWMMYW} {R} 
\tttregle {979} {Y} {BMWWYYYWWW} {B} 
\tttregle {980} {R} {WYBBYWMMMW} {Y} 
\tttregle {981} {B} {RMWWMBYWWW} {Y} 
\tttregle {982} {M} {RYWWWWRWWW} {M} 
\tttregle {983} {W} {RWWWWWMWWW} {W} 
\tttregle {984} {W} {BWWWWMWWWW} {M} 
\tttregle {985} {B} {RMYYWYBYYY} {R} 
\tttregle {986} {M} {BBWWYYRWWW} {M} 
\tttregle {987} {W} {MWWWWWBWWW} {Y} 
\tttregle {988} {Y} {WYYRYWMMMW} {Y} 
\tttregle {989} {Y} {YMWMMRYWWY} {Y} 
\tttregle {990} {W} {YWWWWMMWWW} {W} 
\tttregle {991} {M} {RYWWMYYWWW} {M} 
\tttregle {992} {Y} {WYYWYWMMMW} {Y} 
\tttregle {993} {Y} {YMWYMWYWWY} {Y} 
\tttregle {994} {Y} {WYYWYWMMYW} {Y} 
\tttregle {995} {Y} {YMWYYWYWWY} {Y} 
}
\hfill
\vtop{\leftskip 0pt\parindent 0pt\hsize=80pt
\tttregle {996} {W} {YWYWWYYBWW} {W} 
\tttregle {997} {W} {YWYWWYYRWW} {W} 
\vskip 1pt 
\ligne{\hfill no decrementing\hfill}
\vskip 1pt 
\tttregle {998} {Y} {WYYYGWMMYW} {M} 
\tttregle {999} {M} {WYYYRWMMYW} {M} 
\tttregle {1000} {W} {MWWWMYRWWW} {W} 
\tttregle {1001} {Y} {MMYWMYWWWW} {G} 
\tttregle {1002} {M} {YWWWWYMWWW} {M} 
\tttregle {1003} {Y} {MMWWYYYWWW} {Y} 
\tttregle {1004} {Y} {MYYYWRYYYY} {Y} 
\tttregle {1005} {R} {MWYWWWYWWW} {Y} 
\tttregle {1006} {W} {RYYYWYMYYY} {W} 
\tttregle {1007} {M} {WGYYYWMYYW} {M} 
\tttregle {1008} {W} {MWWWMGYWWW} {W} 
\tttregle {1009} {G} {MMYWYYWWWW} {R} 
\tttregle {1010} {M} {GWWWWYMWWW} {M} 
\tttregle {1011} {Y} {GWWYWWMWWW} {G} 
\tttregle {1012} {W} {GWWWWYYWWW} {W} 
\tttregle {1013} {W} {YMWWWWGWWW} {W} 
\tttregle {1014} {Y} {MYWWYYGWWW} {Y} 
\tttregle {1015} {Y} {YGWWWWMWWW} {M} 
\tttregle {1016} {Y} {MYYYWYYYYY} {Y} 
\tttregle {1017} {Y} {MWYWWWYWWW} {Y} 
\tttregle {1018} {W} {YYYYWYMYYY} {W} 
\tttregle {1019} {M} {WRYYYWMMYW} {Y} 
\tttregle {1020} {W} {MWWWMRYWWW} {W} 
\tttregle {1021} {R} {MMGWMYWWWW} {Y} 
\tttregle {1022} {M} {RWWWWGMWWW} {M} 
\tttregle {1023} {G} {RWWYWWMWWW} {R} 
\tttregle {1024} {W} {RWWWWMGWWW} {W} 
\tttregle {1025} {W} {GMWWWWRWWW} {W} 
\tttregle {1026} {Y} {MMWWYYRWWW} {Y} 
\tttregle {1027} {M} {YRWWWWMWWW} {M} 
\tttregle {1028} {Y} {YMRWMYWWWW} {Y} 
\tttregle {1029} {M} {YWWWWRYWWW} {M} 
\tttregle {1030} {R} {YWWGWWMWWW} {Y} 
\tttregle {1031} {W} {YWWWWMRWWW} {W} 
\tttregle {1032} {Y} {YWWRWWMWWW} {Y} 
\tttregle {1033} {R} {YWYWWWWWWW} {Y} 
\vskip 2pt 
\ligne{\hfill $n = 1$\hfill}
\vskip 2pt 
\ligne{\hfill incrementing\hfill}
\vskip 1pt 
\tttregle {1034} {Y} {WYYWBWMMYW} {B} 
\tttregle {1035} {B} {WYYWRWMMYW} {R} 
\tttregle {1036} {Y} {BMWYYWYWWY} {B} 
\tttregle {1037} {R} {WYBWYWMMYW} {Y} 
}
\hfill}

\ligne{\hfill
\vtop{\leftskip 0pt\parindent 0pt\hsize=80pt
\tttregle {1038} {B} {RMWYYWYWWY} {R} 
\tttregle {1039} {Y} {WYRWYWMMBW} {Y} 
\tttregle {1040} {R} {YMWMBWYWWM} {Y} 
\tttregle {1041} {W} {RWWWWMMWWW} {W} 
\tttregle {1042} {Y} {WYYWYWMMRW} {Y} 
\tttregle {1043} {Y} {YMWYRWYWWM} {Y} 
\tttregle {1044} {Y} {YMWYWWYWWM} {Y} 
\tttregle {1045} {Y} {YMWYWWYWWY} {Y} 
\vskip 1pt 
\ligne{\hfill decrementing\hfill}
\vskip 1pt 
\tttregle {1046} {Y} {WYYWGWMMYW} {G} 
\tttregle {1047} {G} {WYYWRWMMYW} {R} 
\tttregle {1048} {Y} {GMWYYWYWWY} {G} 
\tttregle {1049} {R} {WYGWYWMMYW} {Y} 
\tttregle {1050} {G} {RMWYYWYWWY} {R} 
\tttregle {1051} {Y} {WYRWYWMMMW} {Y} 
\tttregle {1052} {R} {YMWYMWYWWY} {Y} 
\tttregle {1053} {W} {RWWWWYMWWW} {W} 
\tttregle {1054} {Y} {WYYYYWMMMW} {Y} 
\tttregle {1055} {Y} {YMWWMYYWWW} {Y} 
}
\hfill
\vtop{\leftskip 0pt\parindent 0pt\hsize=80pt
\ligne{\hfill $n = 2$\hfill}
\vskip 2pt 
\ligne{\hfill decrementing\hfill}
\vskip 1pt 
\tttregle {1056} {M} {YYWWWWYYWW} {M} 
\tttregle {1057} {Y} {MYMWYWYWWW} {Y} 
\tttregle {1058} {M} {YWWWWWYWWW} {M} 
\tttregle {1059} {Y} {WYWWYWMWWW} {Y} 
\tttregle {1060} {Y} {MWYWGYWWYW} {G} 
\tttregle {1061} {M} {YYWWWWGYWW} {M} 
\tttregle {1062} {Y} {MGMWYWYWWW} {Y} 
\tttregle {1063} {M} {YWWWWWGWWW} {M} 
\tttregle {1064} {G} {MWYWRYWWYW} {R} 
\tttregle {1065} {M} {GYWWWWRYWW} {M} 
\tttregle {1066} {Y} {MRMWYWGWWW} {Y} 
\tttregle {1067} {M} {YWWWWWRWWW} {M} 
\tttregle {1068} {R} {MWGWYYWWYW} {Y} 
\tttregle {1069} {M} {RYWWWWYYWW} {M} 
\tttregle {1070} {Y} {MYMWYWRWWW} {Y} 
\tttregle {1071} {Y} {MWRWYYWWMW} {Y} 
\tttregle {1072} {Y} {MWYYYYWWMW} {Y} 
}
\hfill}
\begin{tab}\label{t_rules}
\leurre
Table of the rules used by the cellular automaton. 
\end{tab}


\end{document}